\documentclass[12pt, a4paper]{article}

\usepackage[english]{babel}
\usepackage{authblk}
\usepackage{setspace}
\usepackage{geometry}
\usepackage{bbm, dsfont}
\usepackage{array,colortbl,xcolor}
\usepackage{lipsum}
\usepackage{tabu}
\usepackage{lscape}
\usepackage{afterpage}
\usepackage{lscape}
\usepackage{placeins}
\usepackage{mathtools}
\usepackage{lineno}
\usepackage{epstopdf}
\usepackage{setspace}
\usepackage{pbox}
\usepackage{amsmath}
\usepackage{booktabs}
\usepackage{rotating}
\usepackage{tabularx}
\usepackage{caption}
\usepackage{subcaption}
\usepackage{array}
\usepackage{graphicx}
\usepackage{epstopdf}
\usepackage{prettyref}
\usepackage{multirow}
\usepackage{lipsum}
\usepackage{endnotes}
\usepackage{longtable,tabulary}
\usepackage{ltablex}
\usepackage[T1]{fontenc}
\usepackage{inputenc}
\usepackage[toc,page]{appendix}
\usepackage[hidelinks]{hyperref}
\usepackage{pifont}
\usepackage{fleqn}
\usepackage{txfonts}
\usepackage{longtable}
\usepackage{makecell}
\usepackage{rotating}
\usepackage{color}
\usepackage{amsmath,amsfonts,epsfig,epstopdf,array}
\usepackage{booktabs,dcolumn}
\newcolumntype{d}{D{.}{.}{2.3}}           % alignment on decimal marker
\usepackage{algorithm}
\usepackage{authblk}
\usepackage{multicol}
\usepackage{algorithmic}

\graphicspath{{Figures/}{TexFiles/}}

\newcommand{\bfw}{\boldsymbol{w} }

\newcommand{\bfmu}{\boldsymbol{\mu} }

\newcommand{\bfSigma}{\boldsymbol{\Sigma} }

\newcommand{\bfv}{\boldsymbol{v}}
\newcommand{\bfe}{\boldsymbol{e}}
\newcommand{\bfalpha}{\boldsymbol{\alpha}}

\newcommand{\bflambda}{\boldsymbol{\lambda}}

\newcommand{\prox}{\text{prox}}

\newcommand{\R}{\mathbb{R}}

\DeclareMathOperator*{\argmin}{arg\,min}

\usepackage{natbib}
 \bibpunct[, ]{(}{)}{,}{a}{}{,}%

\title{Sparse Portfolio Selection via the sorted $\ell_1$ - Norm}

\author[1]{Philipp J.~Kremer\thanks{philipp.kremer@ebs.edu}} 
\author[2]{Sangkyun Lee} 
\author[3]{Ma\l{}gorzata Bogdan}  
\author[1]{Sandra Paterlini}

\affil[1]{EBS Universit\"at f\"ur Wirtschaft und Recht,
Chair of Financial Econometrics and Asset Management, Gustav-Stresemann-Ring 3, 65189 Wiesbaden}
\affil[2]{Division of Computer Science, College of Computing, Hanyang University ERICA, Ansan, Republic of Korea}
\affil[3]{Department of Mathematics, University of Wroclaw, Wroc{\l}aw, Poland}

\begin{document}%
%%%%%%%%%%%%%%%%%

\maketitle
\begin{abstract}
We introduce a financial portfolio optimization framework that allows us to automatically select the relevant assets and estimate their weights by relying on a sorted $\ell_1$-Norm penalization, henceforth SLOPE. Our approach is able to group constituents with similar correlation properties, and with the same underlying risk factor exposures. We show that by varying the intensity of the penalty, SLOPE can span the entire set of optimal portfolios on the risk-diversification frontier, from minimum variance to the equally weighted. To solve the optimization problem, we develop a new efficient algorithm, based on the Alternating Direction Method of Multipliers. Our empirical analysis shows that SLOPE yields optimal portfolios with good out-of-sample risk and return performance properties, by reducing the overall turnover through more stable asset weight estimates. Moreover, using the automatic grouping property of SLOPE, new portfolio strategies, such as SLOPE-MV, can be developed to exploit the data-driven detected similarities across assets.
\end{abstract}%

% Keywords

\newpage
%%%%%%%%%%%%%%%%%%%%%%%
\section{Introduction}%
%%%%%%%%%%%%%%%%%%%%%%%

The development of successful asset allocation strategies requires the construction of portfolios that perform well out-of-sample, provide diversification benefits, and are cheap to maintain and monitor. When setting up quantitative portfolio selection strategies, the problem is then one of statistical model selection and estimation, that is the identification of the assets in which to invest and the determination of the optimal weight for each asset. In 1952, Harry Markowitz laid the foundation for modern portfolio theory by introducing the mean-variance optimization framework. Assuming that asset returns are normally distributed, such model requires only two input estimates: the vector of expected returns and the expected covariance matrix of the assets. By solving a quadratic optimization problem, the investor can then find the optimal portfolio allocation by minimizing the portfolio expected risk, for a given level of expected return. Although Markowitz's model has been widely criticized, it is the backbone of the vast majority of portfolio optimization frameworks that continue to be applied widely and is still largely used in practice, especially in fintech companies as part of their robo-advisory (see e.g. \cite{Kolm2014}).\\
One of the major shortcomings of the mean-variance approach is the fact that optimized weights are highly sensitive to estimation errors and to the presence of multicollinearity in the inputs. In particular, it is acknowledged that estimating the expected returns is more challenging than just focusing on risk minimization and thereby looking for the portfolios with minimum risk, i.e. the so-called global minimum variance portfolios (GMV) \citep{Merton1980, Chopra1993, Jagannathan2003}. But even in the GMV set-up, the sample covariance matrix might exhibit estimation error that can easily accumulate, especially when dealing with a large number of assets \citep{Michaud1989, Ledoit2003, DeMiguel2009b, Fan2012}.\\
Furthermore, as a result of multicollinearity and extreme observations, the Markowitz's set-up often leads to undesirable and unrealistic extreme long and short positions, which can hardly be implemented in practice due to regulatory and short selling constraints \citep{Shefrin2000, DeMiguel2009, Boyle2012, Roncalli2013}. An ideal portfolio then has conservative asset weights, which are stable in time, to avoid high turnover and transaction costs, while still promoting the right amount of diversification and being able to control the total amount of shorting.\\
A natural approach to solving this problem is to extend the Markowitz optimization framework by using a penalty function on the weight vector, typically the norm, whose intensity is controlled by a tuning parameter $\lambda$. Probably, the most recent successful approach using convex penalty functions, that can explicitly control for the total amount of shorting, while avoiding to invest in the entire asset universe, is the Least Absolute Shrinkage and Selection Operator (LASSO) introduced by \citet{Tibshirani1996}. \\
In the portfolio context, the LASSO framework typically relies on adding to the Markowitz formulation a penalty proportional to the $\ell_{1}$-Norm\footnote{Let $\bfw = [w_{1}, w_{2}, ...., w_{k}]'$ be the portfolio weight vector, then the $\ell_{q}$-Norm is defined as: $||\bfw||_{q} = (\sum_{i=1}^{k} |w_{i}|^{q})^{\frac{1}{q}}$, with $0< q < \infty$. If $q=1$, then  $\ell_{1} = \sum_{i=1}^{k} |w_{i}|$ (LASSO), while for $q=2$ we have $||\bfw||_{2} = (\sum_{i=1}^{k} w_{i}^{2})^{1/2}$ (RIDGE). Note that $\ell_{q}$ with $0<q<1$ is not a norm but a quasi norm.} on the asset weight vector \citep{Brodie2009, DeMiguel2009a, Carrasco2012, Fan2012}. \cite{DeMiguel2009a} provide a general framework that nests regularized portfolio strategies based on the $\ell_{1}$-Norm with the approaches introduced by \cite{Ledoit2003} and \cite{Jagannathan2003}, and advocate their superior performance in an out of sample setting. \cite{Brodie2009} and \citet{Fan2012} show that LASSO results in constraining the gross exposures, can be used to implicitly account for transaction costs, and sets an upper bound on the portfolio risk depending just on the maximum estimation error of the covariance matrix. Moreover, the shrinkage covariance estimation of \cite{Jagannathan2003}, obtained by adding a no-short sale constraint (the so-called GMV long only (GMV-LO)), can be considered a special case of the LASSO.\\
Next to the LASSO penalty, \cite{Brodie2009} and \cite{DeMiguel2009} also consider a portfolio with an $\ell_{2}$-Norm penalty on the weight vector, known in statistical literature as RIDGE \citep{Hoerl1988}. Although, the RIDGE penalty stabilizes the mean-variance optimization, as it controls for multicollinearity, the shape of the penalty does not promote sparsity, leading to portfolios with an undesirably large number of active positions \citep{Carrasco2012}.\\
Despite its appealing properties, the LASSO has reported shortcomings of (a) large biased coefficient values \citep{Gasso2010, Fastrich2015}, of (b) reduced recovery of sparse signals when applied to highly dependent data, like crisis periods \citep{Giuzio2016a}, and of (c) randomly selecting among equally correlated coefficients \citep{Bondell2008}. Moreover, it is ineffective in presence of short selling (i.e. $w_{i} \geq 0$) and  budget constraint (i.e., $\sum_{i=1}^{k} w_{i} = 1$), as the  $\ell_{1}$-Norm is then just equal to 1.\\
To overcome these limitations, non-convex penalties, like the $\ell_{q}$-Norm, the log and the SCAD penalty have recently gained increased attention in the portfolio literature \citep{Gasso2010, Fastrich2014, Fastrich2015, Xing2014, Chen2016}.
Among these penalties, the $\ell_{q}$-Norm with $0<q<1$ has shown remarkable theoretical and empirical properties compared to the classic LASSO \citep{Saab2008, Fastrich2014, Chen2016, Giuzio2016a}.\\
In particular, \cite{Fernholtz1998} show that the $\ell_{q}$-Norm can be interpreted as a measure of diversification of a portfolio. As such, the $\ell_{q}$ - Norm attains its maximum for an equally weighted portfolio and its minimum for a portfolio completely invested in one single asset. The resulting allocations outperform those based on the LASSO, especially in the presence of highly dependent data. However, from an optimization standpoint, adding non-convex penalties to the minimum variance framework results in optimization problems that are NP-hard. Thus, solutions are typically computed by relying on heuristics or local optimizer, which might not be efficient and are not guaranteed to converge to the global optimum. %\red{ \cite{Chen2016} are the first who provide a theoretical results for finding the optimal portfolio with regard to the Sharpe Ratio and in the presence of the $\ell_{q}$-penalty.}
In this paper, we focus on the class of convex penalty functions and extend the literature on regularization methods in various ways:\\
First, we introduce the \textit{Sorted $\ell_{1}$ Penalized Estimator} (SLOPE), as a new penalty function within the mean-variance portfolio optimization framework. The SLOPE penalty takes the form of a sorted $\ell_{1}$ - Norm, in which each asset weight is penalized individually using a vector of tuning parameters, $\boldsymbol{\lambda}_{SLOPE} = (\lambda_{1},\lambda_{2}, \ldots,\lambda_{k})$, where $\lambda_{1} \geq \lambda_{2} \geq ... \geq \lambda_{k} \geq 0$. This sequence of $\bflambda_{SLOPE}$ values is decreasing, and the largest weight corresponds to the highest regularization parameter, such that SLOPE penalizes the weights according to their rank magnitude.
In a 2-dimensional setting the penalty takes an octagonal shape (see Figure \ref{Slope_NoBudget}), is singular at the origin and promotes the grouping of variables, that is some asset weights are assigned the same value. The penalty thus combines the two favorable properties of the $\ell_{1}$-Norm and the $\ell_{\infty}$-Norm\footnote{Given a weight vector $\bfw$ with $k$ elements, the $\ell_{\infty} = ||\bfw||_{\infty} = max(w_{1}, .....,w_{k})$.} that promote sparsity and variables grouping, respectively. So far, the study of SLOPE have mostly focused on orthogonal settings and on genetic applications, where it is used to choose relevant genes from a large group of possible explanatory factors. Our work shows that in portfolio optimization, together with the budget constraint (i.e. $\sum_{i=1}^{k} w_{i} = 1$), SLOPE continues to shrink the active weights even when short sales are restricted (i.e. $w_{i} \geq 0,\ \forall \ i=1,...,k$), spanning the diversification frontier from the GMV-LO up to the equally weighted (EW) portfolio.\\
% In fact, in a simulated environment we show that SLOPE shows comparable, or even better risk properties than the LASSO.\\
Second, we introduce a new optimization algorithm to solve the mean-variance portfolio optimization problem with the sorted $\ell_1$ regularization and linear constraints on the asset weights. The algorithm uses the ideas of variable splitting and the Alternating Direction Method of Multipliers (ADMM) framework~\citep{Pow69,Hes69,BoyP11}. Using a mathematically equivalent reformulation of the original problem, the algorithm can use existing implementations of proximal operators~\citep{ParB14} associated with the $\ell_1$, the sorted $\ell_1$, and even other regularizers. Furthermore, Section \ref{AlgoComp} of the Appendix shows that the ADMM provides a more efficient alternative for solving the LASSO optimization problem, then the state-of-art Cyclic Coordinate Descend (CyCoDe) algorithm.\\
Third, we are, to our knowledge, the first to investigate the properties of SLOPE under a realistic factor model, which assumes that all assets can be represented as linear combination of a small number of hidden risk factors, as e.g. in \cite{Fan2008}. In the set-up of classical multiple regression, in which the explanatory variables are assumed independent, \cite{Bogdan2013, Bogdan2015} and \cite{Su2016} provide extensive evidence of SLOPE superior model selection and estimation properties. Superior estimation properties of SLOPE are further supported by the results of \cite{Bellec2} and \cite{Bellec1}, which show that contrary to LASSO, SLOPE is asymptotically optimal for the general class of design matrices satisfying the modified Restricted Eigenvalue condition.\\
Further \cite{Bondell2008} and \cite{Figueiredo2014} investigate the properties of SLOPE and its predecessor OSCAR (Octagonal Shrinkage and Selection Operator, \cite{Bondell2008}) in the situation when regressors are strongly correlated. \cite{Bondell2008} apply OSCAR to agricultural data, showing that the method successfully forms predictive clusters, which can then be analyzed according to their individual characteristics. Further simulations and theoretical results are provided by \cite{Figueiredo2014} illustrating the ``clustering'' properties of the \textit{ordered weighted $\ell_{1}$ - Norm} (OWL) in the linear regression framework with strongly correlated predictors. However, none of these works addresses the interesting situation in which the correlation structure results from the dependency of the explanatory variables on a few hidden factors and on financial real-world data.
Recently, \cite{Xing2014} applied OSCAR to mean-variance portfolio optimization, together with a linear combination of the $\ell_1$- and the $\ell_\infty$-Norms. They advocate the method for its ability to identify portfolios that attain higher Sharpe Ratios and lower turnovers than those resulting from traditional approaches like the GMV and the GMV-LO portfolios, but did not point out the clumping property of the OSCAR. With SLOPE, we consider a generalized framework that nests the GMV, GMV-LO, the LASSO, the $\ell_{\infty}$ - Norm and the approach of \cite{Xing2014}.\\
We analyze the properties of SLOPE both on simulated and on real world data. The purpose of the simulations is to investigate the properties of SLOPE when the data generating mechanism is completely known, so that the results can be compared to the so-called \textit{oracle} solution. These simulations show that SLOPE reduces the estimation errors on the portfolio weights and groups assets depending on the same risk factors. This grouping behaviour might then allow the investor to select individual constituents from these clusters based on her preferences and asset-specific properties, developing new investment strategies as SLOPE-MV, introduced in Section \ref{Data}. \\
For the real world data analysis, we use monthly returns of the 10- and 30-Industry portfolios (Ind), as well as the 100 Fama French (FF) portfolios formed on Size and Book-to-Market from K. French, from 1970 to 2017, and daily returns from the S\&P100 and S\&P500 from 2004 to 2016. 
%We compare SLOPE in a minimum variance framework to state-of-the-art portfolio optimization procedures, including the GMV, the GMV-LO, the GMV with an added $\ell_{1}$-Norm constraint on the weight vector (LASSO), the GMV with an added $\ell_{2}$-Norm constraint (RIDGE), as well as the equal risk contribution portfolio (ERC) and the na\"ive equally weighted portfolio (EW).
%In particular, we introduce two new investment strategies: (a) SLOPE with an added long-only constraint (SLOPE-LO), and (b) a portfolio strategy in which we first identify active assets with SLOPE-LO, and then pick from the identified groups the one with minimum volatility and then, estimate their weights by the GMV-LO portfolio.
%We report results for a rolling window approach for windows of size $\tau =120$ ($\tau=500$) for the monthly (daily) data, rebalancing the portfolio monthly and comparing the resulting allocations using performance and risk diversification metrics. To evaluate whether the resulting portfolios are statistically different from each other, we complement our analysis with the robust hypothesis tests for the Sharpe Ratio (SR) and the Variance, introduced by \citet{Ledoit2003} and \cite{Ledoit2011}.\\
Our results show that the risk of the SLOPE portfolio is comparable to or smaller than the risk of the LASSO portfolio. Also, we observe that SLOPE outperforms the LASSO yielding better risk- and weight diversification measures. In fact, the sorted $\ell_{1}$- Norm is able to span the entire risk-diversification frontier, starting from the GMV, via the GMV-LO up to the EW. The investor can then select the portfolio with the risk-diversification trade-off that best fits her preferences.\\
%Finally, SLOPE's grouping structure opens the door to numerous portfolio strategies in which we can select an arbitrary number of assets from the same group, which is exposed to the same risk factors, e.g. according to any investors preferences or asset characteristics. In our study, these newly created portfolio strategies reduce the overall turnover.\\
The above mentioned characteristics establish SLOPE as a new attractive portfolio construction alternative, capable of controlling short sales and identify groups of assets, offering then the possibility to implement individual views, which goes beyond the standard statistical shrinkage or regularization approaches.\\
The paper is structured as follows: Section 2 introduces our methodology and discusses the properties of SLOPE. %Section 3 describes the newly developed optimization algorithm, based on the ADMM.
Section 3 analyses the behavior of SLOPE in simulated environments, while Section 4 focuses on the empirical results. Section 5 concludes.

%%%%%%%%%%%%%%%%%%%%%%%%%%%%%%%%%%%%%%%%%%%%%%%%%%%%%%%%%%%%%%%%%%
\section{Sparse Portfolio Selection via the Sorted $\ell_1$-Norm}%
%%%%%%%%%%%%%%%%%%%%%%%%%%%%%%%%%%%%%%%%%%%%%%%%%%%%%%%%%%%%%%%%%%
\cite{Markowitz1952} pioneered the idea that investors should consider both risk and return, instead of just focusing on those assets that offered the largest increase in value given today's prices. Central to his argument is the notion of diversification, i.e. not only the individual securities' risk are important, but also their relationship with other assets that is how does the performance of an asset moves with or against the performance of the other assets in the market. Then, given $k$ jointly normally distributed asset returns $R_1, \ldots, R_k$, with expected value vector $\bfmu = [\mu_{1}, ..., \mu_{k}]'$ and covariance matrix $\bfSigma$, the \cite{Markowitz1952} portfolio selection problem can be stated as the following optimization:%
\begin{gather}\label{minvar}%
\min_{\bfw \in \mathbb{R}^k} \frac{\phi}{2} \bfw'\bfSigma \bfw - \bfmu'\bfw \\
\text{s.t.} \sum_{i=1}^k w_i = 1
\end{gather}
\noindent
where $\sigma_{p}^{2} = \bfw\bfSigma \bfw$ is the portfolio risk, $\bfmu' \bfw$ is the portfolio return and $\phi > 0$ is the coefficient of relative risk aversion \citep{Markowitz1952, Fan2012, Li2015}. \\ % If we assume the utility function of the investor to be $U(W) = 1 - exp(-\gamma W)$ with $W$ being the wealth of the investor and $\gamma$ being the coefficient of absolute risk aversion,} then the solution to (\ref{minvar}) maximizes the utility of a mean-variance investor. \\
Despite the appeal of being a quadratic optimization problem, the standard Markowitz model is often criticized, as it leads to extreme and unstable optimal weights. This results from multicollinearity of returns, which increases especially during crisis periods, and from input changes due to extreme data, leading to the consequent accumulation of estimation error. \\
Besides shrinkage and robust estimation methods (see e.g., \cite{Ledoit2004, Welsch2007, Kolm2014}), one of the most recent and interesting approaches is based on statistical regularization. This approach modifies the optimization problem (\ref{minvar}) by adding a penalty function, $\rho_{\lambda}(\bfw)$, that depends on the norm of the asset weights vector and is typically chosen to promote sparsity in the optimal weight vector. An additional parameter $\lambda$ controls the impact of $\rho(\bfw)$ and thereby the amount of shrinkage applied to the weights vector and the level of sparsity. The optimization problem can be stated as:

\begin{gather}\label{eq:minreg}
\min_{\bfw \in \mathbb{R}^k} \frac{\phi}{2} \bfw'\bfSigma \bfw - \bfmu'\bfw + \rho_{\lambda}(\bfw) \\
\text{s.t.} \sum_{i=1}^k w_i = 1
\end{gather}

\noindent
The simplest approach is the LASSO, which considers as a penalty function the $\ell_1$- Norm of the asset weights vector ($\rho_{\lambda}(\bfw)= \lambda \times \sum_{i=1}^{k} |w_i|$, with $\lambda$ being a scalar). Then the larger $\lambda$, the sparser is the solution. The resulting optimization problem is still convex, while promoting model selection and estimation in a single step. %This approach was introduced to the statistical literature by \cite{Tibshirani1996}, who proposed the Least Absolute Shrinkage and Selection Operator (LASSO) for fitting the multiple regression model, and was subsequently (but much later) applied in the finance literature (e.g. see \cite{DeMiguel2009a}, \cite{Brodie2009} and \cite{Fan2012}). Since then, the research is high on the agenda: LASSO allows to introduce sparsity in the optimal portfolio, by selecting only a subset of active weights, and allows to reduce the error accumulation effects as compared to the standard Markowitz setup. Still, LASSO has known shortcomings, such that it tends to produce biased estimates especially for large weights and that it cannot deal with high correlation settings, as asset active weights are then randomly chosen from strongly correlated ones. \\
From a financial perspective, LASSO is interpreted as a gross exposure constraint (i.e. a constraint on the total amount of shorting) or a way to account for transaction costs \citep{Brodie2009}. However, it is not effective in the presence of both a budget ($\sum_{i=1}^{k} w_{i} = 1$) and a no-short selling (i.e., $w_i \geq 0$) constraint, as the $\ell_1$-norm is then simply equal to 1.
We propose a more general approach that within a single optimization algorithm allows us to encompass the original LASSO, the OSCAR of \cite{Bondell2008}, and the methods proposed in \cite{Xing2014} based on the combination of $\ell_1$ and $\ell_\infty$ penalties. In fact, we penalize the weights vector by considering as $\rho_{\lambda}(\bfw)$ the sorted $\ell_{1}$-Norm, defined as:

\begin{gather}
\rho_\lambda(\bfw) := \sum_{i=1}^k \lambda_i |w|_{(i)} = \lambda_{1} |w|_{(1)} + \lambda_{2}|w|_{(2)} + ... + \lambda_{k}|w|_{(k)}\\ \text{s.t.  } \lambda_1 \ge \lambda_2 \ge \dots \lambda_k \ge 0 \text{ and }  |w|_{(1)} \ge |w|_{(2)} \ge \dots |w|_{(k)} \;\;, \label{penalty}
\end{gather}
where $|\bfw|_{(i)}$ denotes the $i$th largest element in absolute value of the vector $\bfw$.
The sorted $\ell_{1}$-Norm was originally introduced in \cite{Bogdan2013, Bogdan2015} to construct the Sorted $\ell_{1}$ Penalized Estimator (SLOPE) for selection of 
%orthogonal 
explanatory variables for the multiple regression model. It was also developed independently by \cite{Zeng2014} as Ordered Weighted $\ell_{1}$ Norm (OWL). To our knowledge, this is the first work in financial portfolio selection that applies SLOPE and discusses its grouping properties, while also introducing a new optimization algorithm.

%%%%%%%%%%%%%%%%%%%%%%%%%%%%%%%%%%%%%%
\subsection{Geometric Interpretation}%
\begin{figure} [h!]
\caption{Geometric Representation of Penalty Functions.}\label{Slope_NoBudget}
\scalebox{0.9}{
\begin{tabular}{ccc}
 $\lambda_1=\lambda_2>0$ & $\lambda_1>\lambda_2=0$ & $\lambda_1>\lambda_2>0$ \\
\includegraphics[scale=.64]{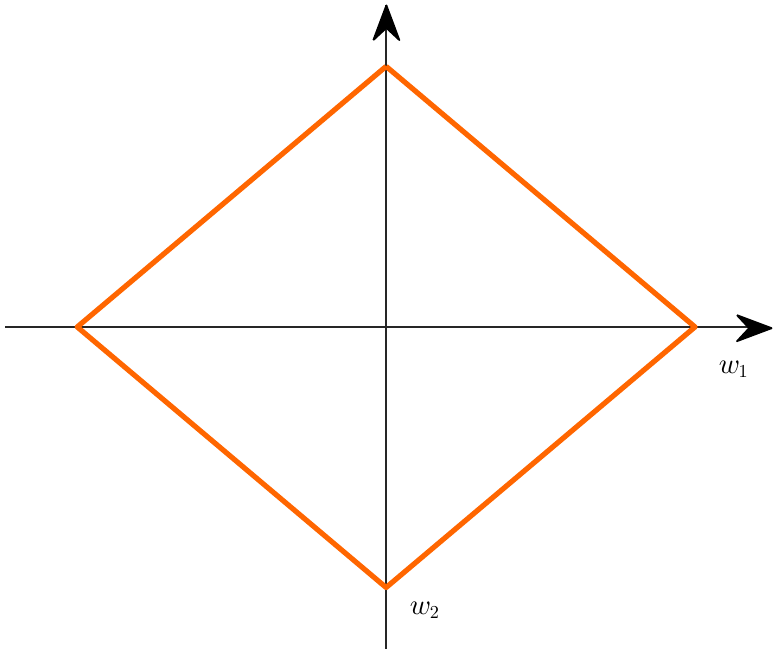}&
\includegraphics[scale=.64]{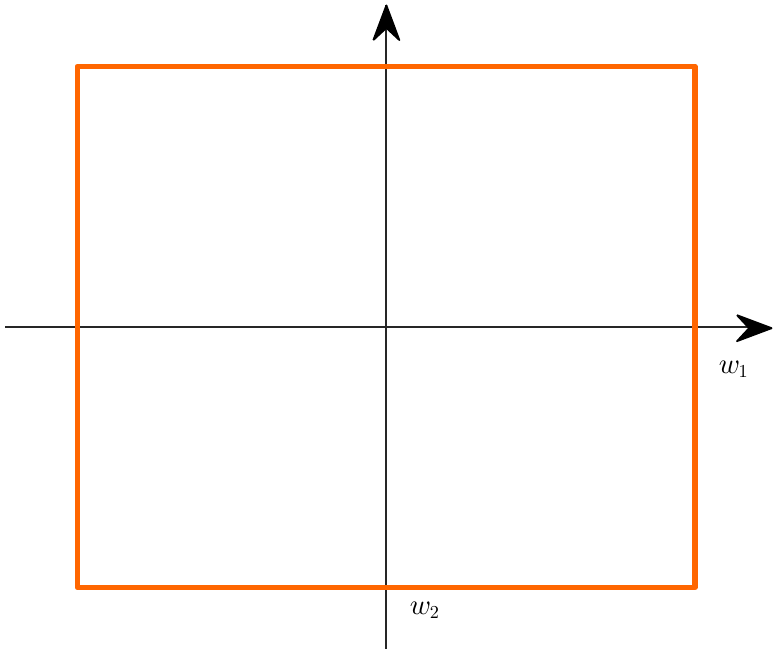} &
\includegraphics[scale=.64]{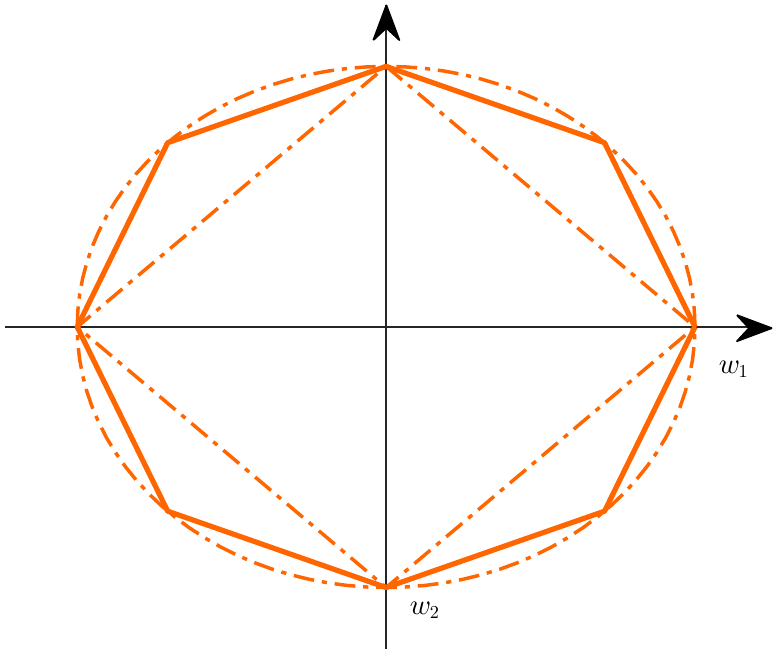} \\
  (a) & (b) & (c)\\
\end{tabular}
}
 \captionsetup{font=scriptsize,labelfont=scriptsize, width=\textwidth}
     \caption*{For two asset weights $\bfw = [w_{1}\ w_{2}]'$, the Figure shows the unit spheres for different SLOPE sequences: (a) the LASSO $\ell_1$ sphere, when $\lambda_{1} = \lambda_{2} > 0$, (b) the $\ell_{\infty}$ sphere, when $\lambda_{1} > \lambda_{2} = 0$ and (c) the SLOPE sphere, when $\lambda_{1} > \lambda_{2} > 0$. The dashed lines in (c) represent LASSO and RIDGE $\ell_2$ balls.}
\end{figure}

%%%%%%%%%%%%%%%%%%%%%%%%%%%%%%%%%%%%%%
Compared to most of the other regularization methods, SLOPE does not rely on a single tuning parameter $\lambda$, but rather on a non-increasing sequence $\bflambda_{SLOPE} = (\lambda_{1}, \lambda_{2}, \ldots,\lambda_{k})$, with $\lambda_{1}\geq \lambda_{2} \geq \ldots \geq \lambda_{k}\geq 0$. This sequence is aligned to the sorted weight vector, such that the largest absolute weight is penalized with the largest tuning parameter. Consequently, the sequence of $\lambda$ parameters gives a natural interpretation of importance to the asset weights, besides providing full flexibility in recapturing the profiles of the $\ell_{1}$- and $\ell_{\infty}$- Norms, as well as of their linear combinations. Figure \ref{Slope_NoBudget} shows a simple set-up with two assets and the respective shapes of spheres (i.e. the set points for which $\rho_\lambda(\bfw)=c$) that we obtain, depending on how the sequence $\bflambda_{SLOPE} = (\lambda_{1},\lambda_{2})$ is chosen. As shown in Panel (a), if all tuning parameters have the same value, while still being larger than zero (i.e. $\lambda_{1} = \lambda_{2} > 0$), the SLOPE sphere coincides with the well studied diamond shape of the LASSO penalty. Through its singularity at the origin, LASSO promotes sparse solutions that set one of the two assets' weights exactly equal to zero. On the other hand, choosing $\lambda_{2}=0$ and $\lambda_{1}>0$, yields the regularization term of the $\ell_{\infty}$-Norm. The respective shape, as shown in Panel (b), takes the form of a square and promotes the grouping of asset weight estimates, i.e. it encourages solutions under which both asset weights are assigned exactly the same value.\\
Given these two extreme cases, Panel (c) of Figure \ref{Slope_NoBudget} shows the octagonal shape of SLOPE, obtained by using a decreasing sequence of lambda values, with $\lambda_{1} > \lambda_{2} > 0$. The penalty thus combines both properties of the LASSO and the $\ell_{\infty}$ penalties. It is able to set some weights exactly equal to zero due to its singularity at the origin, or to assign the same value to some of the other weights. Furthermore, it approximates the already well studied RIDGE penalty, which corresponds to a circle in the 2-dimensional set-up. Although RIDGE is still convex, the shape of the penalty does not promote sparsity among the coefficients, leading to undesirable portfolios with a large number of active positions \citep{Carrasco2012, DeMiguel2009a}. Thus, the choice of the lambda sequence for SLOPE provides the investor with the flexibility to choose any of these shapes of the unit sphere and of the corresponding mode of shrinking the dimension of the weight vector.

%%%%%%%%%%%%%%%%%%%%%%%%%%%%%%%%%%%%%%%
%%%%%%%%%%%%%%%%%%%%%%%%%%%%%%%%%

\begin{figure} [h!]
\centering
\caption{Penalty Function in a two asset universe with imposed budget constraint.}\label{Slope_penalty}
\begin{tabular}{c}
\includegraphics[scale=1]{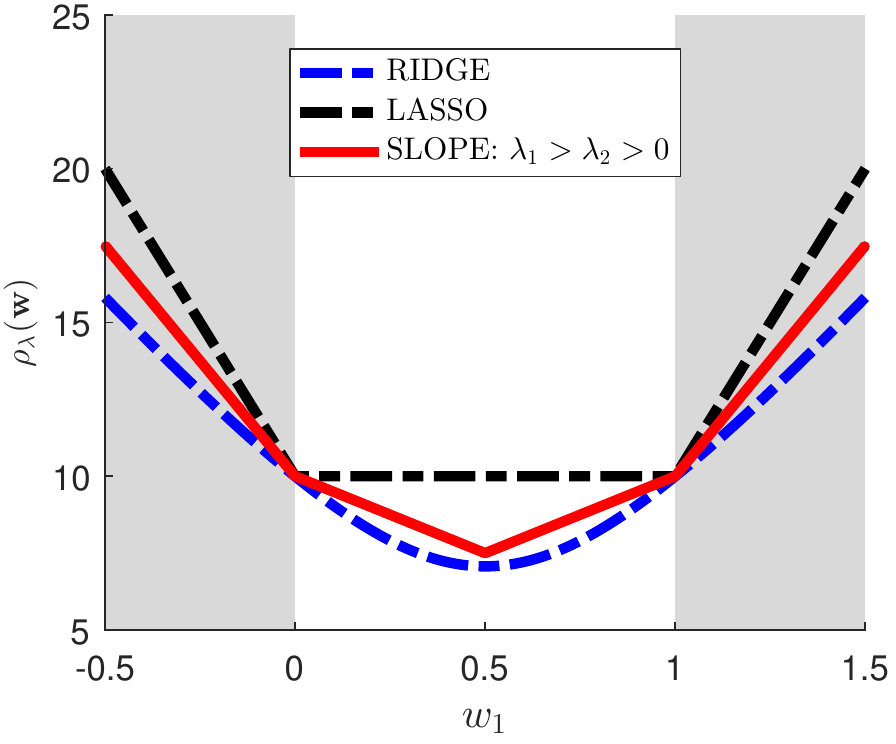} \\
\end{tabular}
 \captionsetup{font=scriptsize,labelfont=scriptsize, width=0.65\textwidth}
     \caption*{The Figure plots the SLOPE coefficient alongside the LASSO ($\ell_{1}-Norm$) and the Ridge penalty ($\ell_{2}-Norm$), for a two asset case and under the condition that $w_{1} + w_{2} = 1$.}
\end{figure}

\noindent
In portfolio optimization, a budget constraint, that requires the weights of the portfolio to sum to one, is imposed. Consequently, we need to understand how the penalties behave in the presence of such an additional constraint. Figure \ref{Slope_penalty} plots the SLOPE penalty, together with the LASSO and the RIDGE penalty for a universe of two assets under the condition that $w_{1} + w_{2} = 1$. We consider the penalty functions in the presence of short sales (gray area) and no short sales (white area).\\
%From the Figure, we can identify that the $\ell_{0.5}$ penalty, in green, reaches its maximum value in the no short sale area for a portfolio that is equally invested in both assets, and attains its minimum for a portfolio completely invested in one of the two assets. This property is well defined in the literature (see i.e. \cite{Fernholtz1998, Fastrich2015, Giuzio2016a}) and gained the $\ell_{q}$-Norm a financial interpretation as a tool to control the amount of diversification in the portfolio, in which diversification is defined as the total number of active positions.
In Figure \ref{Slope_penalty}, we can see that the LASSO penalty (shown in black) is only effective when short sales are permitted, while the presence of the budget constraint makes the penalty ineffective under a short sales restriction.
In contrast, the RIDGE penalty attains its minimum for an equally weighted portfolio when short sales are restricted. Similarly, the SLOPE penalty (shown in red) also reaches its minimum at the equally weighted solution (i.e., $w_{1} = w_{2} = 0.5$). Still, from a financial perspective, we prefer SLOPE over the RIDGE estimator, because it can promote sparsity by exploiting the singularities.

%%%%%%%%%%%%%%%%%%%%%%%%%%%%%%%%%%%%%%%
%%%%%%%%%%%%%%%%%%%%%%%%%%%%%%%%%

\begin{figure}[h!]
\centering
\caption{Sorted $\ell_{1}$-Norm Penalty without budget constraint.}\label{Slope_DesingsBudget}
\begin{tabular}{cc}
\includegraphics[scale=.6]{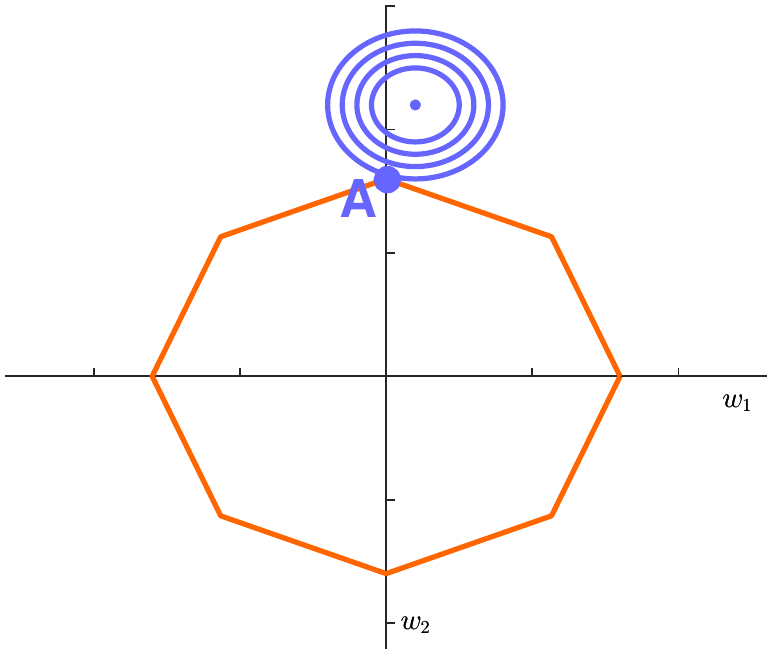}&
\includegraphics[scale=.6]{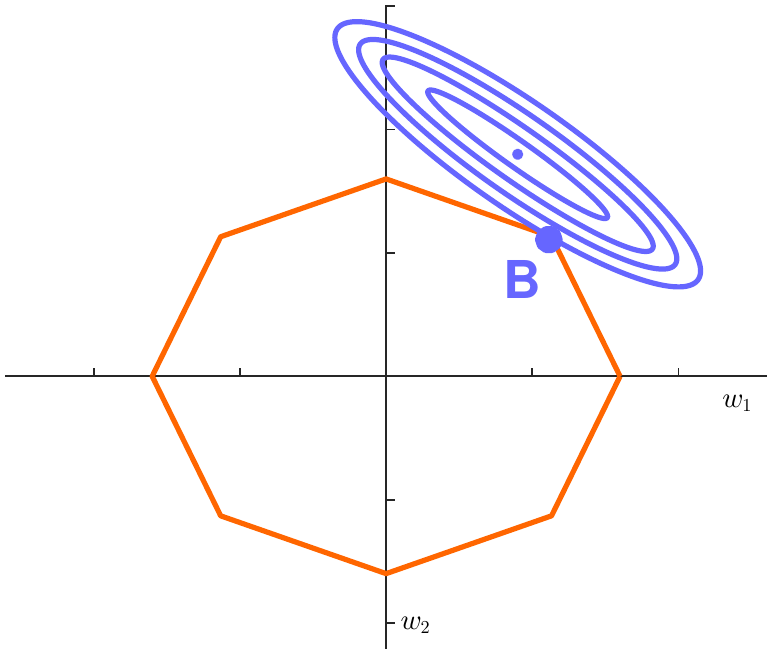}\\
(a) & (b) \\
\end{tabular}
 \captionsetup{font=scriptsize,labelfont=scriptsize, width=0.7\textwidth}
     \caption*{The Figure plots in Panel (a) and (b), the Sorted L1-Norm Penalty (SLOPE) in a 2-dimensional setting, considering orthogonal design and correlated design, respectively.}
\end{figure}

\noindent
Figure \ref{Slope_DesingsBudget} plots the contours of the objective function together with those of SLOPE spheres for the two asset case when we do not impose a budget constraint (i.e. $\sum_{i=1}^{k} w_{i} =1$) and consider orthogonal and correlated designs. As noted before, if only $\lambda_2>0$ %no matter what the sequence of $\bflambda$ is
, SLOPE always has singular points when one of the asset weights is equal to zero, thereby promoting sparsity. When $\lambda_1>\lambda_2>0$, that is, the sequence is monotonically decreasing, then SLOPE has additional singular points, which correspond to $|w_1|=|w_2|$. This is an appealing property in the presence of correlated data. Specifically, as Panel (b) shows, strong correlation between assets lead to the same weights and thereby grouping. This is consistent with portfolio theory, as it is known that, if assets have all the same correlation coefficients, as well as identical means and variances, the EW is the unique optimal portfolio. SLOPE then allows us to automatically group assets with similar properties.
%
%%%%%%%%%%%%%%%%%%%%%%%%%%%%%%
%%%%%%%%%%%%%%%%%%%%%%%%%%%%%%
%

%%%%%%%%%%%%%%%%%%%%%%%%%%%%%%%%%%%%%
\subsection{Optimization Algorithm} %
%%%%%%%%%%%%%%%%%%%%%%%%%%%%%%%%%%%%%

% \red{Sang could you kindly comment here more - many thanks. Would you agree to make the code available on request? If not, please %take out the sentence in red below. Mention ADMM also for LASSO?}
% Sang answered: yes, I'll provide the code on request. 
In this section, we introduce the details of our solution algorithm,
which is based on equivalent reformulations and the Alternating
Direction Method of Multipliers (ADMM)
approach. %This section introduces the Alternating Direction Method of Multipliers (ADMM), which we use to solve the SLOPE mean-variance portfolio optimization problem.
Our algorithm can be used to also solve the LASSO optimization problem, which is a specific instance of SLOPE. In Section \ref{AlgoComp} in the Appendix, we provide a direct comparison of our algorithm to the state-of-art
Cyclic Coordinate Descent (CyCoDe) for LASSO, considering a simulated constant
correlation model.

%%%%%%%%%%%%%%%%%%%%%%%%%%%%%%%%%%%%%%%%%%%%%%%%%%%%%%
\subsubsection*{Reformulation and an ADMM Algorithm.} %
%%%%%%%%%%%%%%%%%%%%%%%%%%%%%%%%%%%%%%%%%%%%%%%%%%%%%%
In order to facilitate applying proximal operators involving
$\rho_\lambda$, we first reformulate \eqref{eq:minreg} - \eqref{penalty} into the following form:
\begin{gather}\label{eq:min2}
\begin{aligned}
\min_{\bfw\in\R^k} &\;\; \frac{\phi}{2} \bfw'\bfSigma \bfw - \bfmu'\bfw + \rho_\lambda(\bfv) \\
\text{s.t.} &\;\; \bfw=\bfv, \\
            &\;\; \sum_{i=1}^k w_i = 1 \;\;,
\end{aligned}
\end{gather}
where $\rho_\lambda(\bfw) := \sum_{i=1}^k \lambda_i |w|_{(i)}$ is the sorted $\ell_1$-Norm corresponding to the sequence $\bflambda_{SLOPE} = (\lambda_1,\dots,\lambda_k)'$ satisfying $\lambda_1 \ge \lambda_2 \ge \dots \lambda_k \ge 0$. To solve \eqref{eq:min2} we design an ADMM algorithm, which is a variant of the augmented Lagrangian scheme that uses partial updates for the dual variables (for detailed discussion of ADMM see e.g. \cite{BoyP11}). In our case the associated augmented Lagrangian is given as:
$$
\begin{aligned}
  \mathcal L_\eta (\bfw,\bfv;\bfalpha,\beta) &= \frac{\phi}{2} \bfw'\bfSigma \bfw - \bfmu' \bfw + \rho_\lambda(\bfv)
  + \bfalpha'(\bfw-\bfv) + \beta (\bfe'\bfw -1)
+ \frac{\eta}{2} \left\{ \|\bfw-\bfv\|^2 + (\bfe'\bfw - 1)^2 \right\},
\end{aligned}
$$
where $\bfalpha \in \R^k$, $\beta \in \R$, $\bfe_{k \times 1} = (1, ...,1)' $, $\boldsymbol{I}_{k \times k}$ is the identity matrix, and $\eta>0$ is a penalty parameter. The ADMM algorithm consists of the updates:\\

\begin{equation}
  \begin{cases}
 w^{j+1} &= \argmin_{\bfw} \;\; \frac{\phi}{2} \bfw'\bfSigma \bfw - (\bfmu - \bfalpha^j - \beta^j \bfe)'\bfw + \frac{\eta}{2} (\|\bfw-\bfv^j\|^2 + (\bfe'\bfw-1)^2) \\
 &= \argmin_{\bfw} \;\; \frac{1}{2} \bfw' ( \phi \bfSigma + \eta (I + \bfe\bfe')) \bfw - (\bfmu - \bfalpha^j - \beta^j \bfe + \eta ( \bfv^j + \bfe))'\bfw \\
 &=  ( \phi \bfSigma + \eta (I + \bfe\bfe'))^{-1} (\bfmu - \bfalpha^j -\beta^j \bfe + \eta ( \bfv^j + \bfe)) \\
 v^{j+1} &= \argmin_{\bfv} \;\; \frac{\eta}{2} \|\bfv-w^{j+1} - 1/\eta \alpha^j \|^2  + \rho_\lambda(\bfv)  \\
 &= \prox_{\rho_{\lambda/\eta}} (w^{j+1} + (1/\eta) \alpha^j )\\
    \alpha^{j+1} &= \alpha^j + \eta( w^{j+1} - v^{j+1} ) \\
    \beta^{j+1} &= \beta^j + \eta( \bfe' w^{j+1} - 1) \;\;,
  \end{cases}
\end{equation}
\vspace{.5cm}
where $\prox_{\rho_{\lambda/\eta}}$ is the proximal operator of the Sorted L-One norm corresponding to the sequence $\lambda/\eta$, provided e.g. in \cite{Bogdan2013,Bogdan2015}. 
%\vspace{.5cm}

%%%%%%%%%%%%%%%%%%%%%%%%%%%%%%%%%%%%%%%%%%%%%%%%%%%%%%%%%%%%
\subsubsection*{A Dual Formulation and the Primal-Dual Gap.} %
%%%%%%%%%%%%%%%%%%%%%%%%%%%%%%%%%%%%%%%%%%%%%%%%%%%%%%%%%%%%

The stopping criterion for our algorithm is based on the Primal-Dual Gap, which we estimate using the following approach. First, taking the infimum over $(\bfw,\bfv)$ of the Lagrangian, we get the dual objective,
$$
\begin{aligned}
  g(\bfalpha,\beta)
 & = \inf_{\bfw} \frac{\phi}{2} \bfw' \bfSigma \bfw - (\bfmu - \bfalpha - \beta \bfe)' \bfw -\beta + \inf_{\bfv} \{ -\bfalpha'\bfv + \rho_{\bflambda}(\bfv) \} \\
 & = \inf_{\bfw} \frac{\phi}{2} \bfw' \bfSigma \bfw - (\bfmu - \bfalpha - \beta \bfe)' \bfw -\beta - \rho_{\bflambda}^*(\bfalpha).
\end{aligned}
$$
From the optimality condition for the infimum over $\bfw$, we have
\begin{gather}\label{eq:opt1}
 w^* = \phi^{-1}\bfSigma^{-1}(\bfmu - \bfalpha - \beta \bfe) .
\end{gather}
Also,
$$
\rho_{\bflambda}^* (\bfalpha) =\sup_{\bfv} \{ \bfalpha^T \bfv - \rho_{\bflambda}(\bfv) \}
= \begin{cases} 0 & \text{ if $ \bfalpha \in C_{\bflambda} $} \\
  +\infty & \text{o.w.}
\end{cases}
$$
where $C_{\bflambda} := \{\bfv : \R^k: \rho_{\bflambda}^D(\bfv) \le 1 \}$ is the unit
sphere defined in the dual norm $\rho_{\bflambda}^D(\cdot)$ of
$\rho_{\bflambda}(\cdot)$.

Plugging-in these, we get the dual problem
$$
\begin{aligned}
  \max_{\bfalpha, \beta} &\;\; - \frac{1}{2\phi} (\bfmu - \bfalpha -\beta \bfe)' \bfSigma^{-1} (\bfmu - \bfalpha -\beta \bfe) - \beta\\
  \text{s.t.} %&\;\; \phi\bfSigma \bfw^* = \bfmu - \bfalpha - \beta \bfe \\
   &\;\; \bfalpha \in C_{\bflambda} .
\end{aligned}
$$

We can estimate the primal-dual gap using \eqref{eq:opt1},
$$
\mathcal G(\bfw^*,\bfv^*,\bfalpha^*,\beta^*) = -(\bfalpha^* + \beta^*\bfe)'\bfw^* + \beta^* + \rho_{\bflambda}(\bfv^*)
%\mathcal G(\bfw^*,\bfv^*,\bfalpha^*,\beta^*) = \frac12 \phi (\bfw^*) \bfSigma \bfw^* -
%\bfmu' \bfw^* + \rho_{\bflambda} (\bfw^*) + \frac{1}{2\phi} (\bfmu - \bfalpha^* -\beta^* \bfe)' \bfSigma^{-1} (\bfmu - \bfalpha^* -\beta^* \bfe) + \beta^*
$$
given the dual feasibility of $\bfalpha^*$, i.e., $\rho_{\bflambda}^D(\bfalpha^*) \le 1$.

\subsubsection*{Bounds on the Objective Function.}
%To investigate the portfolio risk properties of SLOPE, we want to find the optimal weight vector $\bfw = [w_{1}, w_{2}, .... , w_{k}]'$ for the universe of $k$ assets that minimizes (\ref{eq:minreg}) with $\eta_{\lambda}$ equal to the SLOPE penalty given by (\ref{penalty}) and subject to the constraint that the investor is fully invested $(\sum_{i=1}^{k} w_{i} = 1)$.\\
To solve the mean-variance problem, as stated in (\ref{eq:minreg}), the investor needs to provide an estimate of the true covariance matrix of asset returns $\bfSigma$ and of the true mean $\bfmu$, which are in the most simplest form given by the sample covariance matrix $\hat{\bfSigma}$ and the sample mean $\hat{\bfmu}$. However, $\hat{\bfSigma}$ and $\hat{\bfmu}$ might be prone to substantial estimation errors and highly sensitive to outliers. Let us define $M(\bfSigma, \bfmu)= \frac{\phi}{2} \bfw'\bfSigma \bfw - \bfw'\bfmu$, where $\bfw$ is the vector of weights returned by SLOPE. Now, observe that the Sorted L-One Norm satisfies $\rho_{\lambda}(\bfw)\geq \lambda_k ||\bfw||_1$. Thus, as $\lambda_k>0$, we have $||\bfw||_{1} \leq c$, with $c=\frac{\rho_{\lambda}(\bfw)}{\lambda_{k}}$, simple calculations following the results of \cite{Fan2012} for LASSO, easily yield
\begin{gather}
|M(\widehat{\bfSigma}, \hat{\bfmu}) - M(\bfSigma, \bfmu)| \leq \frac{\phi}{2} ||\widehat{\bfSigma} - \bfSigma||_{\infty} \rho_{\lambda}^2(\bfw)/\lambda^2_k + ||\hat{\bfmu} - \bfmu||_{\infty} \rho_{\lambda}(\bfw)/\lambda_k
\end{gather}
where $||\widehat{\bfSigma} - \bfSigma||_{\infty}$ and $||\hat{\bfmu} - \bfmu||_{\infty}$ are the maximum component-wise estimation errors for the covariance matrix and the expected return. This result implies that the difference between the objective functions for the estimated and true vector of parameters decreases as we restrict the Sorted L-One norm of the weight vector.\\
%, and $ w$ is the vector of weights returned by SLOPE. 
It is also important to observe that due to imposing the budget constraint, a higher weight on the penalty sets an upper bound on the total amount of short sales in the portfolio, as $\rho_{\lambda}(\bf w)\geq \lambda_k||\bfw||_{1} = \lambda_k(\bfw^{+} + \bfw^{-})$, with $\bfw^{+} - \bfw^{-}=1$, where $\bfw^{+}= \sum\limits_{w_{i} \geq 0} w_{i}$ and $\bfw^{-} = \sum\limits_{w_{i}<0} w_{i}$ are the gross amount of long and short positions, respectively.
%There exist and inverse proportional relationship between the tuning parameter $\lambda$ and the gross exposure $c$. Choosing $c=\infty$, the constraint is no longer binding, corresponding to a tuning parameter $\lambda = 0$. On the other hand, when $c=1$ the penalty reduces the admissible  there are no short sales allowed and the penalty 

%%%%%%%%%%%%%%%%%%%%%%%%%%%%%% 
\section{Simulation Analysis}%
%%%%%%%%%%%%%%%%%%%%%%%%%%%%%%
In this section, we analyze and explain the effect of SLOPE on the model risk, sparsity and grouping properties by considering simulated data.
Assuming that  $\bfSigma$ is known, we can use the alternative formulation of SLOPE and define 
%$\bfw_{opt} = \min\limits_{\bfw \in \mathbb{R}^{k}} \left\lbrace \frac{\phi}{2} \bfw'\bfSigma \bfw - \bfmu'\bfw \right\rbrace $ 
$\bfw_{opt} = \displaystyle \argmin_{\bfw: \sum_{i=1}^k w_i=1,\; \rho_{\lambda(\bfw)}\leq c}  \frac{\phi}{2} \bfw'\bfSigma \bfw - \bfmu'\bfw$ as the theoretical optimal weights vector, and $\hat{\bfw}_{opt} = \displaystyle \argmin_{\bfw: \sum_{i=1}^k w_i=1,\; \rho_{\lambda(\bfw)}\leq c}   \frac{\phi}{2} \bfw'\hat{\bfSigma} \bfw - \hat{\bfmu}'\bfw$
as the empirical one. Let's then define the \textit{empirical} portfolio risk as $\widehat{Risk}(\hat{\bfw}_{opt}) = \hat{\bfw}_ {opt}' \hat{\bfSigma} \hat{\bfw}_{opt}$, the \textit{actual} portfolio risk as $Risk(\hat{\bfw}_{opt}) = \hat{\bfw}_{opt}' \bfSigma \hat{\bfw}_{opt}$ and the \textit{oracle} portfolio risk as $Risk(\bfw_{opt}) = \bfw_{opt}' \bfSigma \bfw_{opt}$, respectively. Following the proof of Theorem 1 of \cite{Fan2012}, we can easily show that in case when $\lambda_k>0$
%also for SLOPE, 
the pair differences between the three measures are upper bounded by:
% $2||\widehat{\bfSigma} - \bfSigma||_{\infty} \rho^2_{\lambda}(\bfw_{opt})/\lambda^2_k$.\\
 %and the squared $\ell_{1}$ - Norm of the weight vector  (i.e. $||\bfw||_{1}^{2}$) the maximum component-wise estimation error of the covariance matrix $e_{max} = ||\widehat{\bfSigma} - \bfSigma||_{\infty}$ and the squared $\ell_{1}$ - Norm of the weight vector  (i.e. $||\bfw||_{1}^{2}$), \\
%{\it Gosia - it is not clear what is $w$. Is it $w_{opt}$ or $\hat w_{opt}$ ? I modified it below assuming it should be $\hat w_{opt}$.}\\
%are upper bounded by $2 ||\widehat{\bfSigma} - \bfSigma||_{\infty}$
%such that 
\begin{gather}
|Risk(\hat{\bfw}_{opt})  - Risk(\bfw_{opt})| \leq 2 c^2 ||\widehat{\bfSigma} - \bfSigma||_{\infty},  \\
|Risk(\hat{\bfw}_{opt})  - \widehat{Risk}(\hat{\bfw}_{opt})| \leq c^2 ||\widehat{\bfSigma} - \bfSigma||_{\infty}, \\
|Risk(\bfw_{opt})  - \widehat{Risk}(\hat{\bfw}_{opt})|  \leq c^2 ||\widehat{\bfSigma} - \bfSigma||_{\infty}
\end{gather}
The three risk measures allow us to extract different information: The empirical risk is the only one that is known in a practical setting and is estimated from our in-sample data. The actual risk is the one to which investor is truly exposed to, when setting up a portfolio with optimal weights $(\hat{\bfw}_{opt})$, while the oracle risk is the risk the investor could only obtain if she knows $\bfSigma$. As the SLOPE penalty becomes more binding when $\bflambda$ ''increases'', the three risk measures get closer to each other. In the following sections, we consider two simulation set-ups, and show how increasing the SLOPE penalty allows to reduce the estimation error and avoid its accumulation in the portfolio risk.

\noindent
\subsubsection*{Hidden Factor Structure}
\begin{figure}
\centering
\caption{Hidden Factors Correlation Matrix.}\label{Correl_HF}
\begin{tabular}{c}
\includegraphics[scale=.6]{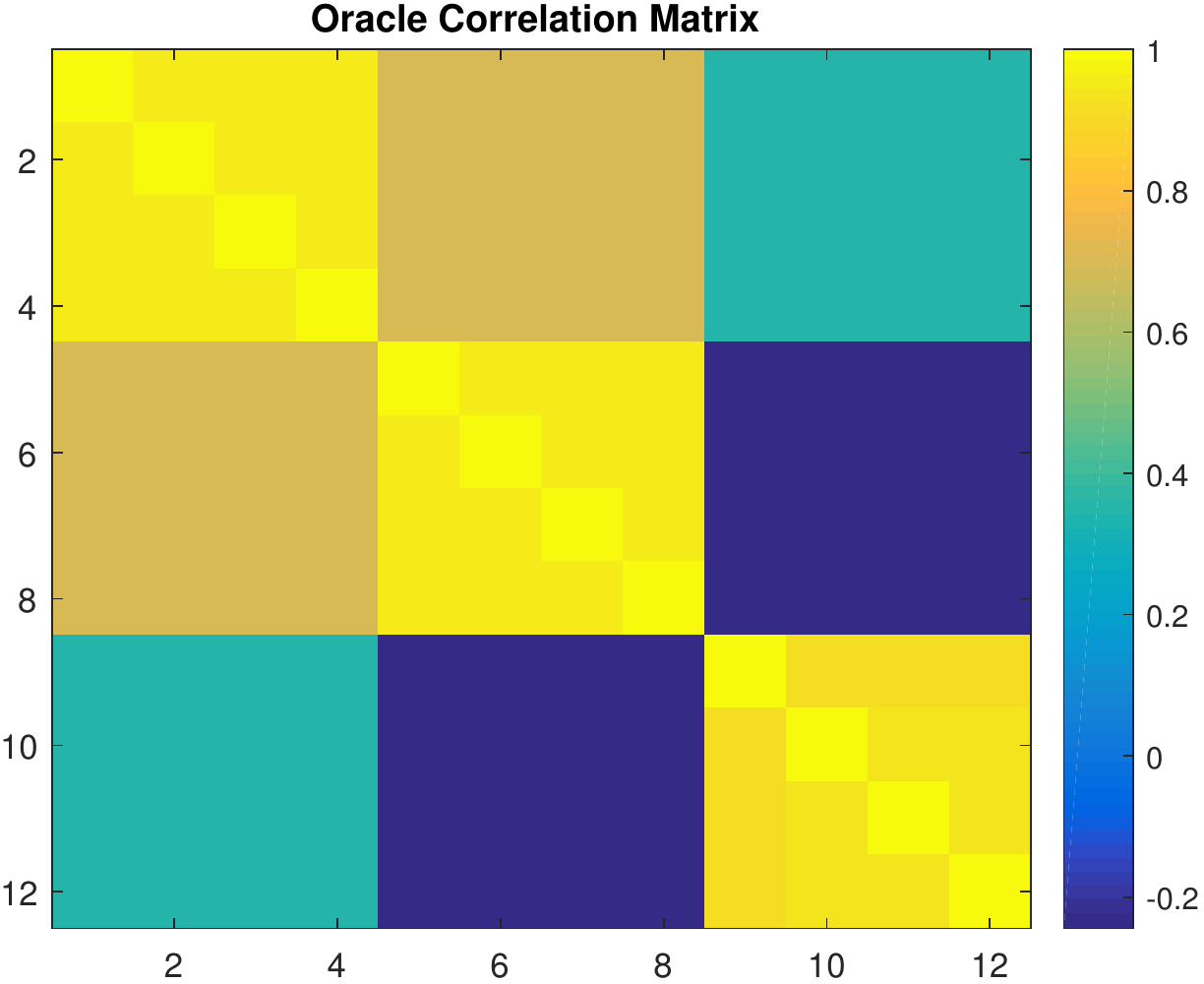}
\end{tabular}
\captionsetup{font=scriptsize,labelfont=scriptsize, width=\textwidth}
     \caption*{The Figure shows the resulting Hidden Factor Correlation Matrix.}
\end{figure}

\begin{figure}[h!]
\centering
\caption{Hidden Factors Minimum-Variance Profile}\label{RiskProfile_HiddenFactors_MinVar}
\scalebox{0.9}{
\begin{tabular}{cc}%
\includegraphics[scale=.6]{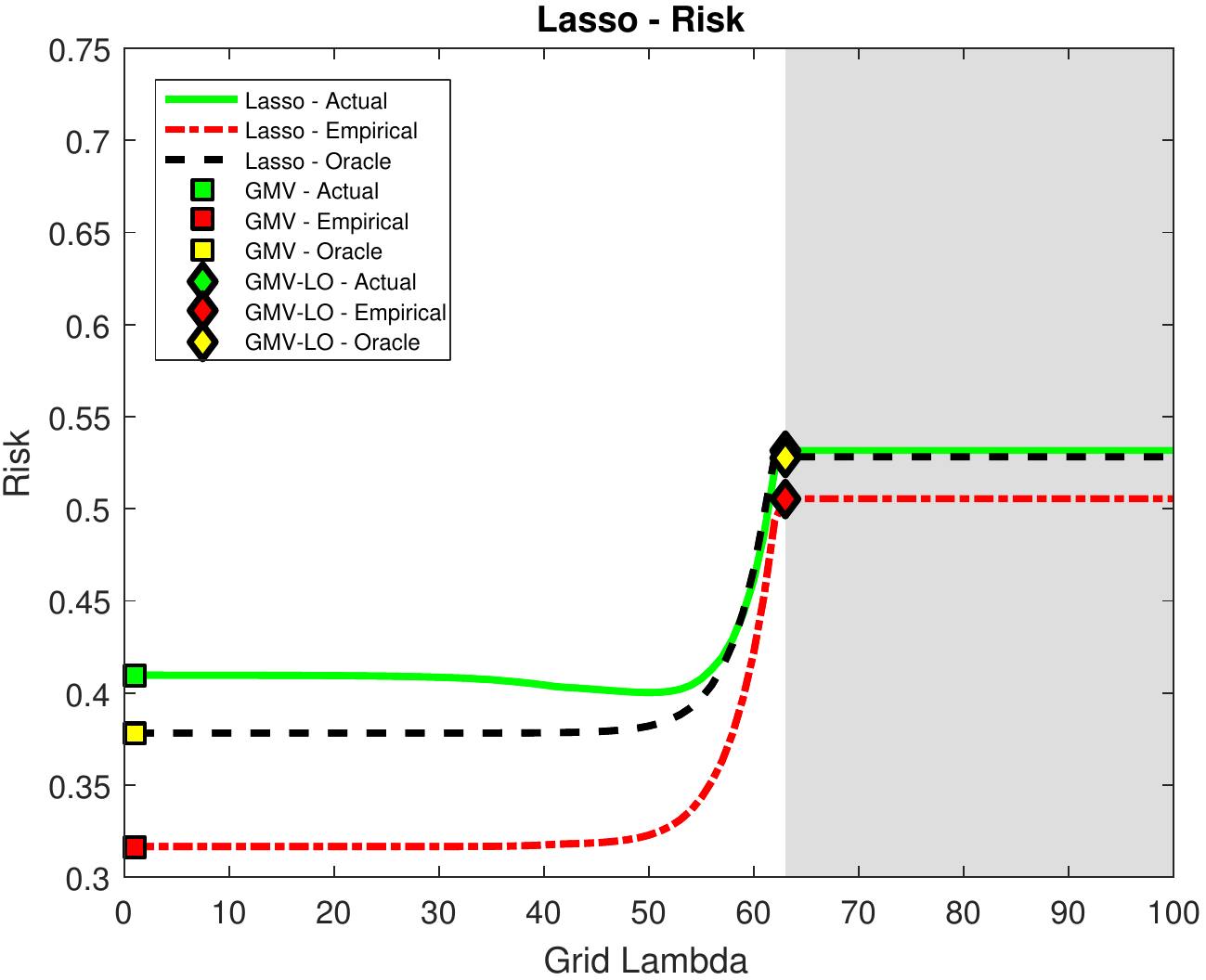} &
\includegraphics[scale=.6]{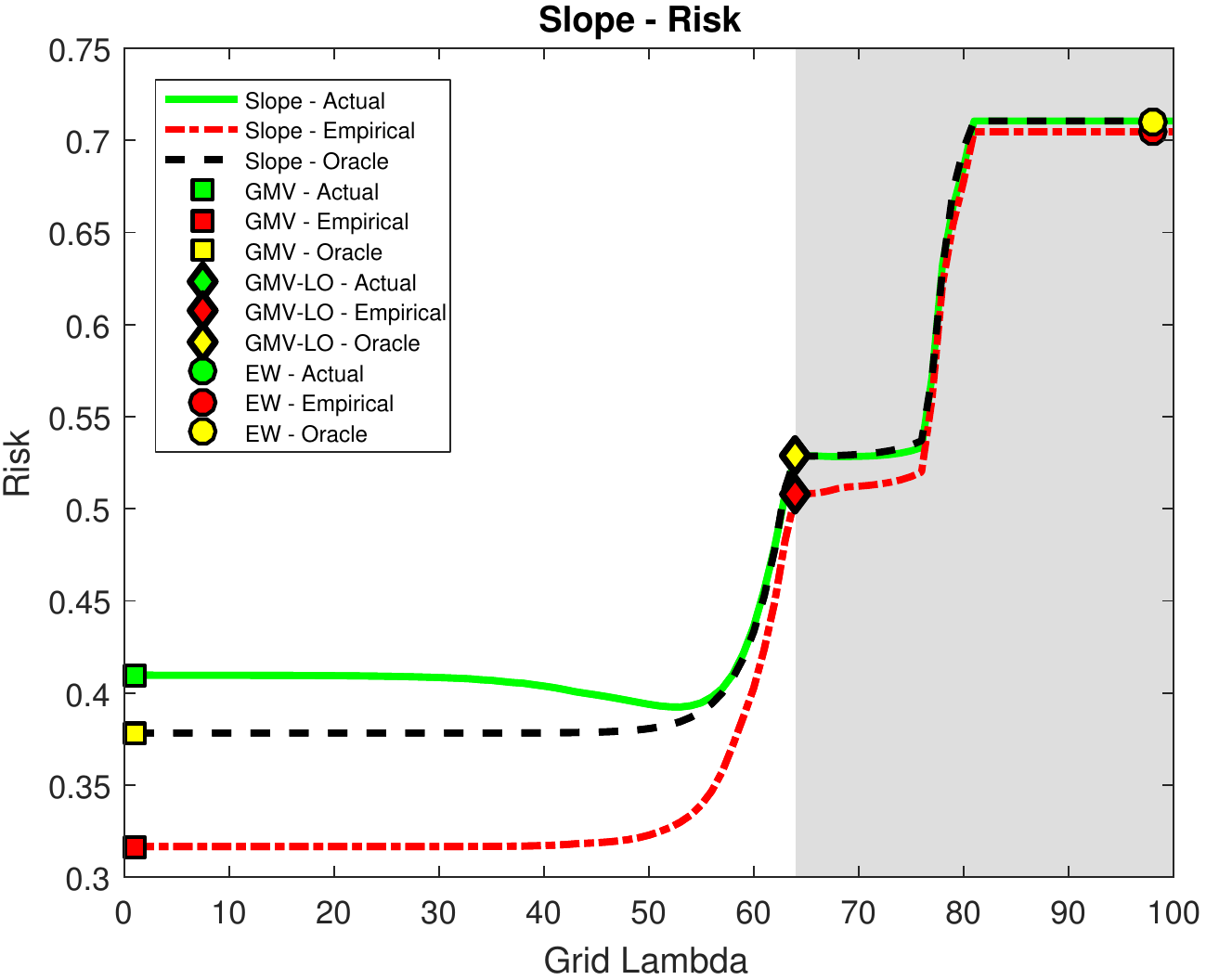}\\
(a) & (b) \\
\includegraphics[scale=.6]{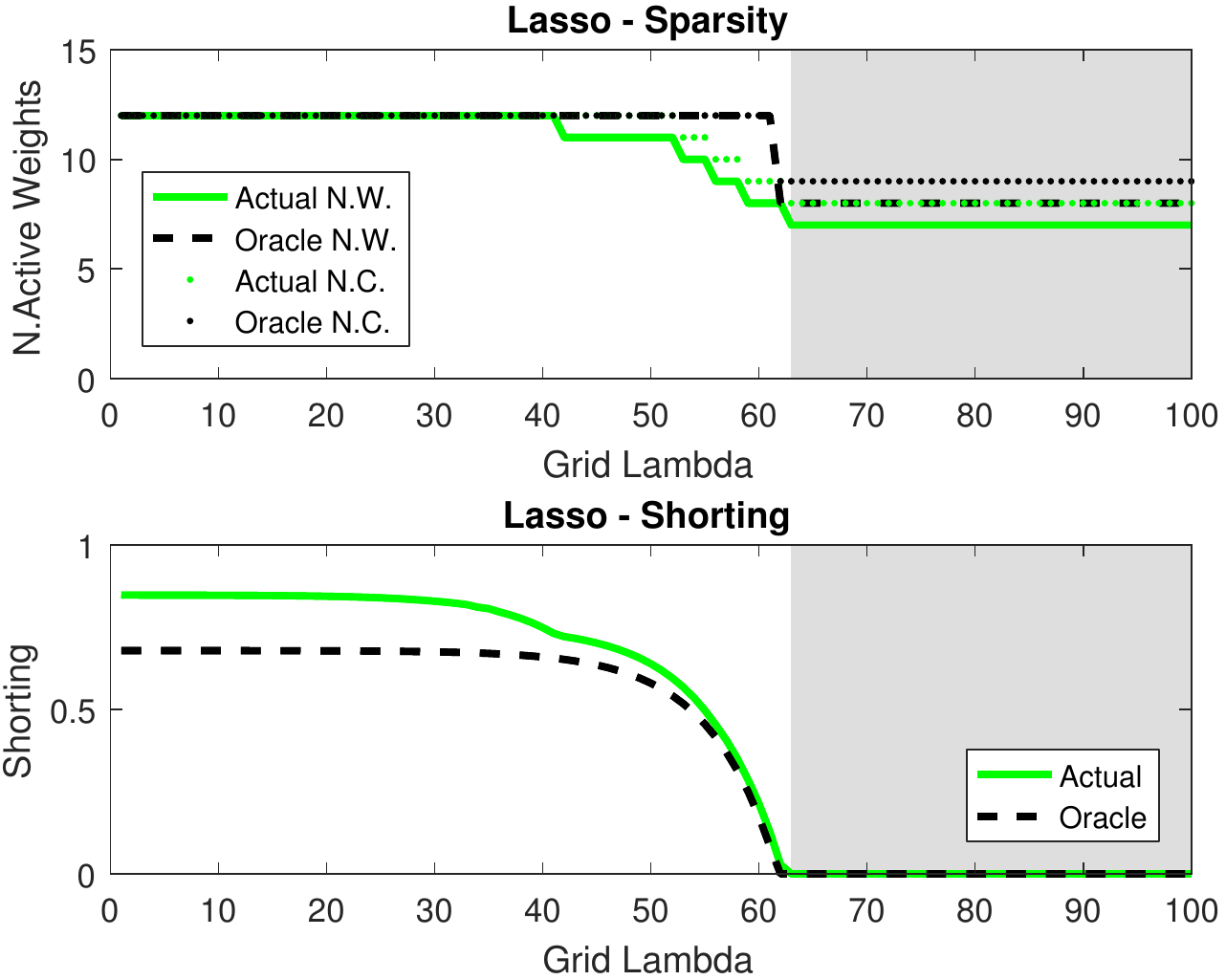} &
\includegraphics[scale=.6]{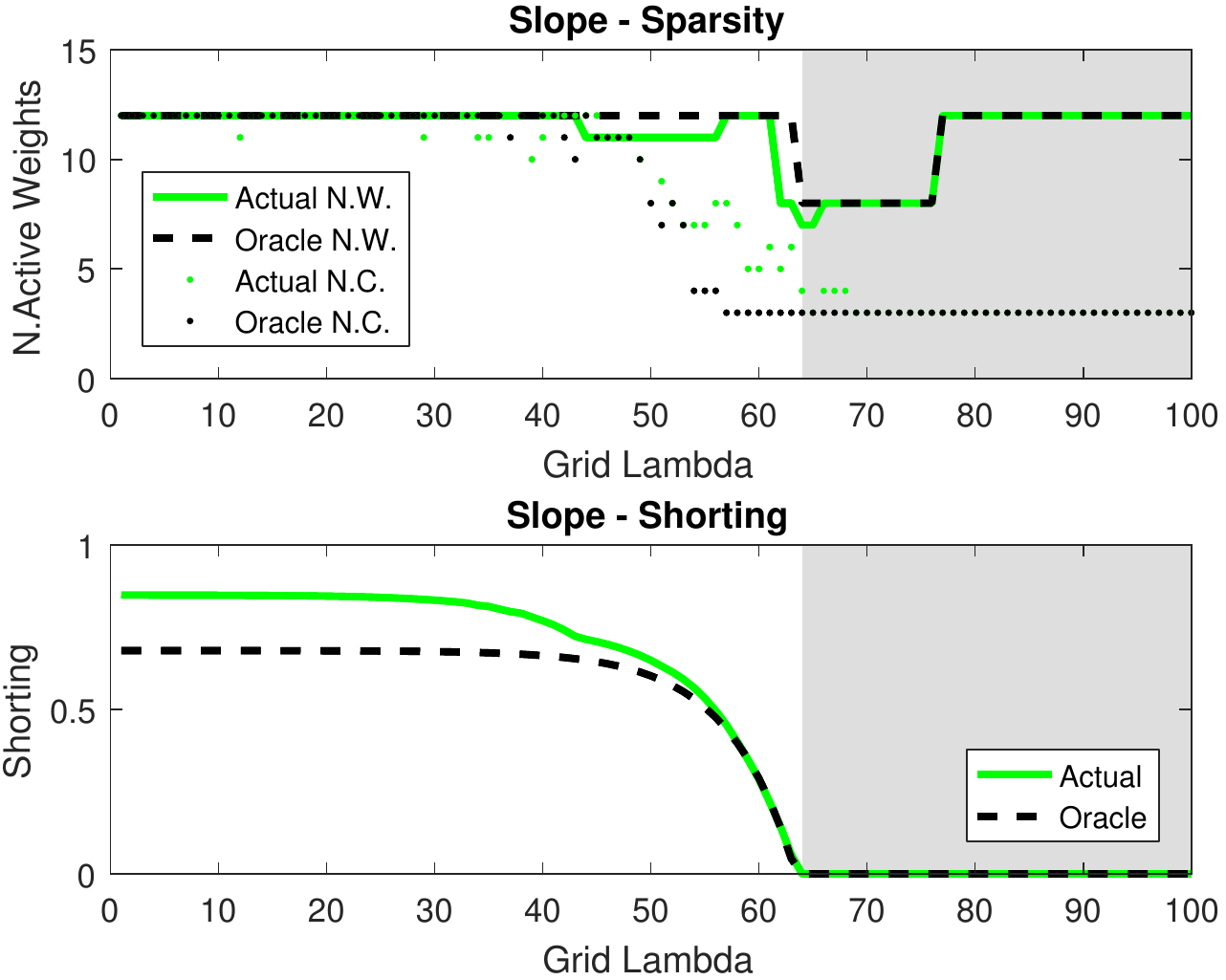}\\
(c) & (d) \\
\end{tabular}
}
\captionsetup{font=scriptsize,labelfont=scriptsize, width=\textwidth}
     \caption*{The Figure shows the Hidden Factor minimum-variance risk profile for the LASSO and  the SLOPE, including in Panel (a) and (b) their actual, empirical and oracle risk profiles together with that of the GMV, GMV-LO and EW solutions, and in Panel (c) and (d) their number of active weights, together with the grouping profile (top) and the total amount of shorting (bottom). All values are computed based on a Hidden Factor Structure, with three risk factors and considering for the exponentially decreasing sequence of lambda parameters, a grid of 100 log spaced starting points, $\lambda_{1}$ from $10^{-5}$ (i.e. x-value = 1) to $10^{2}$ (i.e. x-value = 100).}
\end{figure}

Let us assume that the return of an asset is represented by a linear combination of $r$ risk factors. Furthermore, let  $t$ be the number of observations, $k$ be the number of assets, and $\boldsymbol{F}_{t \times r} = [\boldsymbol{f}_{1} \  \boldsymbol{f}_{2} \ ...\ \boldsymbol{f}_{r}]$, where $\boldsymbol{f}_{i}$ is the $t \times 1$ vector of returns of the $i^{th}$ risk factor. Moreover, let $\boldsymbol{B}_{r \times k}$ be the loading matrix for the individual risk factors. Then, the $t \times k$ matrix of asset returns from the Hidden Factor Model (i.e. $\boldsymbol{R}_{HF}$) can be represented as:
%
%%%%%%%%%%%%%%%%%%%%%%%%%%%%%%
\begin{gather}\label{truemodel}
\boldsymbol{R}_{HF} = \boldsymbol{F} \times \boldsymbol{B} + \boldsymbol{\epsilon}
\end{gather}
%%%%%%%%%%%%%%%%%%%%%%%%%%%%%%
%
where $\boldsymbol{\epsilon}$ is a $t \times k$ matrix of error terms.\\
\noindent
For our first illustration of the performance of SLOPE, we generated the data using the following simplified scenario:
\begin{itemize}
\item $t=50$, $k=12$, $r=3$,
\item the risk factors $f_1$, \ldots, $f_3$ are independent from the multivariate standard normal $N(0, I_{r \times r})$ distribution, with $I_{r \times r}$ being an identity matrix,
\item the vectors of error terms $\epsilon_{i}$, $i=1,\ldots,k$, for each asset are independent from each other and from each of the risk factors and come from the multivariate normal distribution $N(0, 0.05\times I_{r\times r})$
\item the loadings matrix $B_{r\times k}$ is made of exactly four copies of each of the following columns: $[0.77\ 0.64\ 0]'$, $[0.9\ 0\ -0.42]'$ and $[0\ 0.31\ 0.64]'$.
\end{itemize}
\noindent
In this way, we generate three different groups that have the same exposure to the same two risk factors and are thus strongly correlated.\footnote{For the robustness of our results, we tested SLOPE in various set-ups with qualitatively similar results, but restrict ourselves to the most interesting, due to space limitations. The results of the remaining simulations are available from the authors upon request.}\\
Finally, given (\ref{truemodel}), the covariance matrix between different assets $\boldsymbol{\Sigma}_{HF}$ is:
%
%%%%%%%%%%%%%%%%%%%%%%%%%%%%%%
\begin{gather}\label{Covar_Decomp}
\boldsymbol{\Sigma}_{HF} = \boldsymbol{B}' \boldsymbol{B} + 0.05\times I_{k\times k}.
\end{gather}
%%%%%%%%%%%%%%%%%%%%%%%%%%%%%%
%
After generating our $t \times k$ matrix $\boldsymbol{R}_{HF}$ of asset returns from (\ref{truemodel}), we can then estimate $\boldsymbol{\Sigma}_{HF}$, using the sample covariance estimate $\boldsymbol{\hat{\Sigma}}_{HF}$.\\
Figure \ref{Correl_HF} shows the correlation matrix resulting from (\ref{Covar_Decomp}) and assuming that the four assets from each of the three groups are clustered together. 
%That is,
The figure illustrates that our simulation scenario explicitly models a block correlation environment with strong correlation among each of the assets with the same underlying risk factor exposures and low to negative correlations between the assets with a different underlying factor structure.
%
%%%%%%%%%%%%%%%%%%%%%%%%%%%%%%%%%%%%%%%%
%%%%%%%%%%%%%%%%%%%%%%%%%%%%%%%%%%%%%%%%
%
In what follows, we first consider a minimum variance optimization, such that $\hat{\bfmu}=\bfmu= 0$. In a second step, we solve the mean-variance problem, estimating $\bfmu$ 
% assuming that $\hat{\bfmu}$ is the vector of the sample estimates 
by the sample average of the assets returns.
We investigate the behaviour of SLOPE and the LASSO with respect to portfolio risk, when we vary the value of the tuning parameter.\\
Unlike the LASSO or the RIDGE penalties, SLOPE requires us to define a form of decreasing sequence of $\boldsymbol{\lambda}_{SLOPE} = (\lambda_{1},  \lambda_{2}, \ldots,  \lambda_{k})$.
For our analysis, we use the decreasing sequence of quantiles of the standard normal distribution, as in \cite{Bogdan2013} and \cite{Bogdan2015}, with $\lambda_{i} = \alpha \Phi^{-1}(1-q_{i})$, $\forall i = 1,..., k$, where $\Phi$ is the cumulative distribution function of the standard normal distribution and $q_{i} = i\times q/2k$, with $q=0.01$, regulates how fast the sequence of lambda parameters is decreasing. In our simulations, we varied the scaling parameter $\alpha$ so that the first element of the sequence $\lambda_{1} = \alpha \Phi^{-1}(1-q_{1})$ took values from a grid of 100 log-spaced values between $10^{-5}$ and $10^{2}$.  Note that in the case of the LASSO, we only choose one lambda parameter, which then remains constant for all assets. In our simulation and also in the real world analysis, we always choose $\lambda_{LASSO}=\lambda_{1}$. This choice favors sparser solutions for the LASSO, since for the remaining $k-1$ assets its penalty is larger than SLOPE penalty.\\
Figure \ref{RiskProfile_HiddenFactors_MinVar} show the resulting risk and weight profile for the minimum variance optimization, when we solve (\ref{eq:minreg}) separately with the LASSO and the SLOPE penalties for the grid of 100 lambda parameters and considering $\bfSigma_{HF}$ and the sample covariance estimate $\widehat{\bfSigma}_{HF}$. In particular, Panels (a) and (b) show the risk profile of LASSO and SLOPE, respectively, i.e. the actual, the oracle, and the empirical risk, together with the results of the GMV, the GMV-LO and the EW portfolios. For both the oracle and the actual solution, Panels (c) and (d) display on top the number of active weights together with the number of groups, that is the number of distinct coefficients, while shows on the bottom the amount of shorting (i.e. $\bfw^{-}$). The no short sales area (i.e. $w_{i} \geq 0\ \forall \ i=1,.., k$) is indicated by the grey surface.
%
%%%%%%%%%%%%%%%%%%%%%%%%%%%%%%
%%%%%%%%%%%%%%%%%%%%%%%%%%%%%%
%
Figure \ref{RiskProfile_HiddenFactors_MinVar} shows that setting the tuning parameter equal to zero, which corresponds to the GMV solution, the empirical risk is about 1.3 times lower than the actual risk (Panels (a) and (b)), with 12 active positions (Panel (c)) and slightly under 100\% short sales (Panel (d)). This can be interpreted as evidence %for the presence of the estimation error due to
that in over-fitted models the estimation error in $\widehat{\bfSigma}$ strongly affects the estimation of the asset weights.
As here neither the LASSO nor the SLOPE penalty are binding, estimation errors can enter unhindered into the optimization. \cite{Michaud1989} coined the term "error maximization" to describe this phenomena, stating that the ill-conditioned covariance estimates are amplified through the optimization, leading to extreme long and short portfolio weights.\\
Moving along the grid of $\bflambda$ parameters from left to right, Panels (c) and (d) show that the two penalties reduce the total amount of shorting in both the oracle, the actual, and the empirical portfolio.

%
%%%%%%%%%%%%%%%%%%%%%%%%%%%%%%
\begin{figure}[h!]
\centering
\caption{Hidden Factors Minimum-Variance Weight Profiles.}\label{Weights_HiddenFactors_MinVar}
\scalebox{0.9}{
\begin{tabular}{cc}%
\includegraphics[scale=.6]{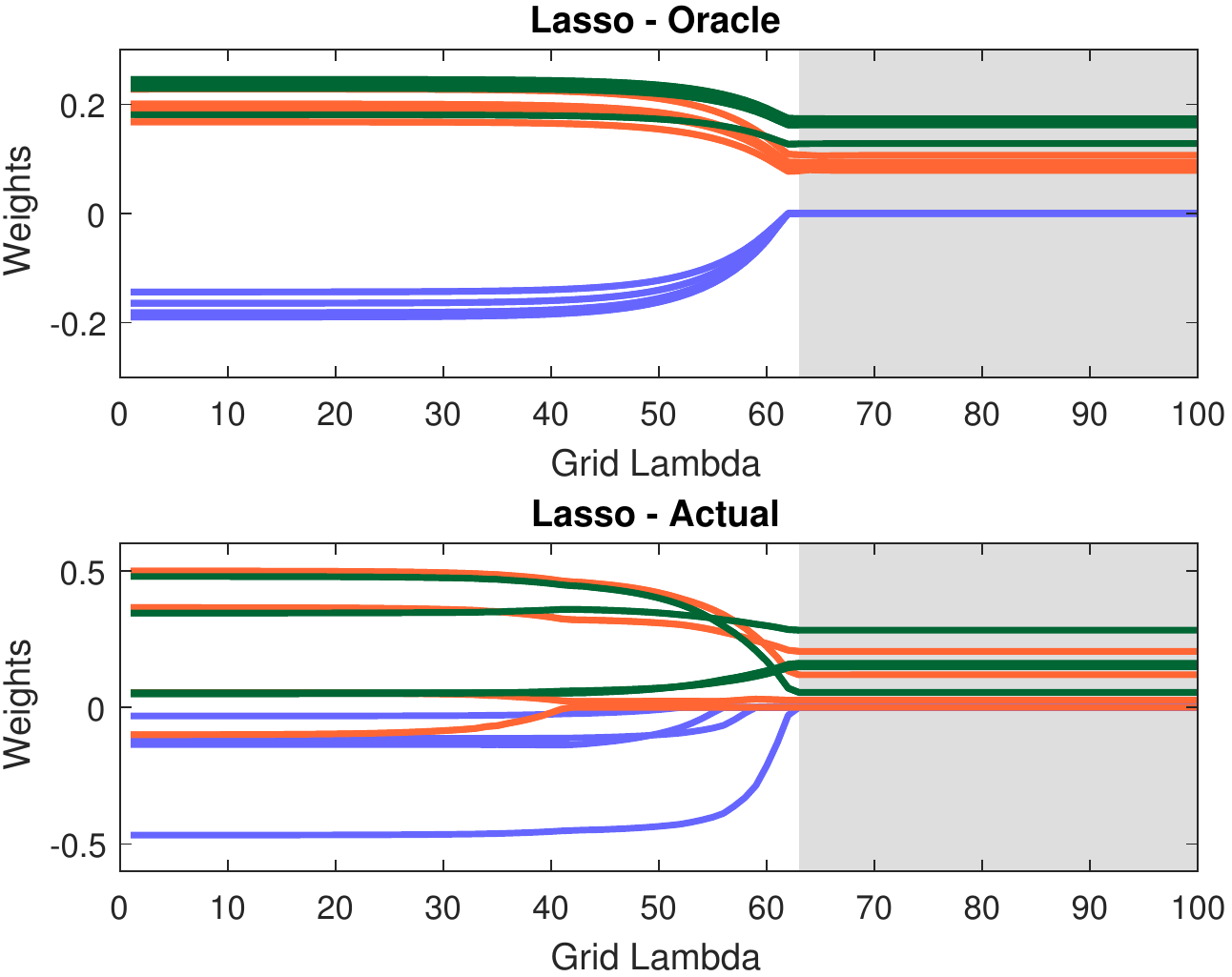} &
\includegraphics[scale=.6]{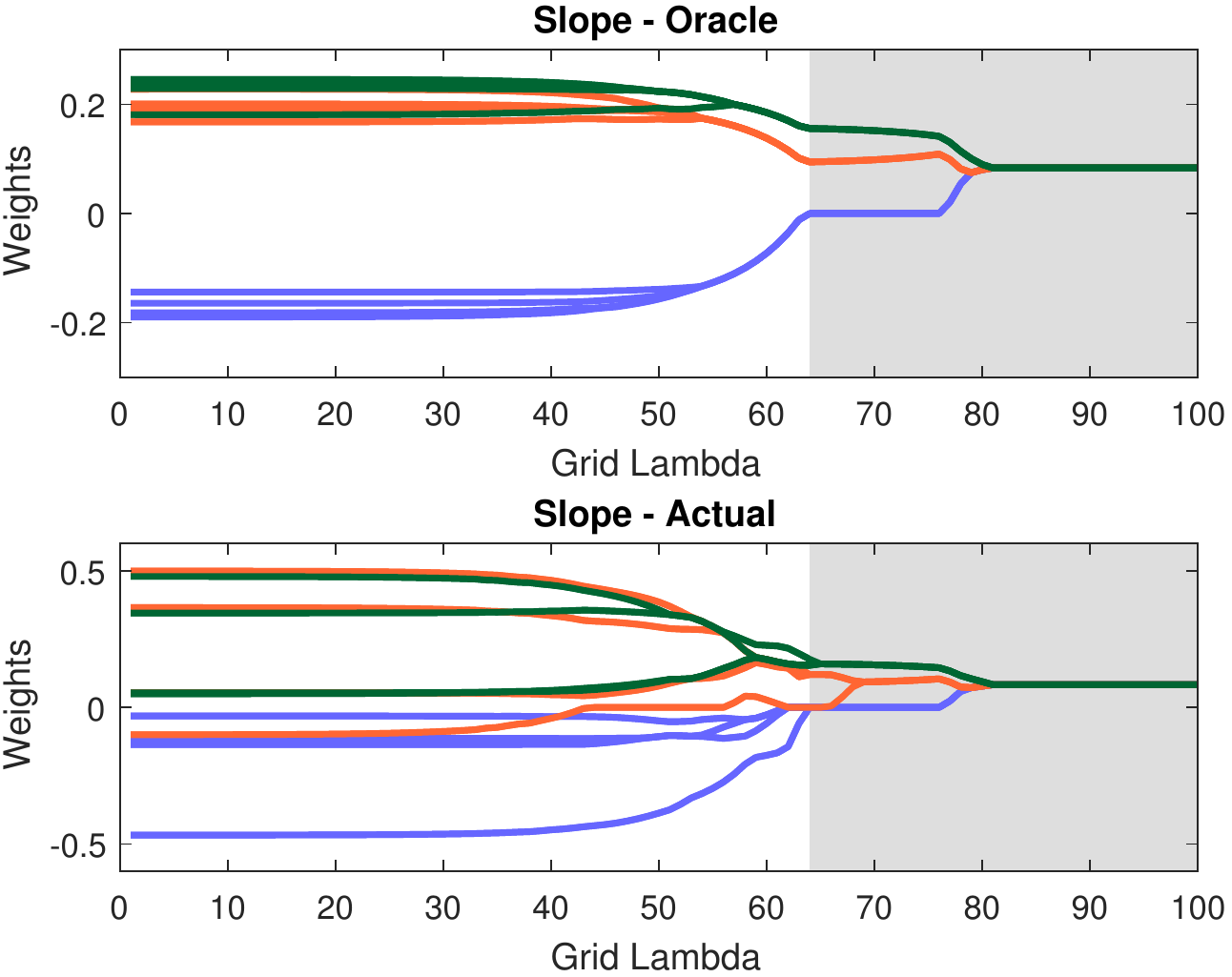}\\
(a) & (b) \\
\end{tabular}
}
\captionsetup{font=scriptsize,labelfont=scriptsize, width=\textwidth}
     \caption*{The Figure shows the weight profile of the oracle (top) and actual (bottom) solution of the LASSO and the SLOPE penalty, considering a minimum variance setup. All values are computed based on a Hidden Factor Structure, with three risk factors and considering for the exponentially decreasing sequence of lambda parameters, a grid of 100 log spaced starting points, $\lambda_{1}$ from $10^{-5}$ (i.e. x-value = 1) to $10^{2}$ (i.e. x-value = 100).Different colors characterize assets with the same factor exposure.}
\end{figure}
%%%%%%%%%%%%%%%%%%%%%%%%%%%%%%
%
\noindent
As we move from the GMV towards the GMV-LO, the actual, oracle, and empirical risks of LASSO and SLOPE align. This effect was first observed and theoretically motivated by \cite{Fan2012}, showing that the portfolio risk evolves in a U-shape, in which risk first decreases before increasing again due to the restriction of short sales. With the observations above, we extend the results of \cite{Fan2012}, showing that the U-shaped behaviour of the portfolio risk is not the only possible. The tighter constraint in terms of short sales shrinks the optimization search space of feasible solutions, preventing diversification benefits, especially when the dependence among the assets is positive. This makes it impossible to exploit the optimal diversification benefits, leading to a higher portfolio risk when reaching the GMV-LO. The investor also reaches the maximum sparsity, that is the maximum number of coefficients equal to zero, at this point. For the LASSO penalty, increasing the tuning parameter beyond this point does not alter the allocation any further, as the regularization penalty is constant and equal to 1. This is different for the SLOPE penalty: in fact, Figure \ref{Weights_HiddenFactors_MinVar} shows the evolution of the portfolio weights for both the oracle and the actual solution, considering both the LASSO and the SLOPE penalty. As before, the no short sale area (i.e. $w_{i} \geq 0\ \forall \ i=1,.., k$) is indicated by the grey surface.\\
From Figure \ref{Weights_HiddenFactors_MinVar}, we can observe two important characteristics of SLOPE: First, while the LASSO shrinks the weights up until the no short sale area, all non-zero coefficients still receive a different weight, independent of their underlying factor exposures. 
%\red{Interestingly, in case when the covariance matrix is estimated, the weights of assets beloNote that although, the coefficients are close to the ones with the same dependence structure, this is typically not the case, as Figure \ref{Weights_HiddenFactors_MeanVar} shows. 
SLOPE, on the other hand, is able to identify the three distinct types of securities consistent with the true model and group them together. The assets in the same group are assigned the same coefficient values, providing information about the dependence structure among assets. This gives the investor the flexibility to select those assets from the groups with the same underlying risk factor exposure, which best fit her individual preferences. Not surprisingly, the oracle risk groups the assets even before entering into the no short sale area, while the actual weights can only capture the underlying structure much later, when we already impose a larger tuning parameter value.
%, as $\bfSigma$ has to be estimated on a finite sample.\\
Second, and different to the LASSO, increasing the lambda parameters past the point of the GMV-LO, the octagonal shape of the penalty pushes the solution towards the equally weighted portfolio. That is, the aforementioned grouping effect increases, and all weights - even those that were shrunken towards zero - are assigned the same coefficient value of $\frac{1}{k}$. Given that the equally weighted portfolio is only optimal when all assets have the same risk and return characteristics, in our example this allocation results in higher portfolio risk when compared to the GMV-LO or GMV portfolios.\\
%, due to the limited sample space for estimation and as the optimization search space gets narrower as $\lambda$ increases.\\
%
%%%%%%%%%%%%%%%%%%%%%%%%%%%%%%
\begin{figure}[h!]
\centering
\caption{Hidden Factors Mean-Variance Profile.}\label{RiskProfile_HiddenFactors_MeanVar}
\scalebox{0.9}{
\begin{tabular}{cc}%
\includegraphics[scale=.6]{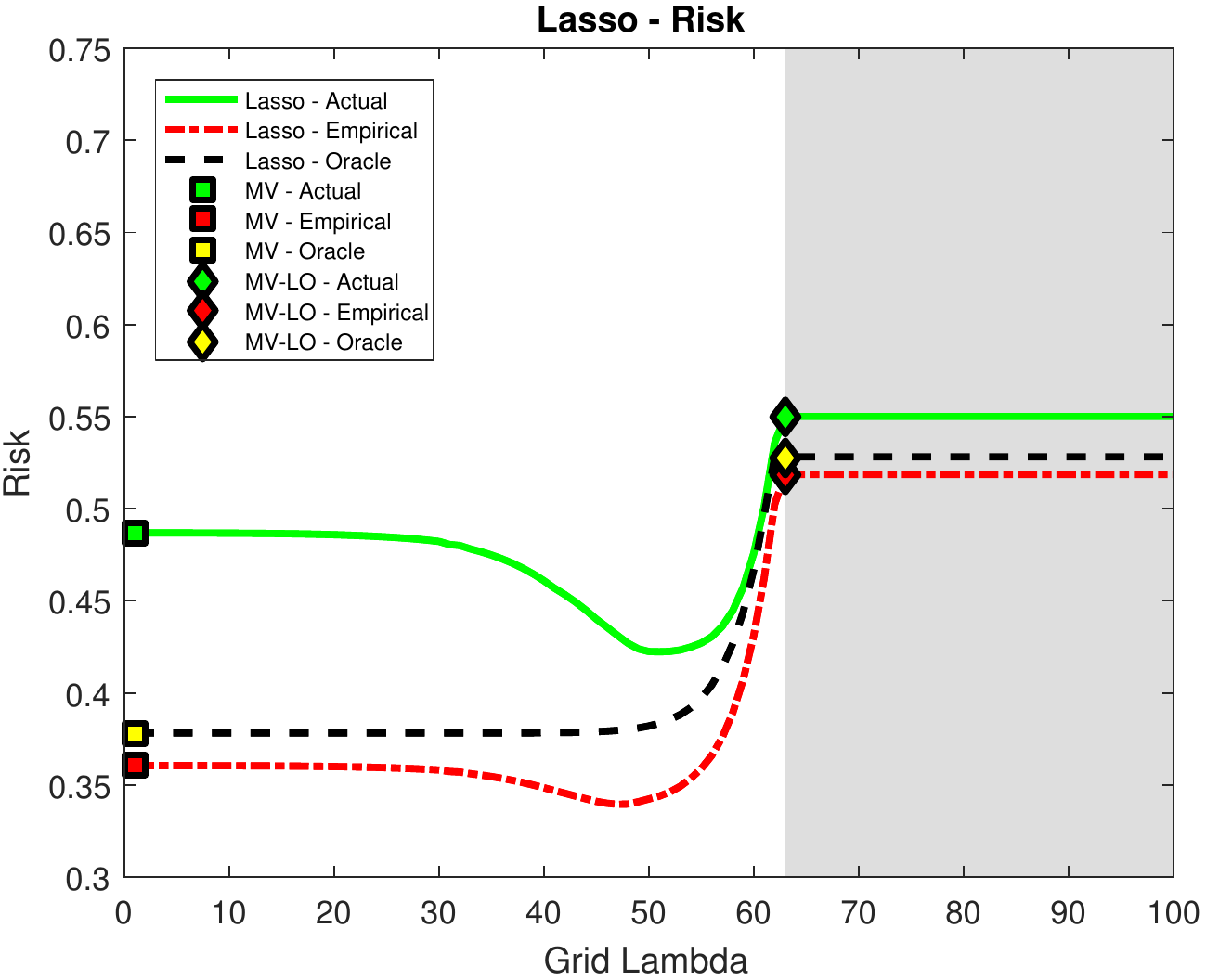} &
\includegraphics[scale=.6]{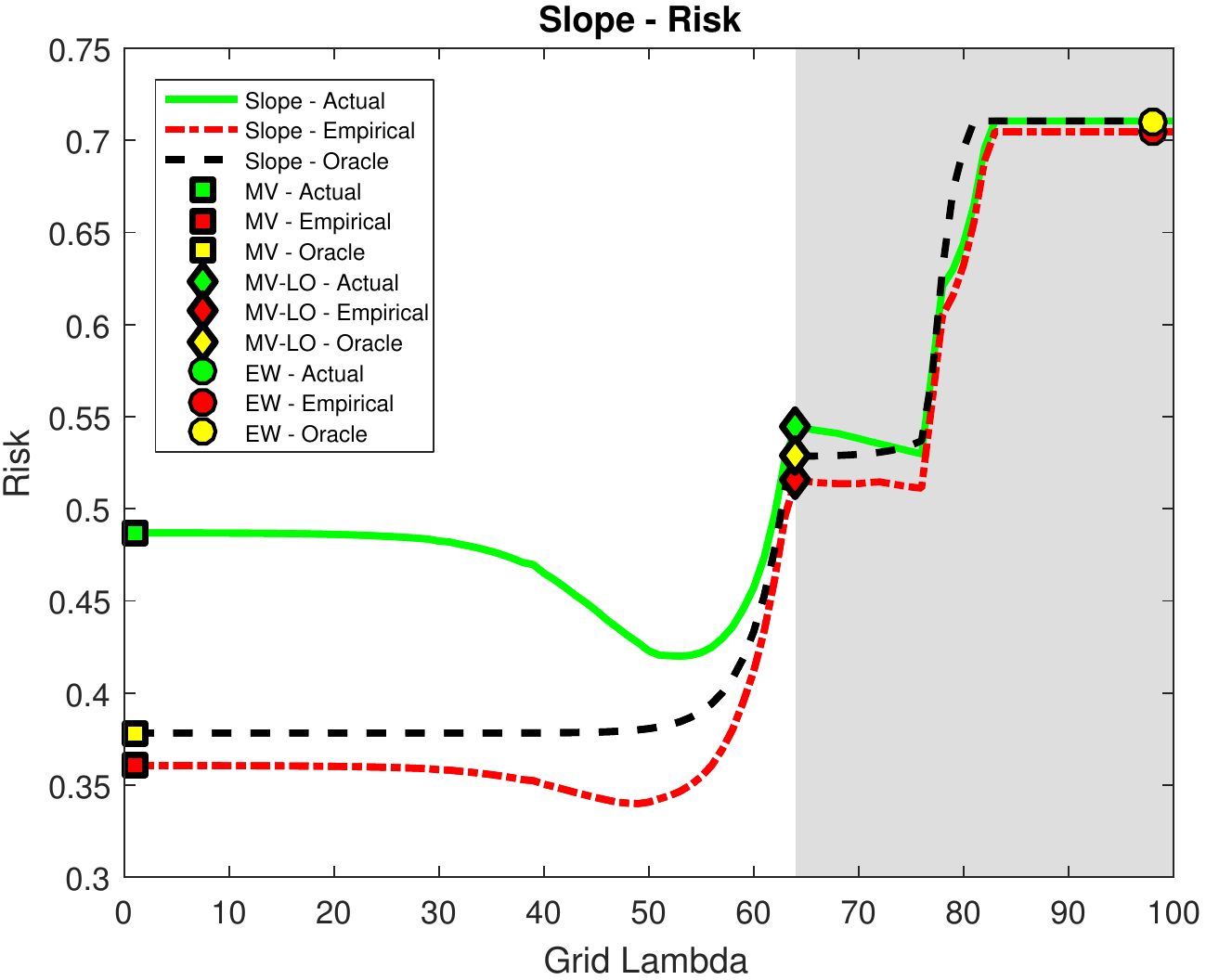}\\
(a) & (b) \\
%\includegraphics[scale=.6]{Lasso2_HiddenFactors_MeanVar} &
%\includegraphics[scale=.6]{Slope2_HiddenFactors_MeanVar}\\
%(c) & (d) \\
\end{tabular}
}
\captionsetup{font=scriptsize,labelfont=scriptsize, width=\textwidth}
     \caption*{The Figure shows the Hidden Factor mean-variance risk profile for the LASSO and the SLOPE, including in Panel (a) and (b) their actual, empirical and oracle risk profiles together with that of the GMV, GMV-LO and EW solutions. % and in Panel (c) and (d) their number of active weights, together with the grouping profile (top) and the total amount of shorting (bottom).
All values are computed based on a Hidden Factor Structure, with three risk factors and considering for the exponentially decreasing sequence of lambda parameters, a grid of 100 log spaced starting points, $\lambda_{1}$ from $10^{-5}$ (i.e. x-value = 1) to $10^{2}$ (i.e. x-value = 100).}
\end{figure}
%%%%%%%%%%%%%%%%%%%%%%%%%%%%%%
%
\noindent
In a next step, we investigate how SLOPE performs in a mean-variance framework, that is we choose $\boldsymbol{\mu} = 0$ and $\boldsymbol{\hat{\mu}} = mean(\boldsymbol{R}_{HF})$, as the sample average of the $t=50$ return observations for the $k=12$ assets. Figure \ref{RiskProfile_HiddenFactors_MeanVar} plots the resulting risk profile of the LASSO and the SLOPE, for 100 logspaced values from $10^{-5}$ to $10^{2}$ for the starting point of the sequence of tuning parameters. The effect of adding the mean estimate to the minimum variance framework can be observed from multiple viewpoints:
First, from the risk plot in Panels (a) and (b), we observe that the estimation errors, which are known to be larger in the mean than in the covariance matrix (see for example \cite{Merton1980, Michaud1989, DeMiguel2009a}), lead to a more pronounced difference among the empirical, oracle, and actual risk.
%, as the optimal weights are now computed for the mean-variance set-up. 
The GMV now has an empirical risk of 0.35 compared to the actual risk of 0.5. Increasing the weight on the tuning parameter for both the LASSO and the SLOPE reduces the estimation error making the three risk measure converge as we move from the GMV to the GMV-LO.\\
%
%%%%%%%%%%%%%%%%%%%%%%%%%%%%%%
\begin{figure}[h!]
\centering
\caption{Hidden Factors Mean-Variance Weight Profiles.}\label{Weights_HiddenFactors_MeanVar}
\scalebox{0.9}{
\begin{tabular}{cc}%
\includegraphics[scale=.6]{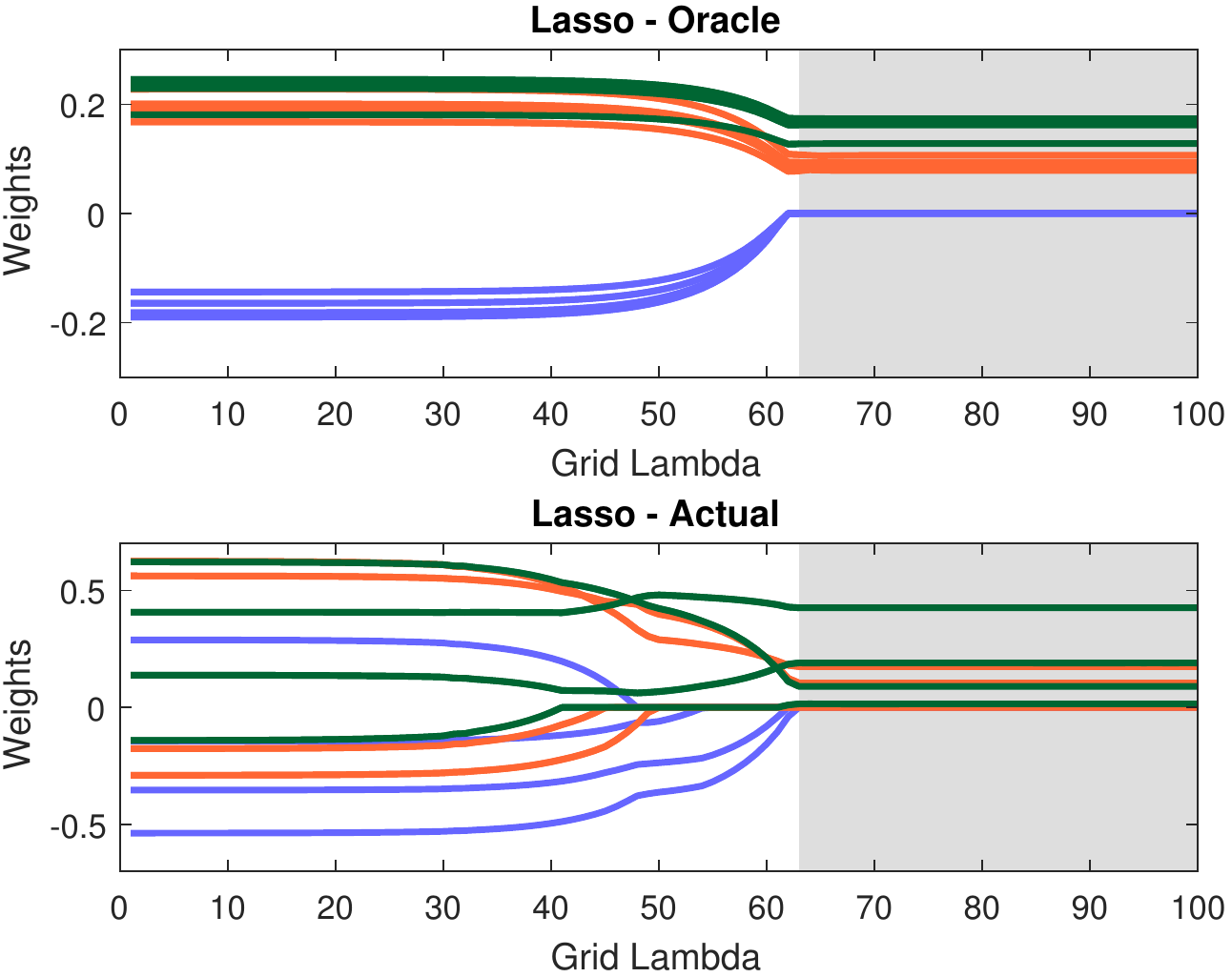} &
\includegraphics[scale=.6]{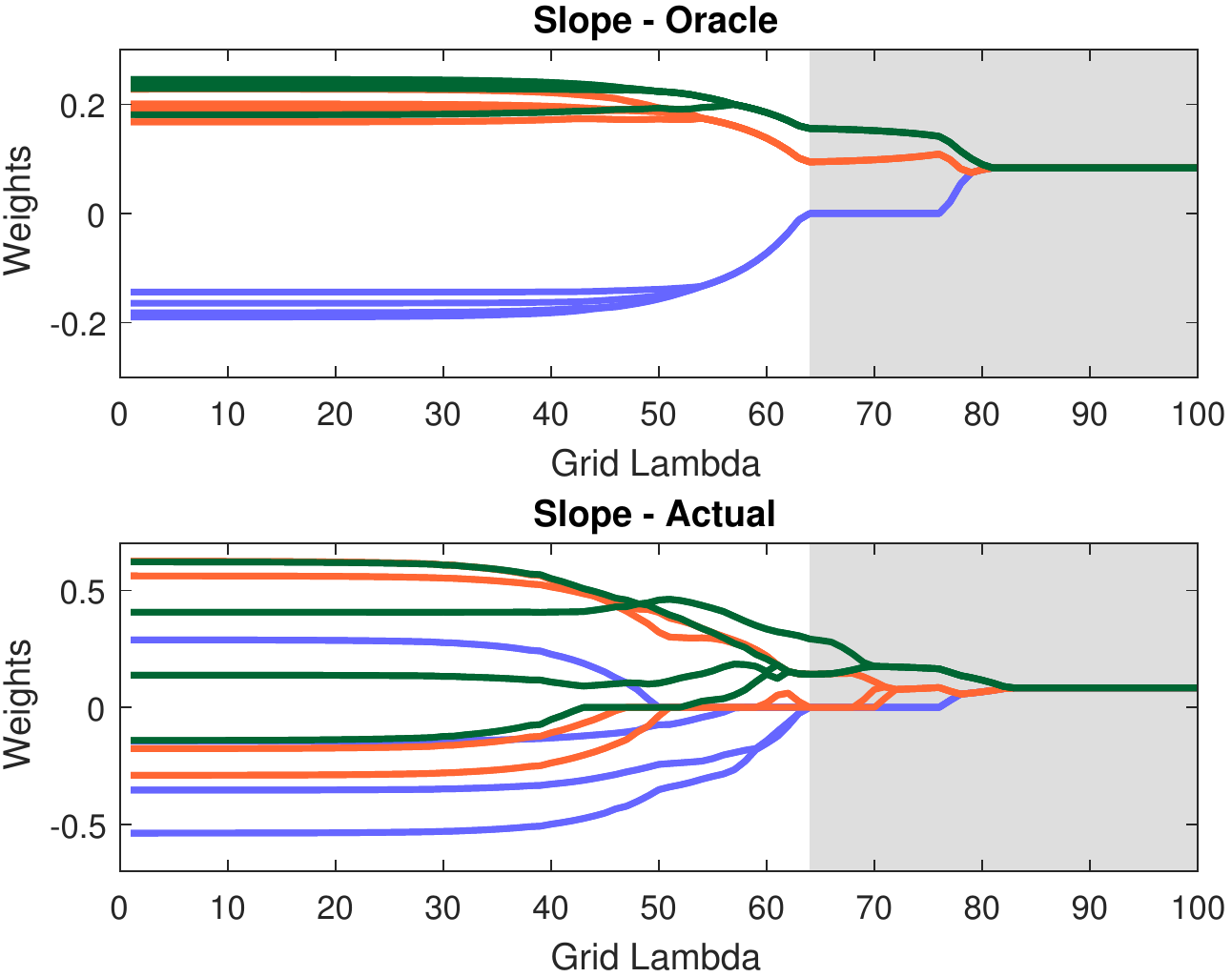}\\
(a) & (b) \\
\end{tabular}
}
\captionsetup{font=scriptsize,labelfont=scriptsize, width=\textwidth}
     \caption*{The Figure shows the weight profile of the oracle (top) and actual (bottom) solution for both the LASSO and SLOPE penalty, considering a mean-variance setup. All values are computed based on a Hidden Factor Structure, with three risk factors and considering for the exponentially decreasing sequence of lambda parameters, a grid of 100 log spaced values for $\lambda_{1}$ from $10^{-5}$ (i.e. x-value = 1) to $10^{2}$ (i.e. x-value = 100). Different colors characterize assets with the same factor exposure.}
\end{figure}
%%%%%%%%%%%%%%%%%%%%%%%%%%%%%%
%
%Second, Panels (c) and (d) show that (i) the estimated allocation is much sparser for lower values of the tuning parameter, while maximum sparsity is again attained for the GMV-LO and (ii) there is a much higher amount of shorting in the oracle and actual portfolio allocations then when we only consider a minimum variance framework.\\
Finally, Figure \ref{Weights_HiddenFactors_MeanVar} shows that, regardless of the penalty function, the oracle and actual weight vectors differ substantially in their values. As before, LASSO weights for assets with different factors are overlapping and are randomly picked, while SLOPE, despite requiring larger $\lambda$ values, is still capable to correctly group the assets with the same underlying risk factor exposure and thereby disentangling signal from noise.\\

\noindent
\subsubsection*{S\&P 500 Simulated Covariance Matrix}
\begin{figure}[h!]
\centering
\caption{S\&P 500 Minimum Variance Profile.}\label{RiskProfile_Fan_MinVar}
{\begin{tabular}{cc}%
\includegraphics[scale=.55]{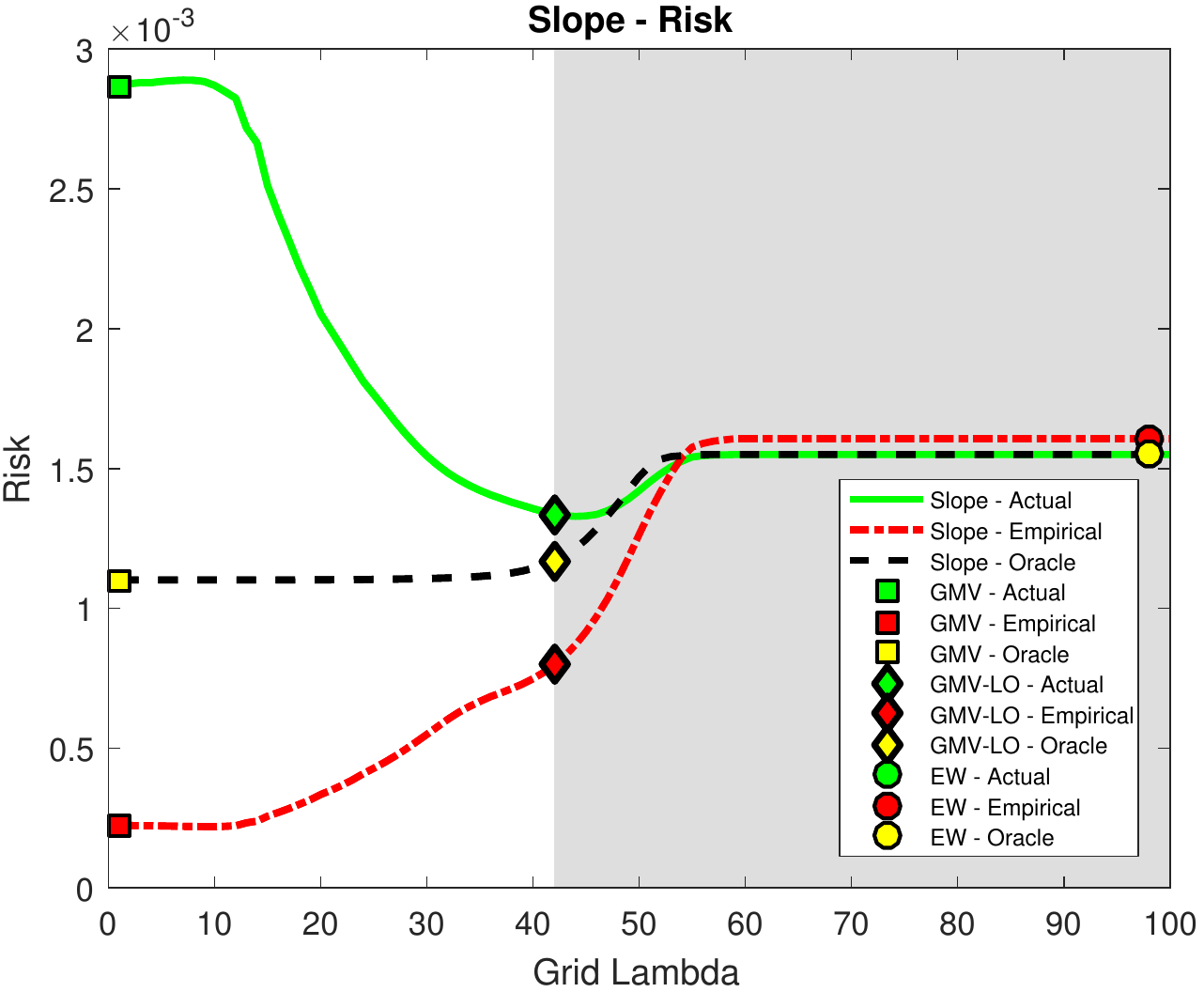} & \includegraphics[scale=.55]{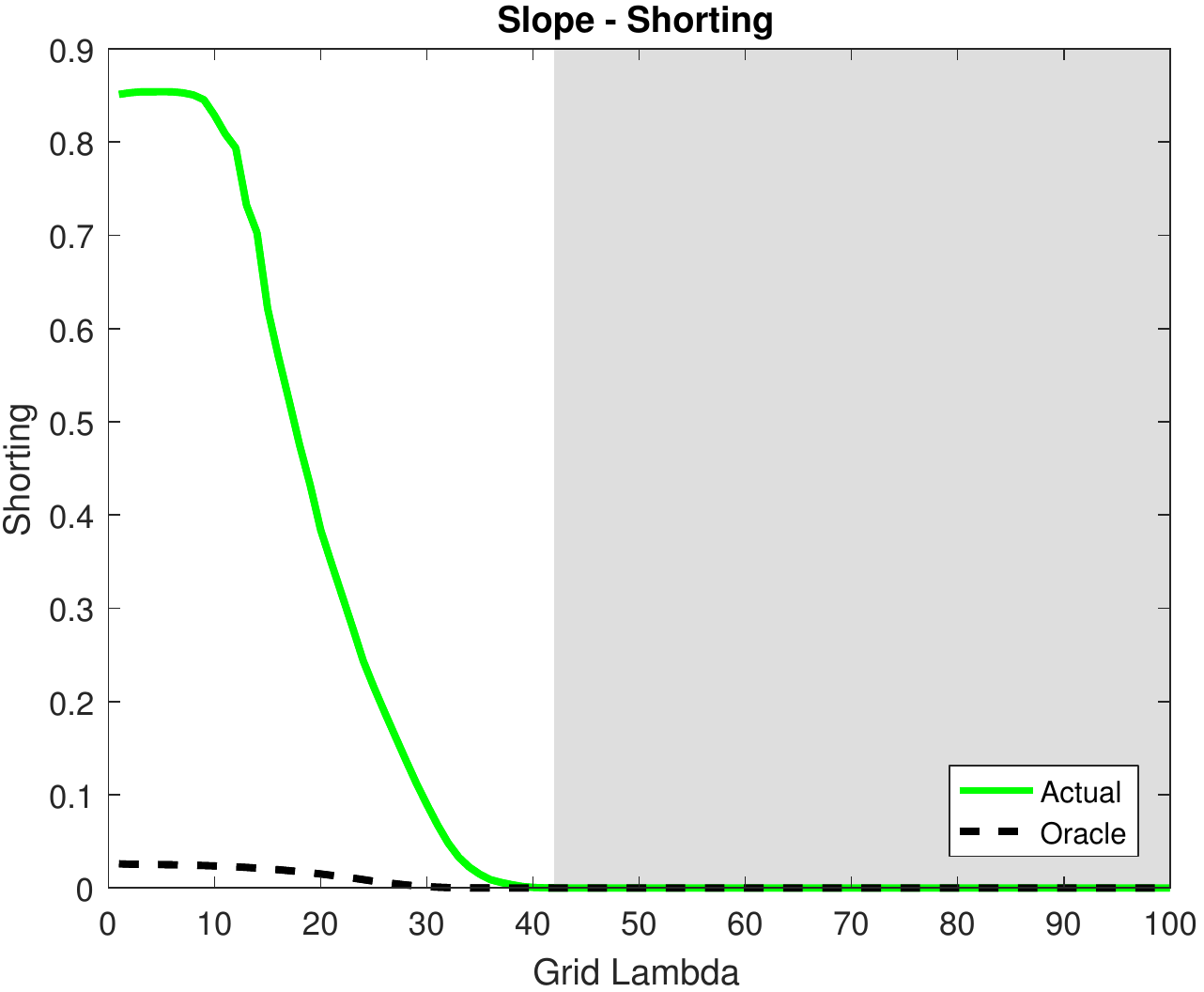}\\
(a) & (b) \\
\end{tabular}}
\captionsetup{font=scriptsize,labelfont=scriptsize, width=\textwidth}
     \caption*{The Figure shows the Fan Simulated S\&P500 minimum-variance risk profile for SLOPE, including in Panel (a) the actual, empirical and oracle risk profile of SLOPE together with that of the GMV, GMV-LO and EW solutions and in panel (b) the number of active weights, together with the grouping profile (top) and the total amount of shorting (bottom). All values are computed based on the \cite{Fan2008} simulated calibrated to the S\&P500 and considering for the exponentially decreasing sequence of lambda parameters, a grid of 100 log spaced values with a starting point, $\lambda_{1}$, from $10^{-8}$ (i.e. x-value = 1) to $10^{-1.5}$ (i.e. x-value = 100).}
\end{figure}

\begin{figure}[h!]
\centering
\caption{S\&P 500 Mean-Variance Profile.}\label{RiskProfile_Fan_MeanVar}
\begin{tabular}{cc}%
\includegraphics[scale=.55]{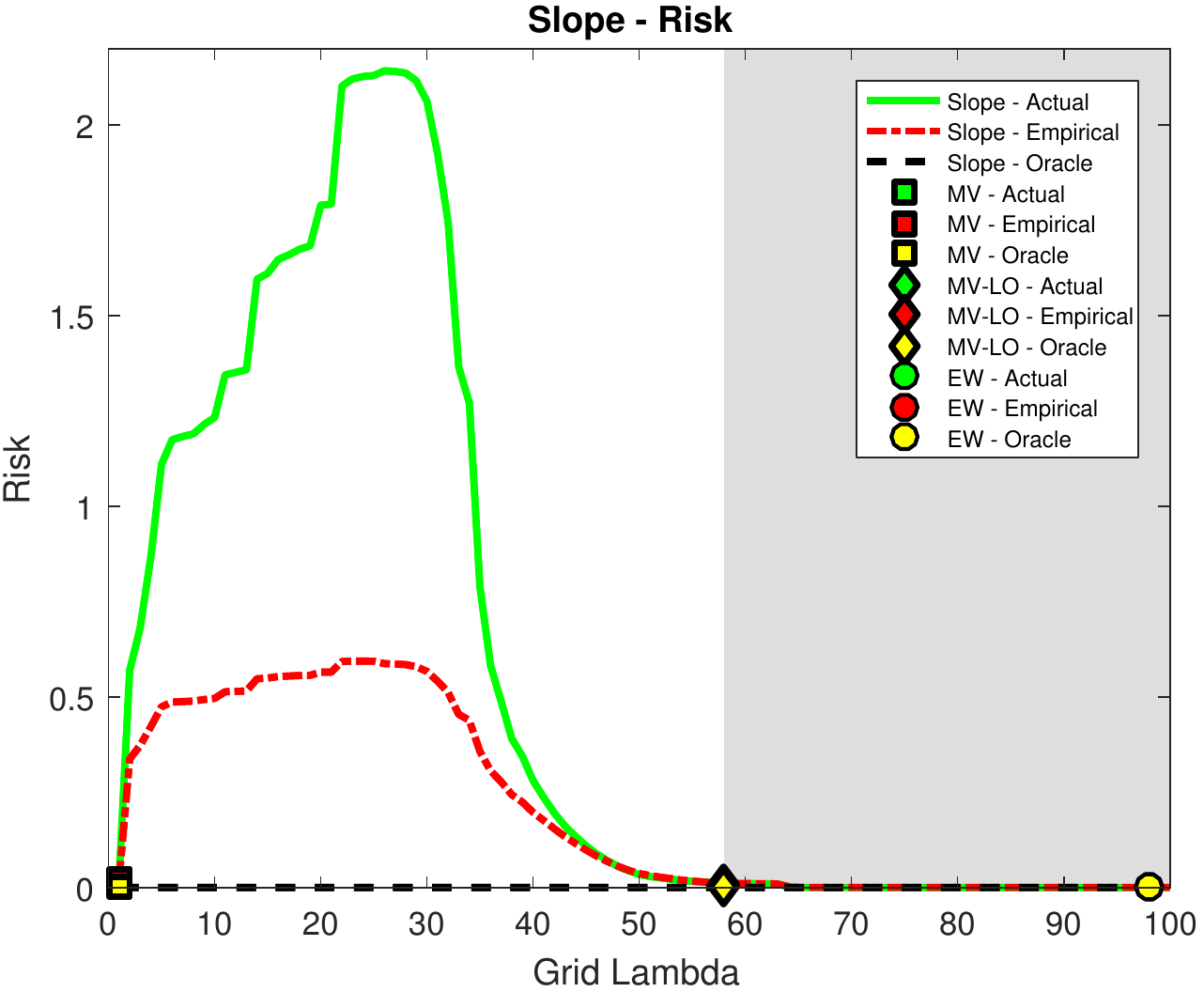} & \includegraphics[scale=.55]{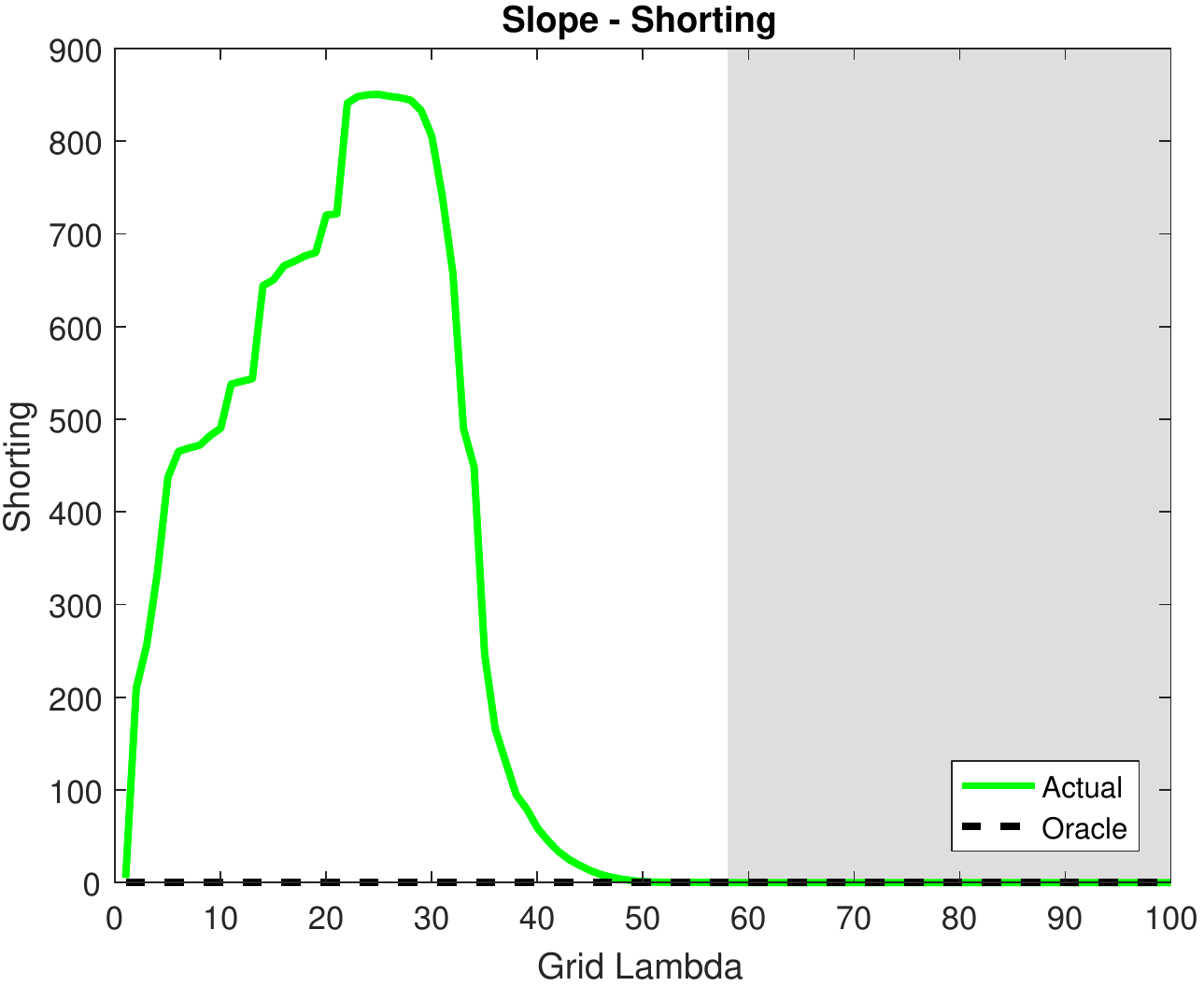}\\
(a) & (b) \\
\end{tabular}
\captionsetup{font=scriptsize,labelfont=scriptsize, width=\textwidth}
\caption*{The Figure shows the Fan Simulated S\&P500 mean-variance risk profile for SLOPE, including in Panel (a) the actual, empirical and oracle risk profile of SLOPE together with that of the GMV, GMV-LO and EW solutions and in panel (b) the number of active weights, together with the grouping profile (top) and the total amount of shorting (bottom). All values are computed based on the \cite{Fan2008} simulated calibrated to the S\&P500 and considering for the exponentially decreasing sequence of lambda parameters, a grid of 100 log spaced values with a starting point, $\lambda_{1}$, from $10^{-4}$ (i.e. x-value = 1) to $10^{-1.5}$ (i.e. x-value=100).}
\end{figure}

Our generic example suffers from the drawback that, in reality, assets do not follow such a strict exposure to only two out of the three underlying risk factors, but are most likely exposed to all factors, with investors often facing an investment universe that is larger than 12 assets.\\
To model a more realistic scenario, we consider a factor setup driven by the estimated covariance matrix between different assets, which was introduced by \citet{Fan2008}. As the behavior of the LASSO in a high dimensional environment is widely studied, and for the sake of brevity, we restrict ourselves to studying the behavior of the new SLOPE procedure. The results for the performance of the LASSO in such an environment are available from the authors upon request.\\
As before, we assume that security returns can be expressed as a linear combination of risk factors, as in (\ref{truemodel}), implying that our true covariance matrix takes the form
$$\bfSigma_{\text{SP500}}=\boldsymbol{B}'\bfSigma_F\boldsymbol{B}+\bfSigma_{\epsilon}\;\;,$$
where $\bfSigma_{F}$ is the covariance matrix of hidden factors and $\bfSigma_{\epsilon}$ is the covariance matrix of error terms.\\
Different to the simulations before, we draw the three factors and the factor loadings from  multivariate normal distributions that are calibrated to real world data. Specifically, our three hidden factors are generated from a trivariate normal distribution $N(\bfmu_{F}, \bfSigma_{F})$, where $\bfmu_{F}$ and $\bfSigma_{F}$ are calculated based on data taken from Kenneth French's Homepage \footnote{\url{http://mba.tuck.dartmouth.edu/pages/faculty/ken.french/data_library.html}}, spanning the time period from 31.12.2004 - 31.01.2016 (see Table \ref{Fan_Calib}). The factor loadings for each of $k=500$ assets were independently drawn from the trivariate normal distribution $N(\bfmu_{B}, \bfSigma_{B})$, in which $\bfmu_{B}$ and $\bfSigma_{B}$ are calculated using data on the S\&P500 from 31.12.2004 - 31.01.2016, as reported in Table \ref{Fan_Calib}. Finally, the idiosyncratic noises are generated from a gamma distribution with shape parameter $\alpha = 7.2609$ and scale parameter $\beta = 0.0028$ (see \citet{Fan2008} for details).
%
%%%%%%%%%%%%%%%%%%%%%%%%%%%%%%
\begin{table}[ht!]
\centering
\caption{Configuration Parameters for \cite{Fan2008} Simulation.}\label{Fan_Calib}
\begin{tabular}{cccccccc}
\hline
\hline
\multicolumn{4}{c}{Parameters for factor loadings} & \multicolumn{4}{c}{Distribution of hidden factors} \\
\cmidrule(lr){1-4}  \cmidrule(lr){5-8}
\multicolumn{1}{l}{$\bfmu_{B}$} & \multicolumn{1}{l}{$\bfSigma_{B}$}& & & \multicolumn{1}{l}{$\bfmu_{F}$} & \multicolumn{1}{l}{$\bfSigma_{F}$}\\
\cmidrule(lr){1-4}  \cmidrule(lr){5-8}
-0.0679& 0.0062  & -0.0016 & 0.0020  & 0.00022 & 0.000157 & 0.000015 & 0.000028\\
0.1505 & -0.0016 & 0.0109  & 0.0012  & 0.00012 & 0.000015 & 0.000033 &-0.000016 \\
-0.0203& 0.0020  & 0.0012  & 0.0173  &-0.00018 & 0.000028 &-0.000016 & 0.000034 \\
\hline
\hline
\end{tabular}
\captionsetup{font=scriptsize,labelfont=scriptsize, width=0.95\textwidth}
     \caption*{The table reports the means ($\bfmu$) and the covariance matrices ($\bfSigma$) used as input parameters for the trivariate normal distributions to sample hidden factors and assets' factor loadings. The first and second moments for the factor loadings are calibrated using daily data from the S\&P 500 from 31.12.2004-31.01.2016, while the parameters for the factor distribution are calibrated, using daily data for the three Fama-French risk factors plus the \cite{Carhart1997} liquidity factor, covering again the period from 31.12.2004-31.01.2016.}
\end{table}
%%%%%%%%%%%%%%%%%%%%%%%%%%%%%%
%
We then set $t=500$  and solve (\ref{eq:minreg}) both with (a) $\hat{\bfmu}_{\text{SP500}} = \bfmu_{\text{SP500}} = 0$, which is then the minimum variance problem, and with (b) $\bfmu_{\text{SP500}} = \boldsymbol{B}'\bfmu_{F}$, while $\hat{\bfmu}$ is equal to the sample mean of the last 500 return observations obtained from (\ref{truemodel}). We used the same $\bflambda$ sequences as in the previous section, with the first element of the sequence $\lambda_{1}$ taking  $100$ log spaced values from $10^{-8}$ to $10^{-1.5}$ in the minimum variance framework, and  from $10^{-4}$ to $10^{-1.5}$ for the mean-variance framework.\\
Figure \ref{RiskProfile_Fan_MinVar} shows the risk and sparsity profiles for the minimum variance optimization, while Figure \ref{RiskProfile_Fan_MeanVar} displays the results for the mean-variance set-up. Both plots are consistent with the findings from the Hidden Factor Model. Again we observe that in the GMV solution (i.e., $\lambda = 0$), there is a substantial difference among the three risk measures, with the empirical risk (given in red) highly underestimating the actual risk.
%, resulting from over-fitting the portfolio weight vector to the sample covariance matrix estimate.
Still, with a larger $\lambda$, the difference between the risk measures decreases and we move closer to the GMV-LO solution. Furthermore, given a larger investment universe and the realistic dependence structure, the impact of a reduction in the search space due to a larger penalty is smaller and the actual risk decreases as we move closer to the no short selling area and towards the EW solution. The same is true for the mean-variance set-up. In fact, Figure \ref{RiskProfile_Fan_MeanVar} shows that including the estimate of the mean $\hat{\bfmu}_{\text{SP500}}$, introduces as expected even more estimation error, than compared to the minimum variance optimization. This is different for the oracle solution, which considers the true mean, $\bfmu_{\text{SP500}}$, and such is not prone to any extreme mean estimates. Panel (a) of Figure \ref{RiskProfile_Fan_MeanVar} shows that the effect of the mean estimate is so large that even in the EW solution the actual risk is smaller than in the GMV solution. Still, the three risk measures align and the overall risk reduces as we increase the $\bflambda$ sequence. Most importantly, while the LASSO is not effective in the no-short sale area, SLOPE provides this risk reduction effect even past the GMV-LO, when moving towards the EW solution.
%
%%%%%%%%%%%%%%%%%%%%%%%%%%%%%%
%%%%%%%%%%%%%%%%%%%%%%%%%%%%%%
%
This is especially valuable for investors who face short sale constraints and deal with large portfolios and many candidate assets. Indeed, the LASSO would be stuck in the GMV-LO solution, as previously discussed. With SLOPE, investors facing short sale constraints are then able to further reduce their risk and possibly even set up new strategies to exploit the grouping property to select among assets that best correspond to their financial investment objectives.\\
%Panel (c) of Figure \ref{RiskProfile_Fan_MinVar} and Figure \ref{RiskProfile_Fan_MeanVar} show that sparsity increases as we move along the grid of lambda values. Maximum sparsity is again reached in the GMV-LO solution and at the beginning of the no short sale area.\\
An interesting difference between the minimum variance and mean-variance optimization is the behaviour of the short sales, as shown in Panel (b) of Figure \ref{RiskProfile_Fan_MinVar} and Figure \ref{RiskProfile_Fan_MeanVar}. While in the minimum variance set-up, we start with an initial solution that shows a certain amount of shorting, which is then reduced with an increasing value of $\boldsymbol{\lambda}$, the total amount of shorting first increases in the mean variance optimization, before decreasing it again and when we consider the estimated mean. For the actual risk, this behaviour can be explained by the presence of positively correlated assets, which is typical in financial markets, and the optimization exploiting the diversification, by taking at the beginning extreme positive and negative weights. The penalty then has an effect on shrinking large weights and reducing the number of active weights. Hence, to exploit diversification benefits, other assets with smaller weights end up in increasing their negative exposure.\\
%
%%%%%%%%%%%%%%%%%%%%%%%%%%%%%%
%%%%%%%%%%%%%%%%%%%%%%%%%%%%%%
%
\noindent
Finally, even not displayed, we observe that the desired feature of grouping assets together persists, not only in a directly modelled Hidden Factor Structure, but even in a high dimensional scenario with real world calibrated covariance matrices. This might be valuable for investors who can then set up sophisticated asset allocation strategies, exploiting SLOPE grouping property, such as SLOPE-MV, introduced in Section \ref{EmpiricalData}.

%%%%%%%%%%%%%%%%%%%%%%%%%%%%%%%%%%%%%%%%%%%%%%%%%%%
\section{Empirical Analysis}\label{EmpiricalData} %
%%%%%%%%%%%%%%%%%%%%%%%%%%%%%%%%%%%%%%%%%%%%%%%%%%%

%%%%%%%%%%%%%%%%%%%%%%%%%%%%%%%%%%%%%%%%%%%
\subsection{Set up and Data}\label{Data} %
%%%%%%%%%%%%%%%%%%%%%%%%%%%%%%%%%%%%%%%%%%%

This section studies the out-of-sample performance of the SLOPE procedure, considering a minimum variance framework, as typical of most studies (see i.e. \cite{Jagannathan2003, Brodie2009, DeMiguel2009a, Giuzio2016a}). Our analysis compares SLOPE with state-of-the-art and portfolio selection methods, such as the EW, the GMV, the GMV-LO, the equal risk contribution (ERC), the RIDGE and the LASSO portfolio. We examine two extensions to our standard SLOPE procedure: (1) SLOPE with an added long only constraint (SLOPE-LO) and (2) a portfolio in which we utilize SLOPE's selection ability, by first running SLOPE-LO on the whole time period, identifying groups of similar assets and pick out of each group the one with minimum variance. We then roll through the dataset and solve for these active securities the GMV-LO portfolio (SLOPE-MV).\\
We consider five data sets, including the monthly log-return observations for the 10- and 30 Industry Portfolios (Ind) and the 100 Fama French (FF) portfolios formed on Size and Book-to-Market Ratio, as well as the daily returns of the S\&P100 and S\&P500. The monthly portfolio values are from Kenneth French's Homepage\footnote{http://mba.tuck.dartmouth.edu/pages/faculty/ken.french/} and span the period from January 1970 to January 2017 ($T=565$ monthly observations). The daily return data were obtained from Datastream, covering the period from 31.12.2004 to 31.01.2016 ($T= 2890$ daily observations). Table \ref{SumStat} reports the descriptive statistics for the data sets. All of them exhibit the typical return time series characteristics, including fat tails and slight asymmetry, as shown by skewness and kurtosis values.

%
%%%%%%%%%%%%%%%%%%%%%
\begin{table}[ht!]
\centering
\caption{Descriptive Statistics of the Dataset.}\label{SumStat}
\scalebox{0.74}{
\begin{tabular}{l|cc|ccccccccc}
\hline
\hline
Dataset & $T$ & $k$& $\hat{\mu}$ &$\hat{\sigma}$ &\textit{med}  &\textit{min} &\textit{max} &\textit{skew} &\textit{kurt} & \textit{period} & \textit{freq.}\\
\hline
10Ind &565& 10 &0.099&0.043 &0.012 &-0.211  &0.156 &-0.476 &5.077 &01/1970 - 01/2017 &Monthly\\
30Ind &565& 30 &0.010 &0.048 &0.012 & -0.255 &0.179  &-0.507 &5.749 &01/1970 - 01/2017 &Monthly\\
100FF &565& 100 &0.011 &0.052 &0.015 &-0.262  &0.241 &-0.551  &5.600 &01/1970 - 01/2017 &Monthly\\
SP100 &2890& 93 &0.000  &0.013 &0.000&-0.098 &0.116 &\ -0.240 & 14.816 &12/2004 - 01/2016 &Daily\\
SP500 &2890& 443 &0.000 &0.014 &0.000 &-0.107 & 0.109 & -0.418 &13.234 &12/2004 - 01/2016 &Daily\\
\hline
\hline
\end{tabular}
}
\captionsetup{font=scriptsize,labelfont=scriptsize, width=\textwidth}
\caption*{The table reports descriptive summary statistics for the 10 Industry Portfolios, the 30 Industry Portfolios, the 100 Fama French Portfolios, the S\&P 100 and the S\&P 500, respectively. Reported are for the daily (monthly) data: the number of observations (\textit{T}), the number of constituents($k$), the mean ($\hat{\mu}$), the standard deviation ($\hat{\sigma}$), the median (\textit{med}), the minimum (\textit{min}), the maximum (\textit{max}), the skewness (\textit{skew}), the kurtosis (\textit{kurt}), the period that the data set covers (\textit{period}) and the frequency (\textit{freq.}).}
\end{table}
%%%%%%%%%%%%%%%%%%%%%
%

\noindent

\begin{figure}[h!]
\centering
\caption{Condition Numbers.} \label{Cond_Numbers}
\scalebox{0.9}{
\begin{tabular}{cc}
\includegraphics[scale=.56]{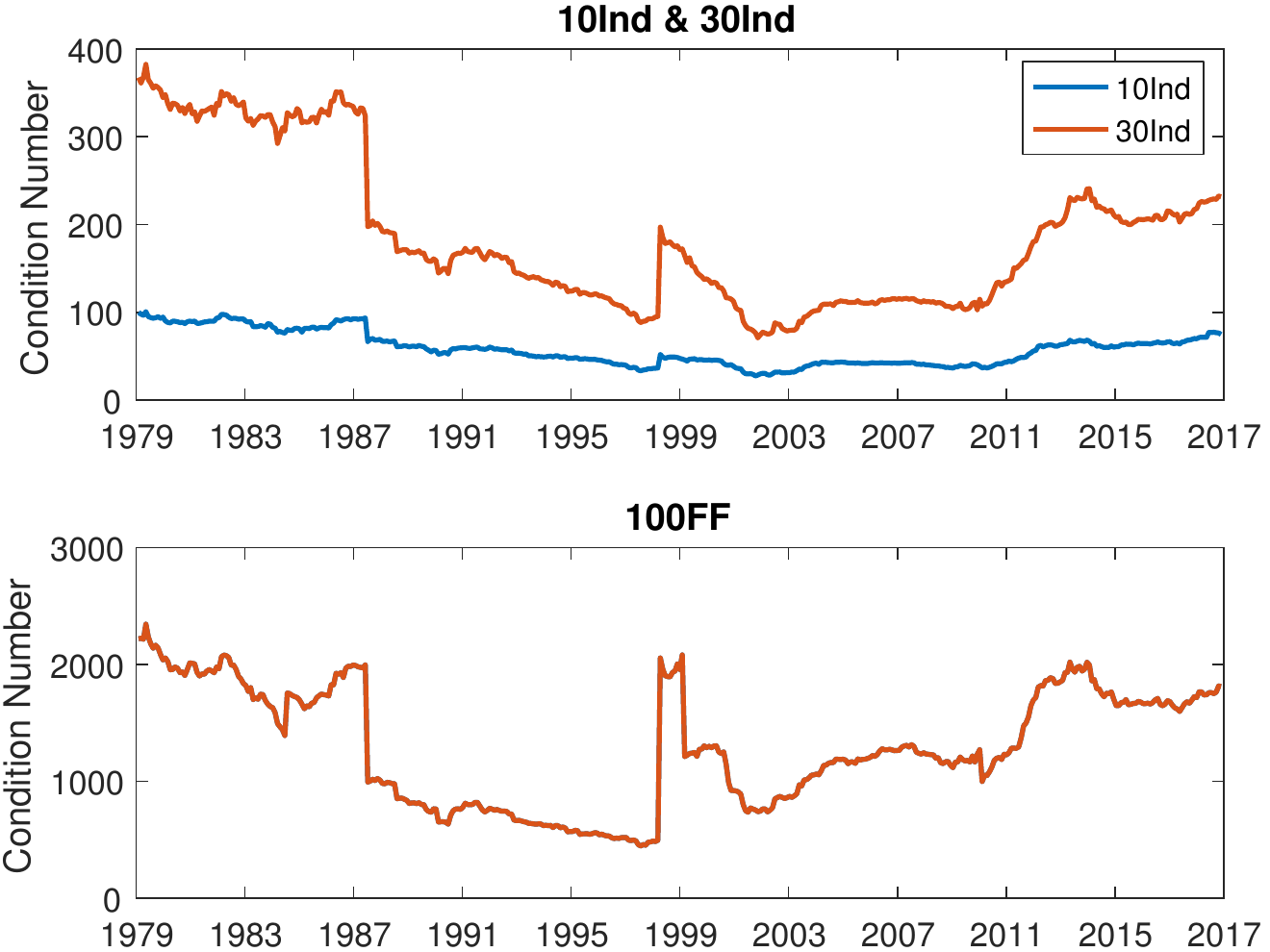} & \includegraphics[scale=.56]{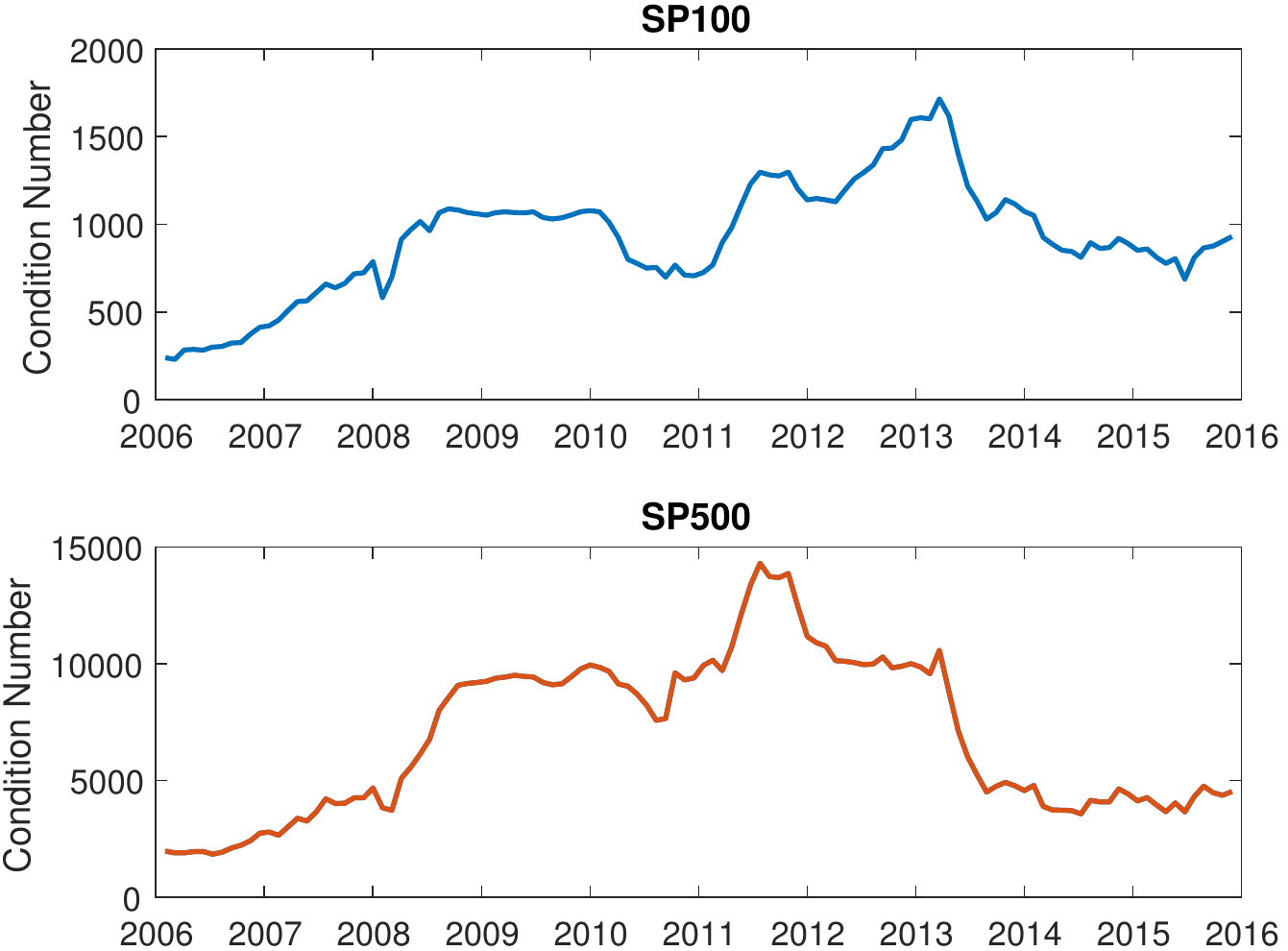} \\
\end{tabular}
}
\captionsetup{font=scriptsize,labelfont=scriptsize, width=0.95\textwidth}
\caption*{The figure shows the evolution of the condition number for the 10- and 30 Industry Portfolios, the 100 Fama French Portfolios on Size and Book-to-Market, as well as the S\&P100 and S\&P500. The condition number was computed based on the shrunken covariance matrix from \cite{Ledoit2004}, considering a window size of $\tau=500$ ($\tau =120$) observations for the S\&P Indices and the 10-, 30- and 100- Portfolios, respectively, and rebalancing the portfolio every month.}
\end{figure}

To evaluate our portfolios in an OOS setting, we rely on a rolling window approach with a window size of $\tau=120$ monthly observations for the 10- and 30Ind, the 100FF, and $\tau=500$ daily observations for the S\&P100 and S\&P500.\footnote{To test the robustness of our results, we account for different window sizes of $\tau =250, 750$ and $1000$ daily observations, and make the results available upon request. Results are qualitative similar.} All portfolios are re-balanced monthly, discarding the oldest and including the most recent observations, allowing for a total of $t=445$ ($t=115$) OOS returns for the monthly (daily) data. The rolling window approach for the daily data works as follows: the first $\tau$ return observations are used to estimate $\hat{\bfSigma}_{t}$, by using the shrinkage approach by \cite{Ledoit2004}. Then, $\hat{\bfSigma}$ is used as the input to compute the optimal weight vector $\hat{\bfw}_{t}$. The resulting portfolio is assumed to be held for the following 21 days. At $t+1$, the $k$ constituents' returns over this monthly period, $\boldsymbol{R}_{t+1}$, are used to compute the OOS portfolio return as: $R_{p, t+1} = \hat{\bfw}_{t} \boldsymbol{R}_{t+1}$. In the next step, we roll the data window forward, dropping the last and adding the most recent 21 observations to our training set. We then estimate a new weight vector, which determines our portfolio holdings and the OOS return for the next month. This process is repeated until the end of the data set is reached. The same procedure applies to the Industry and Fama French portfolios, though the window is rolled forward by one monthly observation instead of 21 daily observations.\\
Figure \ref{Cond_Numbers} plots the condition numbers \footnote{The condition number is defined as the ratio of the largest to the smallest singular value.} for the covariance matrix estimate: large values indicate that our estimate is very sensitive to changes in the underlying data structure. These large condition numbers often stem from multicollinearity  between the assets. As we rebalance the portfolio, the changes in the input parameter then lead to extreme changes in the portfolio weights and to high turnover levels.

%
%%%%%%%%%%%%%%%%%%%%%
%%%%%%%%%%%%%%%%%%%%%
%

\begin{figure}[h!]
\centering
\caption{Correlation Coefficients in Time.} \label{CorrelCoeffs_Ind}
\scalebox{0.9}{
\begin{tabular}{ccc}
\includegraphics[scale=.39]{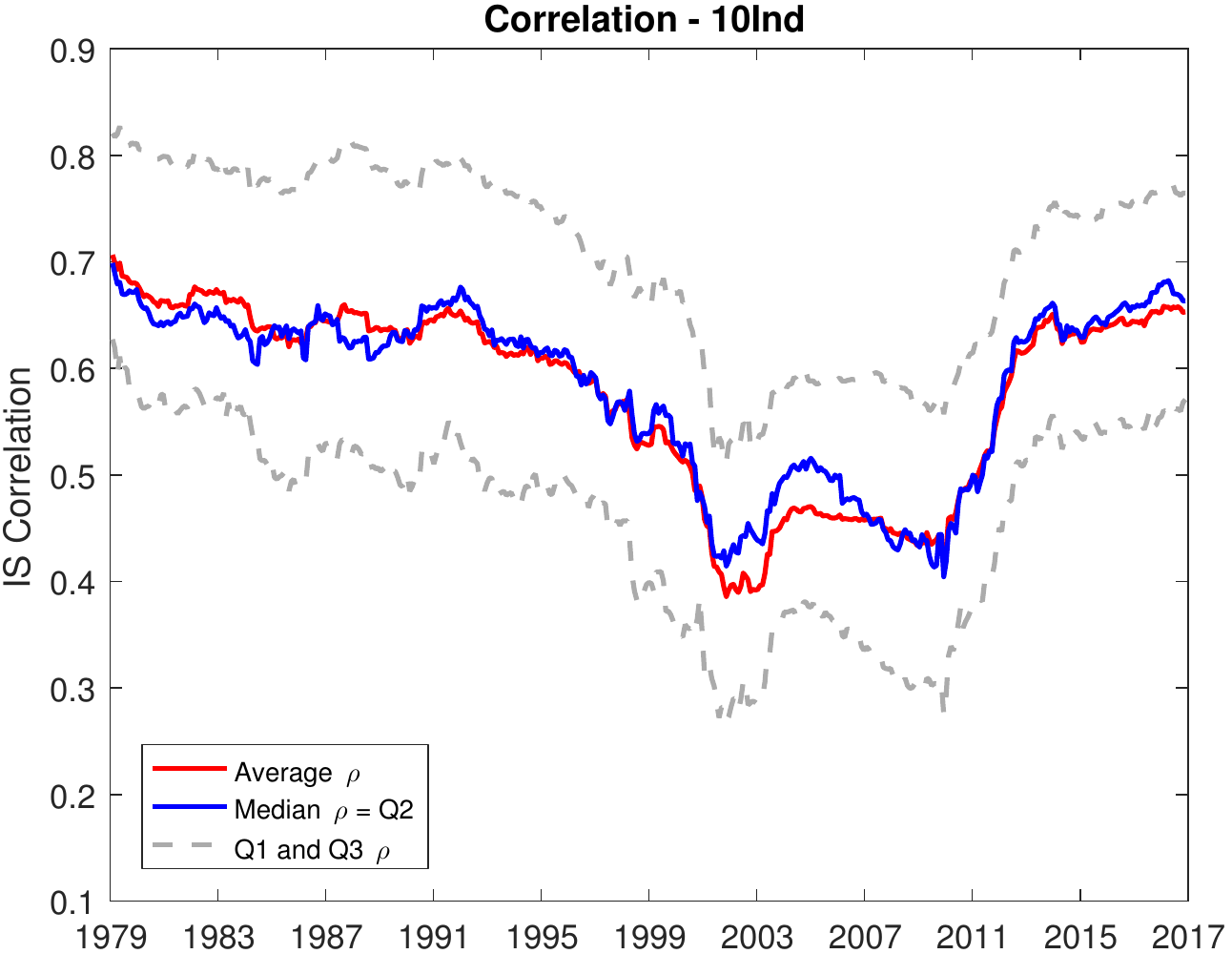} &
\includegraphics[scale=.39]{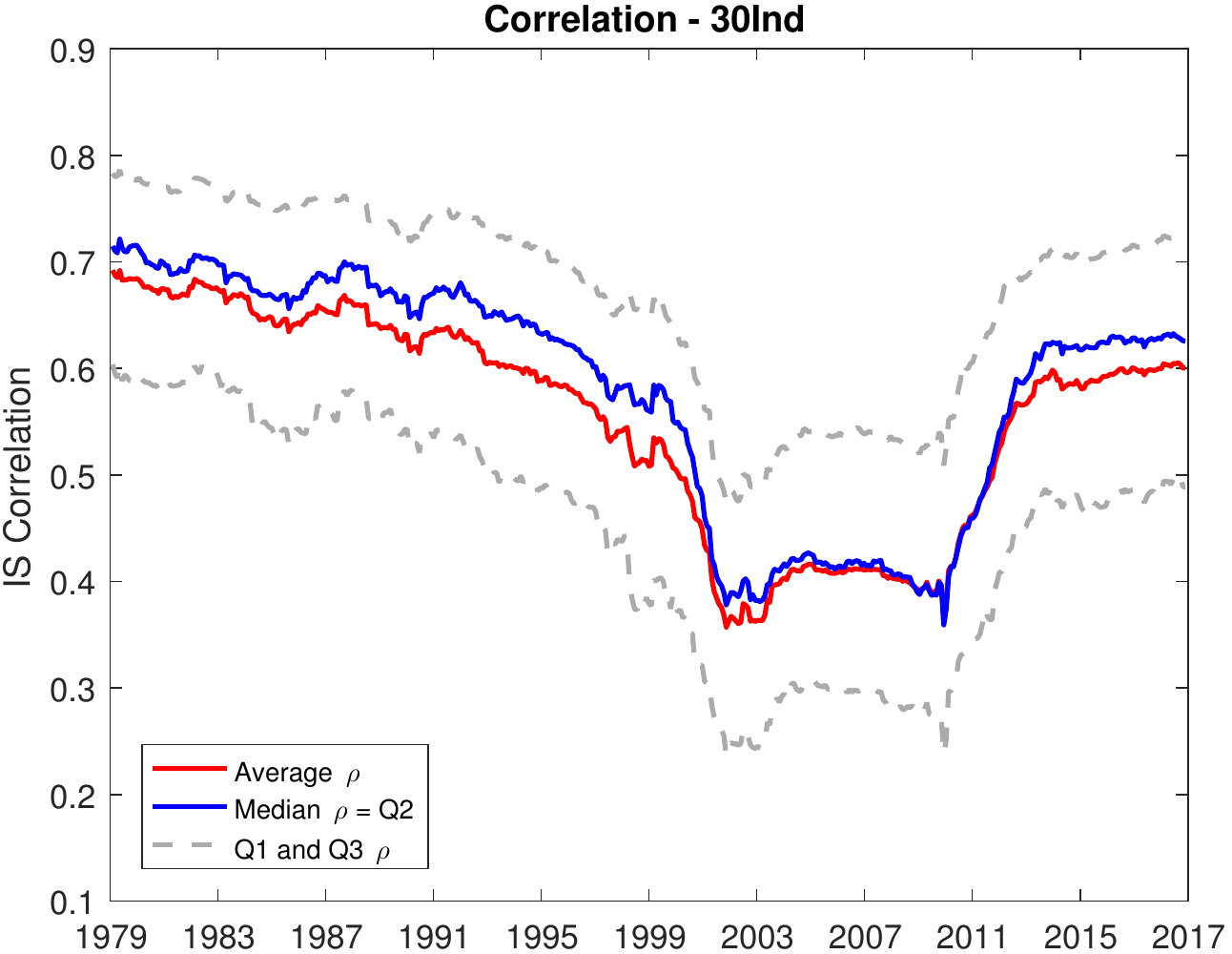} &
\includegraphics[scale=.39]{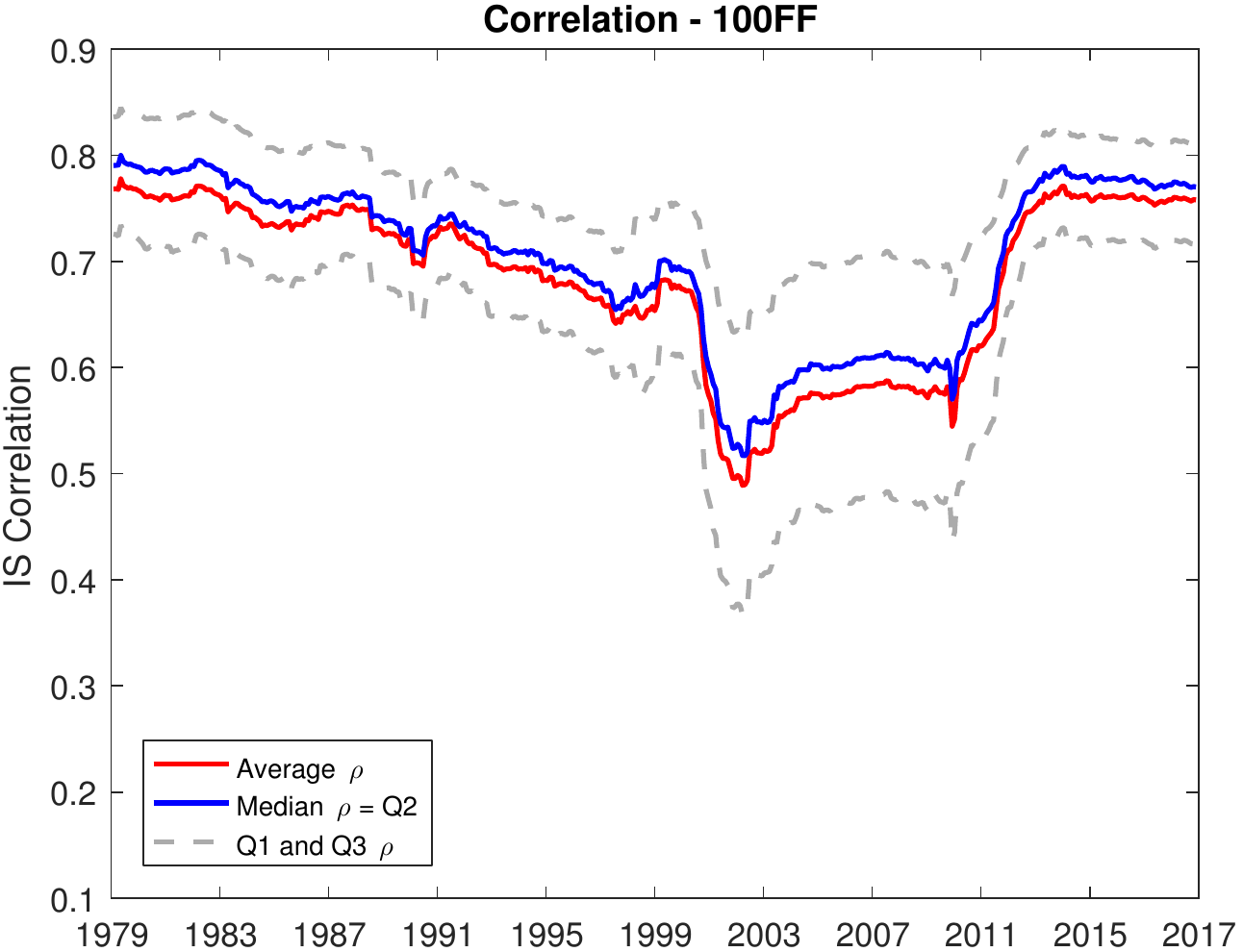} \\
\includegraphics[scale=.39]{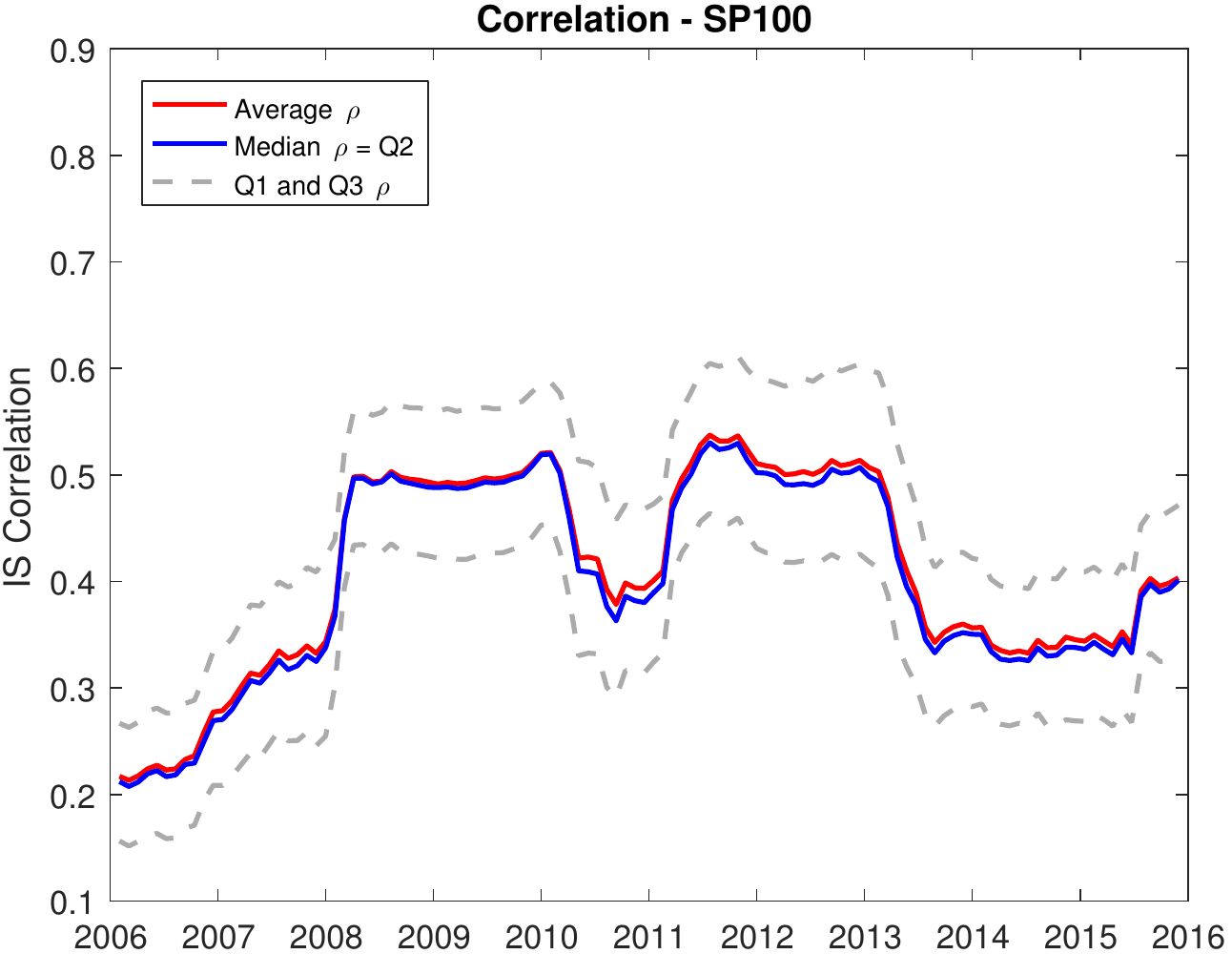} &
\includegraphics[scale=.39]{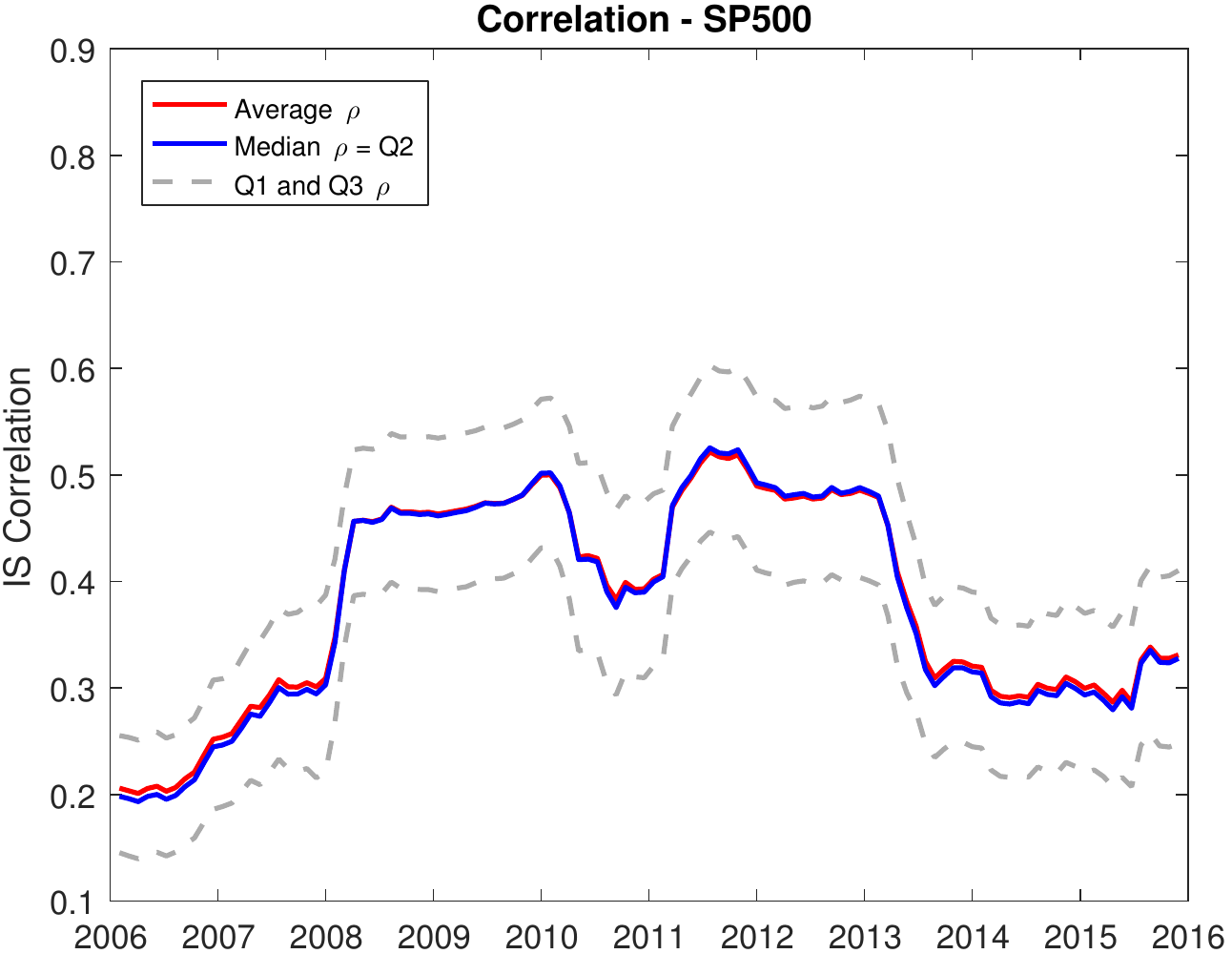} \\
\end{tabular}
}
\captionsetup{font=scriptsize,labelfont=scriptsize, width=\textwidth}
\caption*{The Figure plots the mean, the median, the first and the third quantile of the correlation coeffcients across all constituents for the 10-, 30-, and 100-Portfolios, as well as the SP100 and SP500 Indices, considering a windowsize of $\tau = 120$ month and $\tau =500$ daily observations, and re-balancing the portfolio every month over the period from 01/1970 - 01/2017 and 12/2004 to 01/2016, respectively. The correlation coefficients are computed based on the robust covariance matrix estimate by \cite{Ledoit2004}.}
\end{figure}

\noindent
Figure \ref{CorrelCoeffs_Ind} shows that the mean, the median, as well as the first and third quartiles of the correlation coefficient, across all constituents, are strictly positive for all data sets. In fact, the correlations of the industry and Fama French portfolios are high and positive in the late 1970s, decreasing slightly in the beginning, probably due to the new economy boom and the offset of the technology sector. However, with the burst of the DotCom Bubble and the beginning of the financial crisis in 2007, the correlation rises sharply for all three datasets. As for the Industry and Fama French portfolios, the values for the SP100 and SP500 peak with the onset of the financial crisis in 2007. After falling slightly, the correlation increases again in 2012, during the European sovereign debt crisis. The correlation coefficient plays an important role for our following analysis, as increased positive correlation among the constituents is reported to reduce the effects of diversification \citep{Choueifaty2008, You2010, Giuzio2016a}.
%
%%%%%%%%%%%%%%%%%
%%%%%%%%%%%%%%%%%
%
\noindent
For all portfolios, the optimal weights vector, $\hat{\bfw}_{t}$, depends on the choice of the optimal $\lambda$ parameter value. To select the optimal tuning parameter, we consider a grid of $100$ log-spaced values of $\lambda$ between $10^{-7.5}$ and $10^{1}$, from which we choose $\lambda_{LASSO} = \lambda_{1}=\alpha \Phi^{-1}\left(1-\frac{0.01}{2k}\right)$. The remaining elements $i=2,...,k$ of the $\lambda$ sequence for SLOPE are as before, i.e., $\lambda_{i}=\alpha \Phi^{-1}\left(1-\frac{0.01 i}{2k}\right)$.\\ %Note that by choosing $\lambda_{LASSO} = \lambda_{1}$, the lambda values for the remaining $k-1$ assets for the SLOPE penalty are lower than for the LASSO. Consequently, given that $\lambda_{LASSO} = \lambda_{1}$, the effect of shrinking the weights towards zero is lower for SLOPE than LASSO.\\
Among the 100 lambda values, we select the optimal tuning parameter for the various strategies in the following ways: As the RIDGE produces portfolios in between the GMV and the EW, we created a grid of six portfolios between these two points and choose the portfolio right before the EW. For the LASSO and the SLOPE we choose the portfolio that lies between the GMV and GMV-LO solution and which provides us with approximately 30\% of active positions. Note that as we increase the tuning parameter, beyond the GMV-LO, SLOPE moves towards the EW solution. As such, the portfolio could also lie in the no-short sale area. Still, we do not consider this interval, as we solve the optimization for SLOPE also with the long only constraint. Consequently, we explicitly want to exploit the grouping feature with SLOPE-LO and select from an interval of six portfolios in between the GMV-LO and EW solution, the one with the largest number of groups. To guarantee that all our portfolios can also be implemented in practice, all weights that are smaller in absolute value than the threshold of $0.05\%$ are set to zero.\\
Given the optimal portfolio vector $\hat{\boldsymbol{w}}$, we compute the OOS mean and the OOS standard deviation, defined as:
\begin{gather}
\hat{\mu}_{p} = \frac{1}{t} \sum_{i=1}^{t} \hat{\bfw}_{t} \boldsymbol{R}_{t+1} \\
\hat{\sigma}_{p} = \sqrt{\frac{1}{t-1} \sum_{i=1}^{t} (\hat{\bfw}_{t} \boldsymbol{R}_{t+1} - \hat{\mu}_{p})^{2}}
\end{gather}
from which we construct the OOS Sharpe Ratio (SR) as:
\begin{gather}
\widehat{SR} = \frac{\hat{\mu}_{p}}{\hat{\sigma}_{p}}
\end{gather}
To test whether the $\widehat{SR}$ and $\widehat{\sigma_{p}^{2}}$ of any portfolio is statistically different from our SLOPE procedure, we use the test developed by \cite{Ledoit2008} and \cite{Ledoit2011}.\\
As frequent re-balancing of a portfolio is costly, we complement our analysis by computing the turnover of each portfolio, defined as:
\begin{gather}
\widehat{TO} = \frac{1}{t} \sum_{i=1}^{t} ||\hat{\bfw}_{t+1}- \hat{\bfw}_{t}||_{1}
\end{gather}
Furthermore, we include the following diversification measures: the Diversification Ratio (DR), the weight (WDiv) and the risk diversification (RDiv) measures.
The DR is defined as the ratio of the weighted asset volatilities to the overall portfolio volatility:
\begin{gather}\label{DR}
\widehat{\text{DR}}=\frac{\sum_{i=1}^{k} \hat{w}_{i} \hat{\sigma}_{i}}{\hat{\sigma}_{p}}\;\;,
\end{gather}
where $\hat{\sigma}_{i}$ is the $i$-th asset's estimated volatility, and $\hat{\sigma}_{p}$ is the estimated portfolio volatility. This ratio is by construction always larger or equal to 1 and the investor typically prefers a higher value of the DR \citep{Choueifaty2008}.
%The Gini coefficient was initially introduced in the area of macroeconomics, to measure inequalities and is based on the famous Lorentz curve. Given the weight vector $\boldsymbol{\bfw}$ whereas each $w_{i} \in [0,1] \forall i=1,...,k$ with distribution function $\bf{F}$, the Lorenz curve is given by:
%\begin{gather}
%L(\boldsymbol{\bfw}) = \frac{\int_{0}^{w}\theta d F(\theta)}{\int_{0}^{1}\theta d F(\theta)}
%\end{gather}
%If $w_{i} = w_{j}, \forall i \neq j$, the Lorentz curve forms a $45^{\circ}$ line. When $w_{i} \neq w_{j}, \forall i \neq j$, the Lorenz curve falls below this $45^{\circ}$ line. Given this, the Gini coefficient indicates the amount of inequality:
%\begin{gather}
%\text{G} = 1 - 2\int_{0}^{1} L(\boldsymbol{w})
%\end{gather}
%By definition the Gini coefficient takes the value one for a perfectly concentrated portfolio (i.e.~one that is only invested in one security) and the value zero for an equally weighted representation.\\
Finally, both the WDiv and RDiv measure the concentration of the portfolio in terms of weights and risk \citep{Maillard2010, Roncalli2013}. The WDiv ranges from $\frac{1}{k}$ for a perfectly concentrated portfolio to $1$ for the equally weighted portfolio and can be computed as:
\begin{gather}
\widehat{\text{WDiv}}= \frac{1}{k \times \sum_{i=1}^{k} {\hat{w}_{i}^{2}}}
\end{gather}
We obtain the RDiv, by substituting the weights for the risk contribution, defined as $\widehat{RC}_{i} = \hat{w}_{i} \times \partial_{w_{i}}\sigma(\hat{\bfw}_{i})$, where $\partial_{w_{i}}\sigma(\hat{\bfw}_{i})$, defines the marginal contribution to risk (MRC) of asset $i$, that is the first derivative of the portfolio variance with respect to portfolio weight $w_{i}$. The MRC measures the sensitivity of the portfolio variance, given a change in asset $i$-th weight. The RDiv takes a value of $1$ for the equally-weighted risk contributions (ERC) portfolio, which is least concentrated in terms of risk contributions and $\frac{1}{k}$ for a portfolio which is fully concentrated on one asset, with regard to risk contribution:
\begin{gather}
\widehat{\text{RDiv}}=\frac{1}{k \times \sum_{i=1}^{k}{\widehat{RC}_{i}^{2}}}
\end{gather}
Summing up, we prefer large values of DR and values close to one for the WDiv and the RDiv \citep{Cazalet2014}.

%%%%%%%%%%%%%%%%%%%%%%%%%%%%%%%%%
\subsection{Empirical Results} %
%%%%%%%%%%%%%%%%%%%%%%%%%%%%%%%%%
\subsubsection*{Industry and Fama French Portfolios}
Table \ref{RiskReturn_Ind} reports the annualized OOS volatility, the annualized OOS SR, the number of active positions, the turnover, and the Value-at-Risk (VaR), evaluated at the 5\% significance level, for the 10Ind, 30Ind and the 100FF, using a window size of $\tau = 120$ observations with monthly re-balancing. We indicate portfolios that are statistically different from our SLOPE procedure at the 10\%, 5\% and 1\% level, given the test for the difference in the SR and the volatility, following \cite{Ledoit2008} and \cite{Ledoit2011}.\\
Looking at the values for the OOS volatility in Table \ref{RiskReturn_Ind}, we observe that no portfolio is statistically different from our new SLOPE procedure, across any of the three data sets. Still, SLOPE yields consistently lower variance than any of the EW, ERC, RIDGE or GMV-LO portfolios. Similar performance can be seen for our two portfolio strategies, SLOPE-LO and SLOPE-MV, with regard to the EW, the ERC and the RIDGE. 

%%%%%%%%%%%%%%%%%%%%%%%%%%%%%%%%%%%%%%%%%%%%%%%
% Table generated by Excel2LaTeX from sheet 'WS=120'
\begin{table}[h!]
  \centering
  \caption{Risk- and Return Measures - Industry and Fama French Portfolios.}\label{RiskReturn_Ind}
  {\scalebox{0.55}{
   \begin{tabular}{l ddd dddddddddddd}
    \toprule
    \toprule
      & \multicolumn{3}{l}{Vol. (in \%)} & \multicolumn{3}{l}{Sharpe Ratio} & \multicolumn{3}{l}{AP} & \multicolumn{3}{l}{Turnover} & \multicolumn{3}{l}{VaR 5\% (in \%)} \\
      \cmidrule(lr){2-4} \cmidrule(lr){5-7} \cmidrule(lr){8-10} \cmidrule(lr){11-13} \cmidrule(lr){14-16}
      & \multicolumn{1}{c}{10Ind} & \multicolumn{1}{c}{30Ind} & \multicolumn{1}{c}{100FF} & \multicolumn{1}{c}{10Ind} & \multicolumn{1}{c}{30Ind} & \multicolumn{1}{c}{100FF} & \multicolumn{1}{c}{10Ind} & \multicolumn{1}{c}{30Ind} & \multicolumn{1}{c}{100FF} & \multicolumn{1}{c}{10Ind} & \multicolumn{1}{c}{30Ind} & \multicolumn{1}{c}{100FF} & \multicolumn{1}{c}{10Ind} & \multicolumn{1}{c}{30Ind} & \multicolumn{1}{c}{100FF} \\
    \midrule
    EW & 14.489 & 16.255 & 17.507 & 0.780^{**} & 0.661^{***} & 0.682^{***} & 10.000 & 30.000 & 100.000 & 0.000 & 0.000 & 0.000 & -5.880 & -6.418 & -7.182 \\
        GMV & 10.911 & 9.153 & 6.028 & 1.116^{*} & 1.319^{**} & 3.309^{***} & 9.982 & 29.885 & 99.470 & 0.086 & 0.221 & 0.759 & -4.476 & -3.576 & -1.390 \\
    GMV-LO & 11.474 & 11.214 & 13.129 & 1.018 & 1.005 & 0.963^{***} & 5.371 & 8.569 & 9.292 & 0.042 & 0.055 & 0.085 & -4.881 & -4.657 & -5.592 \\
    ERC & 13.578 & 15.029 & 16.905 & 0.845 & 0.735^{***} & 0.718^{***} & 10.000 & 30.000 & 100.000 & 0.006 & 0.007 & 0.005 & -5.247 & -5.898 & -6.920 \\
    RIDGE & 12.288 & 13.504 & 15.304 & 0.945 & 0.840^{***} & 0.857^{***} & 9.962 & 29.798 & 97.737 & 0.027 & 0.046 & 0.037 & -4.809 & -5.213 & -6.132 \\
    LASSO & 11.423 & 10.876 & 9.664 & 1.021 & 1.043 & 1.765^{***} & 6.162 & 11.856 & 28.265 & 0.052 & 0.107 & 0.293 & -4.839 & -4.570 & -3.385 \\
    SLOPE & 11.429 & 10.939 & 9.961 & 1.024 & 1.045 & 1.665 & 6.303 & 12.551 & 30.155 & 0.049 & 0.099 & 0.254 & -4.864 & -4.541 & -3.744 \\
    SLOPE - LO & 11.686 & 11.870 & 13.704 & 0.963^{*} & 0.968 & 0.953^{***} & 7.299 & 18.465 & 34.616 & 0.104 & 0.206 & 0.403 & -4.826 & -4.753 & -5.639 \\
    SLOPE - MV & 12.726 & 12.511 & 14.038 & 0.854^{**} & 0.864^{**} & 0.901^{***} & 2.609 & 3.389 & 5.002 & 0.018 & 0.022 & 0.056 & -5.282 & -4.955 & -5.381 \\
    \bottomrule
    \bottomrule
    \end{tabular}%
    }}
\captionsetup{font=scriptsize,labelfont=scriptsize, width=\textwidth}
\caption*{The table reports the OOS Risk and Return Measures for the 10-, 30-, and 100-Portfolios, considering a windowsize of $\tau = 120$ monthly observations and re-balancing the portfolio every month over the period from 12/2004 to 01/2016. Reported are: The annualized OOS Volatility, the annualized OOS Sharpe Ratio, the number of active positions (AP), the average total turnover, and the Value at Risk (VaR) evaluated at the 5\% significance. Furthermore, we report the significance for the difference in the Volatility and the SR, with regard to SLOPE, and where $^{*}$, $^{**}$, $^{***}$, indicates a significance at the $10\%$, $5\%$ and $1\%$ level for tests on variances and Sharpe Ratios.}
\end{table}%
%%%%%%%%%%%%%%%%%%%%%%%%%%%%%%%%%%%%%%%%%%

\noindent
Simultaneously, the values for the OOS SR, establish SLOPE among the best performing portfolios, with some results being statistically significant. In detail: we observe the highest SR after the GMV for the industry portfolios, while having statistically higher SR than all strategies, except from the GMV and the LASSO, for the Fama-French portfolios. Finally, across all data sets, SLOPE is able to statistically significantly outperform the EW, challenging its widely reported characteristic of a tough benchmark to beat \citep{DeMiguel2009}. This also holds for SLOPE-LO and SLOPE-MV. \\
Beside reducing the overall portfolio variance, our goal is to construct sparse portfolios with a low turnover. For that, reconsider that the EW always invests naively in all constituents and thus reports a turnover value of zero by definition, but at the same time the highest possible number of active positions. Similar values are reported for the ERC, which aims at equalizing the risk contribution of each asset to overall portfolio risk. These two portfolios are closely followed by the GMV and the RIDGE penalty. The GMV has an exposure to all assets, and estimation error can enter unhindered into the optimization leading to unstable weights (see i.e., \cite{Ledoit2004}). Looking at Figure \ref{Cond_Numbers}, the large values of the condition number imply a high instability due to the presence of multicollinearity, leading to changes in the portfolio composition. Consequently, the GMV has the highest turnover values among the non-regularization strategies. The RIDGE, on the other hand, results in more stable asset allocations, despite not setting any asset weight exactly equal to zero. Although both strategies should invest in all assets, the number of active positions are slightly reduced, due to our imposed threshold. \\
Compared to the strategies above, our new SLOPE procedure is able to promote sparse solutions and to reduce the overall portfolio turnover. In fact, we consistently report a lower turnover than the LASSO portfolio across all three data sets.\\
Especially interesting is the performance of SLOPE-MV: although the investor suffers an increase in volatility, he is still able to outperform the ERC, RIDGE or EW. Furthermore, this portfolio has the smallest number of active positions, together with the smallest turnover value, among all sparse portfolio methods.  In fact, our new SLOPE procedure provides the investor with a large amount of flexibility, as with a larger lambda value the penalty starts to form groups among assets, assigning them the same weight.
%Figure \ref{Groups} plots the number of groups for the SLOPE-LO solution. From the Figure, we can observe that with a larger lambda value SLOPE starts to form groups among assets, assigning them the same weight.
%%
%%%%%%%%%%%%%%%%%%%%%%%%%%%%%%%%%%%%%%%%%%%
%\begin{figure} [h!]
%\FIGURE
%{\begin{tabular}{cc}
%\includegraphics[scale=.61]{Groups_Ind} & \includegraphics[scale=.61]{Groups_SP} \\
%\end{tabular}}
%{Grouping Structure.\label{Groups}}
%{The Figure shows over the whole time period and considering a windowsize of $\tau = 500$ observations the number of Groups, i.e. the number of coefficients which are assigned to two or more assets, for the 10Ind, 30Ind and 100FF, on the left, as well as for the SP100 and SP500 on the right.}
%\end{figure}
%%%%%%%%%%%%%%%%%%%%%%%%%%%%%%%%%%%%%%%%%%%
%%
This is of special interest for investors who want to move beyond the property of statistical shrinkage, to include any form of financial measure, like among others fundamental multiples (i.e. Price/Earnings, Dividends/Earnings), accounting values (i.e., net income, Free Cash Flow) or other quantitative measures (i.e., Value-at-Risk or Expected Shortfall), in their portfolio construction process. We show here just a simple strategy, SLOPE-MV, that selects the one asset with minimum volatility out of the groups, while other strategies could be easily developed. 

\begin{figure}[h!]
\centering
\caption{Risk and Weight Diversification Frontier.}\label{Frontier_Ind}
\scalebox{0.94}{
\begin{tabular}{cc}
\includegraphics[scale=.57]{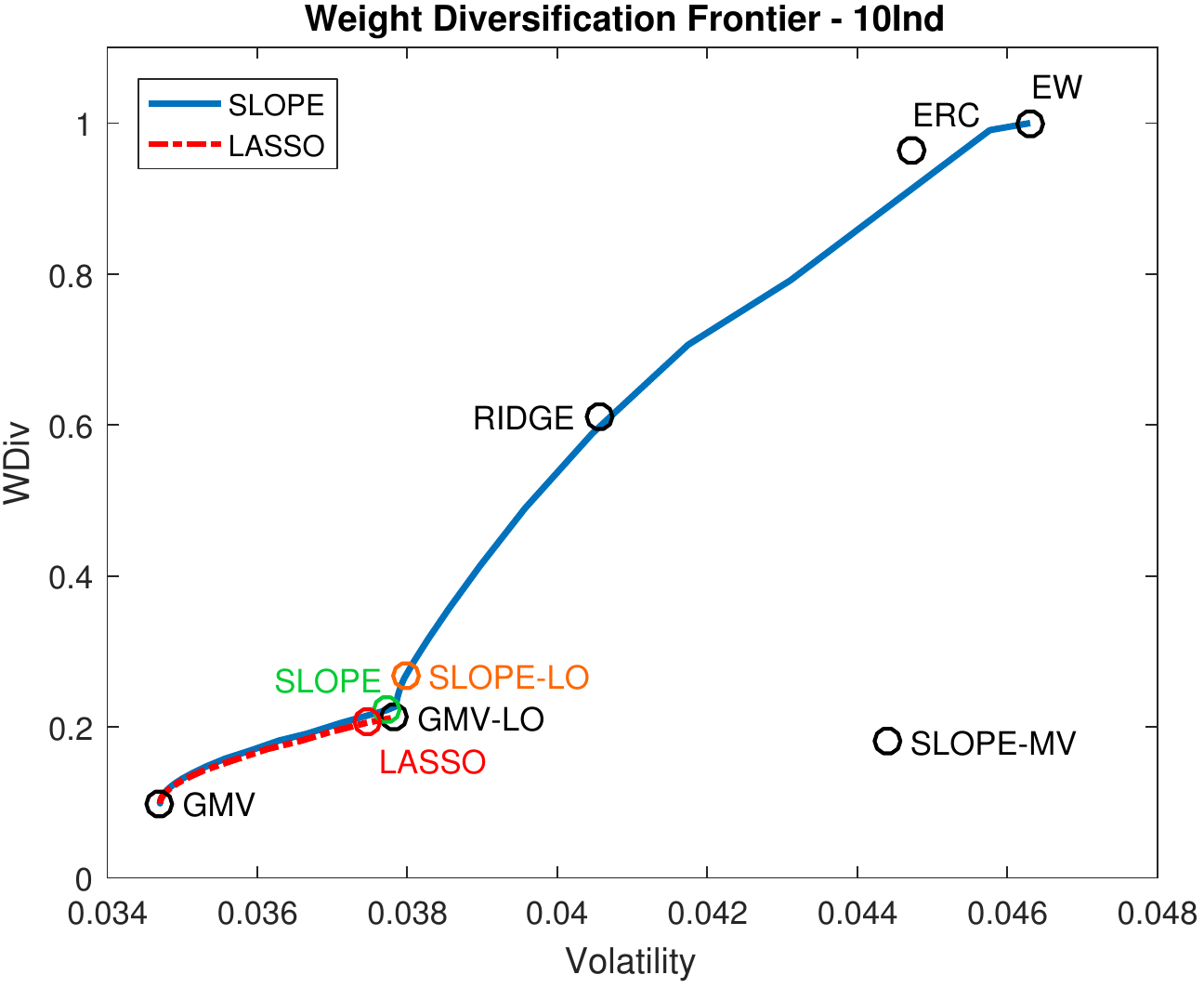} &
\includegraphics[scale=.57]{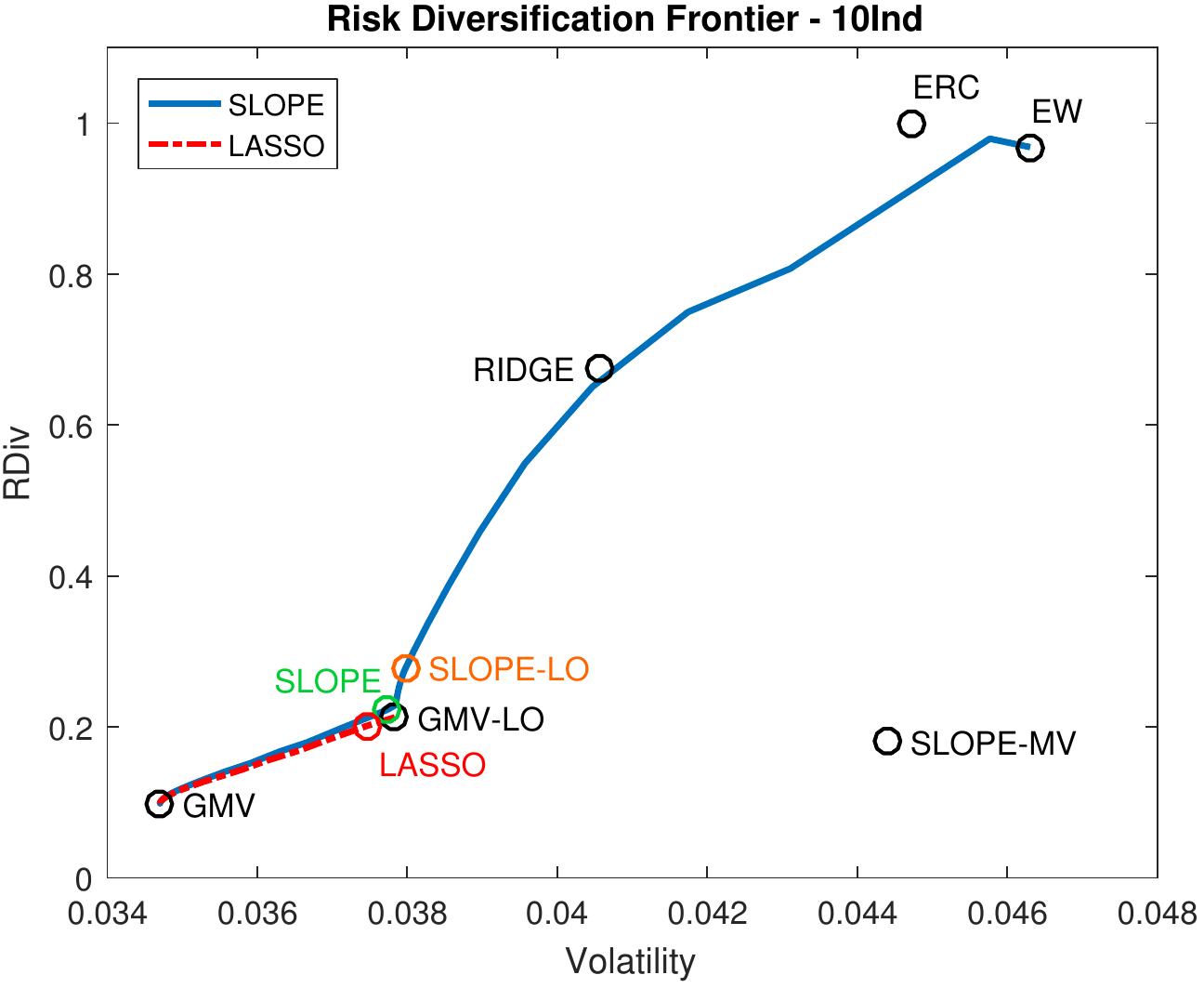}
\end{tabular}
}
\captionsetup{font=scriptsize,labelfont=scriptsize, width=\textwidth}
\caption*{The Figure shows on the left the weight diversification and on the right the risk diversification frontier, both reporting on the x-axis the portfolio volatility and on the y-axis the risk and weight diversification measure, respectively. Considered are the first window size of $\tau =120$ months for the 10Ind. Plotted are the resulting combinations for the GMV, the GMV-LO, the EW, the ERC, as well as the different combinations for the LASSO and the SLOPE procedure, considering a range of lambda values from $10^{-7.5}$ to $10^{1}$.}
\end{figure}

\begin{table}[htbp]
  \centering
  \caption{Diversification Measures - Industry and Fama French Portfolios.} \label{RiskMes_Ind}
  \scalebox{0.85}{
    \begin{tabular}{l ddddddddddddddd}
    \toprule
    \toprule
      & \multicolumn{3}{l}{DR}  & \multicolumn{3}{l}{WDiv} & \multicolumn{3}{l}{RDiv}   \\
 \cmidrule(lr){2-4} \cmidrule(lr){5-7} \cmidrule(lr){8-10}
      & \multicolumn{1}{c}{10Ind} & \multicolumn{1}{c}{30Ind} & \multicolumn{1}{c}{100FF} & \multicolumn{1}{c}{10Ind} & \multicolumn{1}{c}{30Ind} & \multicolumn{1}{c}{100FF} & \multicolumn{1}{c}{10Ind} & \multicolumn{1}{c}{30Ind} & \multicolumn{1}{c}{100FF} \\
    \midrule
EW & 1.270 & 1.343 & 1.212 & 1.000 & 1.000 & 1.000 & 0.933 & 0.935 & 0.958 \\
    GMV & 1.255 & 1.362 & 0.958 & 0.197 & 0.078 & 0.013 & 0.197 & 0.078 & 0.013 \\
    GMV-LO & 1.289 & 1.414 & 1.299 & 0.320 & 0.150 & 0.062 & 0.320 & 0.150 & 0.062 \\
    ERC & 1.300 & 1.382 & 1.225 & 0.935 & 0.914 & 0.963 & 1.000 & 1.000 & 1.000 \\
    RIDGE & 1.328 & 1.430 & 1.244 & 0.655 & 0.629 & 0.622 & 0.735 & 0.722 & 0.617 \\
    LASSO & 1.289 & 1.411 & 1.180 & 0.316 & 0.145 & 0.035 & 0.313 & 0.135 & 0.022 \\
    SLOPE & 1.293 & 1.424 & 1.200 & 0.328 & 0.157 & 0.043 & 0.326 & 0.148 & 0.027 \\
    SLOPE - LO & 1.315 & 1.457 & 1.295 & 0.417 & 0.287 & 0.209 & 0.437 & 0.319 & 0.221 \\
   SLOPE - MV & 1.163 & 1.220 & 1.205 & 0.179 & 0.072 & 0.036 & 0.179 & 0.072 & 0.036 \\
    \bottomrule
    \bottomrule
    \end{tabular}%
    }
\captionsetup{font=scriptsize,labelfont=scriptsize, width=\textwidth}
\caption*{The table reports the diversification measures for the 10-, 30-, and 100- Portfolios, considering a windowsize of $\tau = 120$ monthly observations and re-balancing the portfolio every month over the period from 01/1970 to 01/2017. Reported are: The Diversification Ratio (DR), the Weight Diversification (WDiv) and the Risk Diversification (RDiv) measures.}
\end{table}%
%%%%%%%%%%%%%%%%%%%%%%%%%%%%%%%%%%%%%%%%%%
%
\noindent
Finally, Table \ref{RiskMes_Ind} complements our risk and return analysis for the 10Ind, 30Ind and 100FF, and reports the DR, the WDiv, and the RDiv. As the EW invest equally in all assets, it achieves the best values for the WDiv by definition, with similar values reported for the ERC. As the ERC aims to equalize the contribution to portfolio risk from each asset, it also reports the highest values for the RDiv. SLOPE-LO shows the best diversification measures, followed by SLOPE, while both consistently dominating the LASSO across all data sets. \\
As before, SLOPE does not only outperform the LASSO, but also provides flexibility with regard to the diversification measures. For that, Figure \ref{Frontier_Ind} plots the weight- and risk diversification measure against the attainable portfolio volatility for the LASSO and the SLOPE, together with the EW, the ERC, the GMV and the GMV-LO, by considering the first window size of $\tau=120$ observations for the 10Ind.
%
%%%%%%%%%%%%%%%%%%%%%%%%%%%%%%%%%%%%%%%%%%
%%%%%%%%%%%%%%%%%%%%%%%%%%%%%%%%%%%%%%%%%%
%
For both frontiers, the full grid of lambda parameters for the LASSO enables the investor to select only a combination between the GMV and the GMV-LO solution. SLOPE, on the other hand, is able to span a much larger set of portfolios, beginning from the GMV, via the GMV-LO to the EW. The investor can thus control the trade-off between diversification and volatility out of a much larger set of portfolios, to find the allocation that best fits her individual preferences.

\subsubsection*{S\&P Indices}
Table \ref{RiskReturn_SP} reports the annualized OOS portfolio volatility, the annualized OOS SR, the number of active positions, the turnover, and the $5\%$ VaR, for the S\&P100 and S\&P500 Indices, using a window size of $\tau = 500$ daily observations with monthly re-balancing (i.e., 21 observations). %Again we use asterisks to indicate strategies that are statistically different from our SLOPE procedure at the $10\%$, $5\%$ and $1\%$ significance levels, following the tests of \cite{Ledoit2008} and \cite{Ledoit2011}.

%%%%%%%%%%%%%%%%%%%%%%%%%%%%%%%%%%%%%%%%%%
% Table generated by Excel2LaTeX from sheet 'WS=500'
\begin{table}[h!]
  \centering
  \caption{Risk- and Return Measures - S\&P Indices.} \label{RiskReturn_SP}
  \scalebox{0.8}{
    \begin{tabular}{ldd dd dd dd dd dd}
    \midrule
    \midrule
      & \multicolumn{2}{l}{Vol. (in \%)} & \multicolumn{2}{l}{Sharpe Ratio} & \multicolumn{2}{l}{AP} & \multicolumn{2}{l}{Turnover} & \multicolumn{2}{l}{VaR 5\% (in \%)} \\
    \cmidrule(lr){2-3} \cmidrule(lr){4-5} \cmidrule(lr){6-7} \cmidrule(lr){8-9} \cmidrule(lr){10-11}
 & \multicolumn{1}{c}{SP100} & \multicolumn{1}{c}{SP500} & \multicolumn{1}{c}{SP100} & \multicolumn{1}{c}{SP500} & \multicolumn{1}{c}{SP100} & \multicolumn{1}{c}{SP500} & \multicolumn{1}{c}{SP100} & \multicolumn{1}{c}{SP500} & \multicolumn{1}{c}{SP100} & \multicolumn{1}{c}{SP500}  \\
 \midrule
   EW & 18.855 & 20.238 & 0.254 & 0.210 & 93.000 & 443.000 & 0.000 & 0.000 & -7.431 & -8.155 \\
       GMV & 10.829 & 11.522 & 0.479 & 0.340 & 92.035 & 434.377 & 0.695 & 2.669 & -5.800 & -6.963 \\
    GMV-LO & 12.338 & 10.822 & 0.231 & 0.415 & 18.421 & 30.298 & 0.159 & 0.227 & -6.558 & -4.929 \\
    ERC & 16.574 & 17.947 & 0.295 & 0.238 & 93.000 & 443.000 & 0.020 & 0.022 & -6.819 & -7.366 \\
    RIDGE & 14.243 & 14.732 & 0.406 & 0.345 & 91.070 & 399.509 & 0.063 & 0.070 & -6.716 & -6.854 \\
    LASSO & 11.276 & 9.401 & 0.279 & 0.615 & 36.307 & 146.184 & 0.256 & 0.447 & -6.990 & -4.426 \\
    SLOPE & 11.202 & 9.481 & 0.316 & 0.584 & 38.789 & 153.860 & 0.238 & 0.418 & -6.945 & -4.609 \\
    SLOPE - LO & 12.474 & 11.991 & 0.392 & 0.402 & 44.991 & 129.632 & 0.464 & 0.584 & -6.197 & -6.221 \\
    SLOPE - MV & 13.513 & 13.804 & 0.465 & 0.305 & 10.307 & 13.289 & 0.111 & 0.131 & -5.635 & -5.828 \\
    \bottomrule
    \bottomrule
    \end{tabular}%
    }
\captionsetup{font=scriptsize,labelfont=scriptsize, width=\textwidth}
\caption*{The table reports the OOS Risk and Return Measures for the S\&P100 and S\&P500 Indices, considering a windowsize of $\tau = 500$ daily observations and re-balancing the portfolio every month over the period from 12/2004 to 01/2016. Reported are: The annualized OOS volatility , the annualized OOS Sharpe Ratio, the number of active positions (AP), the average total turnover, and the Value at Risk (VaR) evaluated at the 5\% significance. Furthermore, we report the significance for the difference in the Volatility and the SR, with regard to SLOPE, and where $^{*}$, $^{**}$, $^{***}$, indicates a significance at the $10\%$, $5\%$ and $1\%$ level for tests on variances and Sharpe Ratios.}
\end{table}%
%%%%%%%%%%%%%%%%%%%%%%%%%%%%%%%%%%%%%%%%%%
%
\noindent
Table \ref{RiskReturn_SP} shows that, with regard to the OOS variance and the SR, no strategy is statistically significantly different form each other. 
Still, SLOPE and LASSO performs best for the SP500, reporting the smallest variance among all strategies and the highest SR. This improved performance can be explained twofold: first, we have a situation in which the number of observations is only marginally bigger than the size of our investment universe. Thus, our estimates are very prone to estimation error. We can see that together with the shrunken covariance matrix, SLOPE and LASSO are able to shrink these extreme estimates. Furthermore, we explicitly select for the LASSO and the SLOPE, a portfolio with a moderate amount of short sales, making it possible to further exploit diversification benefits. The resulting portfolios show then a smaller variance as compared to the GMV-LO. At the same time, and compared to the LASSO, SLOPE is again able to reduce overall turnover and the cost of implementing such strategy. \\
The results for our two portfolio strategies are mixed. As we explicitly restrict short sales for the SLOPE-LO and SLOPE-MV, the resulting variances are higher, but still outperform some of the other strategies. Still, SLOPE-MV has again the smallest number of active position and the smallest turnover for both strategies. Depending on the investors objectives, he is then able to further reduce the transaction and monitoring costs. 
%To provide an overview of the performance of all methods, Figure \ref{Performance_Plots} compares the evolution of wealth of the various portfolio strategies for the 10- and 30Ind, as well as the SP500 index, starting at a normalized value of 100. Across the three data sets we consider, SLOPE or any of the constructed strategies consistently perform among the best of the portfolio strategies, only being inferior to one or two strategies. Furthermore, they even show a lower draw-down in wealth for the SP500 during the recent financial crisis.\\
%
%%
%%%%%%%%%%%%%%%%%%
%\begin{figure}[htbp!]
%\FIGURE
%{\begin{tabular}{ccc}
%\includegraphics[scale=.36]{OOS_PerformancePlot_Sample_4_WS_120} &
%\includegraphics[scale=.36]{OOS_PerformancePlot_Sample_5_WS_120} &
%\includegraphics[scale=.36]{OOS_PerformancePlot_Sample_2_WS_500} \\
%\end{tabular}}
%{Portfolio Strategy Performance Plots.\label{Performance_Plots}}
%{The Figure plots for the 10- and 30Ind, as well as the SP500 the evolution of the wealth, considering all nine portfolio strategy and starting with a normalized value of 100.}
%\end{figure}
%%%%%%%%%%%%%%%%%%
%%

\noindent
Finally, Table \ref{DivMes_SP} reports the risk diversification measures for the SP100 and SP500. For the DR the evidence is mixed with SLOPE-LO doing better for the SP100 and SLOPE for the SP500. The difference might be due to short sales in the portfolio, making it possible to exploit more diversification benefits. For the WDiv and RDiv, both SLOPE and SLOPE-LO again dominate LASSO, confirming previous results.
%
%%%%%%%%%%%%%%%%%%%%%%%%%%%%%%%%%%%%%%%%%%
\begin{table}[h!]
  \centering
  \caption{Diversification Measures - S\&P Indices.} \label{DivMes_SP}
  \begin{tabular}{l dddddddddd}
    \toprule
    \toprule
      & \multicolumn{2}{l}{DR}  & \multicolumn{2}{l}{WDiv} & \multicolumn{2}{l}{RDiv}  \\
      \cmidrule(lr){2-3} \cmidrule(lr){4-5} \cmidrule(lr){6-7}
      & \multicolumn{1}{c}{SP100} & \multicolumn{1}{c}{SP500}  & \multicolumn{1}{c}{SP100} & \multicolumn{1}{c}{SP500} & \multicolumn{1}{c}{SP100} & \multicolumn{1}{c}{SP500} \\
     \toprule
    EW & 1.586 & 1.675 & 1.000 & 1.000 & 0.892 & 0.894 \\
    GMV & 1.576 & 3.147 & 0.050 & 0.011 & 0.050 & 0.012 \\
    GMV-LO & 1.631 & 1.944 & 0.090 & 0.032 & 0.090 & 0.032 \\
    ERC & 1.630 & 1.728 & 0.892 & 0.880 & 1.000 & 1.000 \\
    RIDGE & 1.662 & 1.797 & 0.661 & 0.567 & 0.660 & 0.487 \\
    LASSO & 1.630 & 2.265 & 0.104 & 0.061 & 0.081 & 0.027 \\
    SLOPE & 1.650 & 2.235 & 0.117 & 0.070 & 0.091 & 0.031 \\
    SLOPE-LO & 1.706 & 1.936 & 0.285 & 0.206 & 0.312 & 0.219 \\
    SLOPE-MV & 1.474 & 1.528 & 0.062 & 0.013 & 0.062 & 0.013 \\
    \bottomrule
    \bottomrule
    \end{tabular}
    \captionsetup{font=scriptsize,labelfont=scriptsize, width=0.8\textwidth}
\caption*{The table reports the diversification measures for the S\&P100 and S\&P500 Indices, considering a windowsize of $\tau = 500$ daily observations and re-balancing the portfolio every month over the period form 12/2004 to 01/2016. Reported are: The Diversification Ratio (DR), the Weight Diversification (WDiv) and the Risk Diversification (RDiv) measures.}
\end{table}%
%%%%%%%%%%%%%%%%%%%%%%%%%%%%%%%%%%%%%%%%%%
%
\noindent
\subsubsection*{Robustness Tests}
\begin{figure} [h!]
\centering
\caption{Transaction Cost Regimes.}\label{TC}
\scalebox{0.9}{
\begin{tabular}{cc}
\includegraphics[scale=.52]{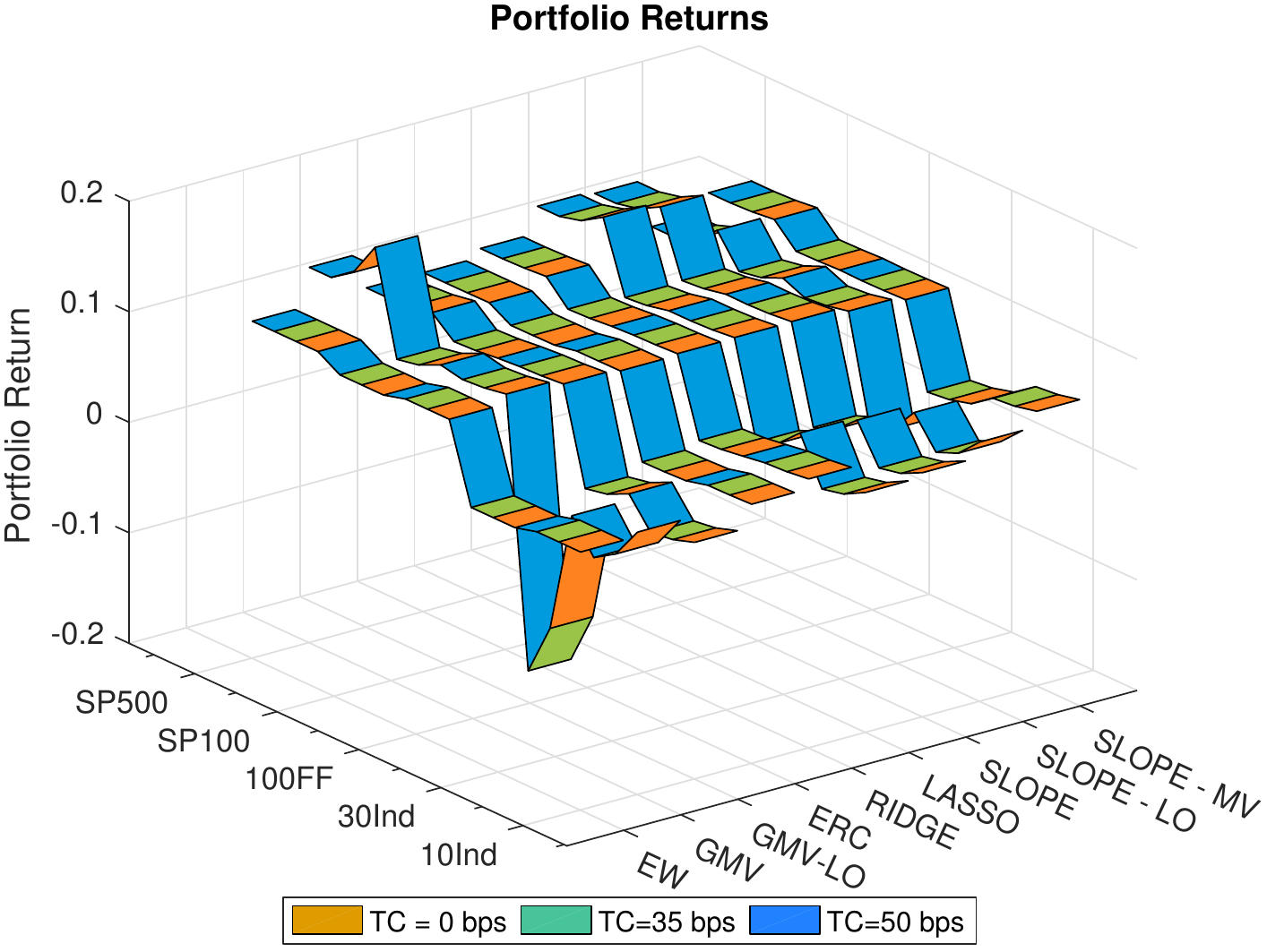} & \includegraphics[scale=.52]{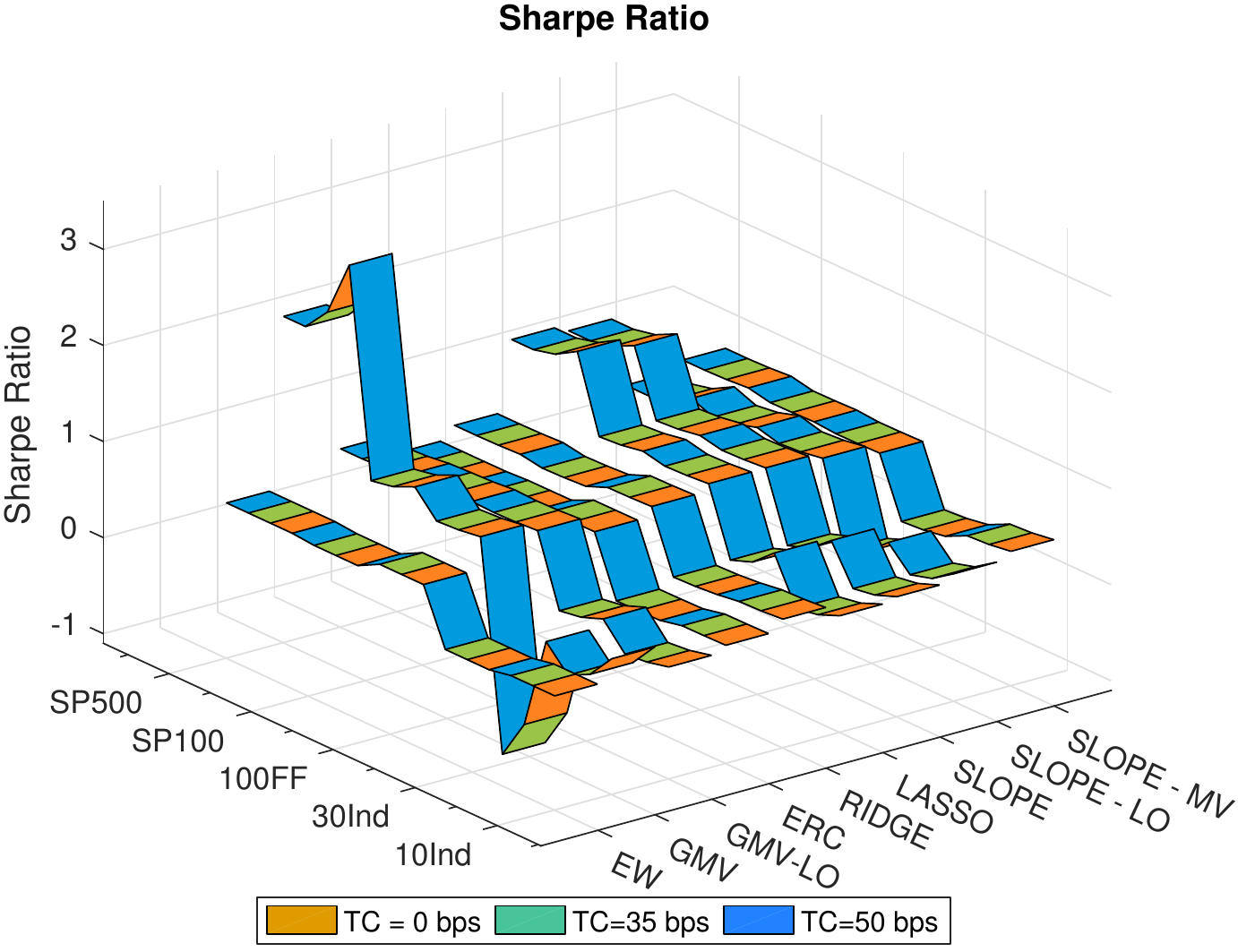} \\
\end{tabular}
}
\captionsetup{font=scriptsize,labelfont=scriptsize, width=\textwidth}
\caption*{The figure reports the portfolio return and Sharpe Ratio for the different transaction cost regimes across all considered strategies and data sets. We report results for three linear cost regimes, considering (1) no costs ($\text{TC} = 0 \text{bps}$), (2) low costs ($\text{TC}=35 \text{bps}$) and (3) high cost ($\text{TC}=50 \text{bps}$), which all are assumed to be linear and the same for buying and selling securities.}
\end{figure}

We investigate the robustness of our results by (a) imposing a linear transaction cost proportional to the turnover and (b) investigating the behavior of the SP indices for different window sizes.\\
For our transaction cost (TC) analysis, we consider three cost regimes: (1) no costs ($\text{TC} = 0 \text{bps}$), (2) low costs ($\text{TC}=35 \text{bps}$) and (3) high cost ($\text{TC}=50 \text{bps}$).\footnote{Note that, one basispoint (bps)$ = 0.01\%$.} Our set-up assumes that the transaction costs are linear in the turnover and are the same for selling and buying securities. Figure \ref{TC} shows the impact of an imposed market transaction cost for the five data sets, across the different portfolio strategies and considering all three transaction cost regimes. High turnover strategies like the GMV suffer with regard to returns and SR in the higher cost regimes and even report negative returns for the 30Ind. On the other hand, SLOPE portfolios show a nearly steady performance for all data set and when considering the different TC regimes.\\
%
%%%%%%%%%%%%%%%%%%%%%%%%%%%%%%%%%%%%%%%%%%
%%%%%%%%%%%%%%%%%%%%%%%%%%%%%%%%%%%%%%%%%%
%
Finally, we also investigate the performance of the portfolio methods across different window sizes. In detail, we consider a length of $\tau = 250$, $\tau = 750$, and $\tau = 1000$ for the SP Indices, as well as $\tau = 60$ for the 10Ind, 30Ind and 100FF portfolios. The results are robust to changes in the window size. Due to space limitations, we do not report the results here but make them available from the authors upon request.

%%%%%%%%%%%%%%%%%%%%%%
\section{Conclusion} %
%%%%%%%%%%%%%%%%%%%%%%

Regularization methods have gained increased attention in the financial literature, because they allow to reduce the influence of the estimation errors and to stabilize the resulting asset weight vector. In this paper, we extend the literature on financial regularization by introducing SLOPE to the mean-variance portfolio optimization discussing its properties and testing its performance with regard to risk and return on simulated and real world data.\\
SLOPE relies on the sorted $\ell_{1}$-Norm, whose intensity is controlled by a decreasing sequence of $\boldsymbol{\lambda}$ parameters. As the largest tuning parameter is assigned to the largest weight, SLOPE penalizes the assets by their rank, and provides a natural interpretation of importance. To solve the penalized mean-variance optimization, we propose a novel algorithm based on the Alternating Direction Method of Multipliers (ADMM). When applied for LASSO, which is a specific instance of SLOPE, this algorithm provides the same accuracy as the state-of-the-art CyCoDe, but is superior with regard to computing time, especially when the asset universe is large.\\
We study the properties of our new penalty function in a simulated environment and show that SLOPE has the advantage of still being active in a setting in which there are no short sales in the portfolio. Furthermore, SLOPE can automatically identify assets with the same underlying risk factor exposure and group them together, assigning them the same weight. This property is especially desirable for investor planning to incorporate their individual views into the optimization by selecting assets from the groups according to a specific financial characteristic or individual preference. We employ such method by introducing a simple investment strategy, SLOPE-MV, selecting from the groups the asset with the lowest volatility.\\
Moreover, we investigate the performance of SLOPE in an OOS setting, considering a rolling window approach, re-balancing the portfolio every month. The empirical analysis includes five major data sets, considering the 10Ind, 30Ind and 100FF, as well as the constituents of the S\&P100 and S\&P500. Our results show that SLOPE is able to achieve equal and even better OOS portfolio volatilities and SR, when compared to the LASSO. Although, only part of the differences are statistically significant, SLOPE is able to construct sparse portfolios with reduced turnover. This especially holds for situations in which we have a large amount of estimation error, like for the SP500. Furthermore, the grouping of equally correlated assets enables the investor to create investment strategies by picking assets according to certain characteristics from these groups. Establishing the SLOPE-MV portfolio, we provide a very sparse portfolio with even lower turnover than state-of-the art methods and at the same time maintaining a comparable performance.\\
With regard to weight and risk diversification measures, SLOPE outperforms the LASSO, reporting improved values for the DR, the WDiv and the RDiv. The shape of the penalty extends the frontier, ranging from the GMV via the GMV-LO, to the EW portfolio and to select among them the one that provides her with the desired volatility- and diversification trade-off.\\
The results support SLOPE as a valid alternative to the standard LASSO for creating sparse portfolios with a reduced turnover rate, improved risk- and weight diversification, and a high degree of flexibility in the portfolio construction process.\\
A natural extension to our study is to investigate, how different sequences of lambda parameters would impact the risk and portfolio allocation, and whether the investor should choose them according to the underlying correlation regime of the stock market or his own prior beliefs on the assets.
%%%%%%%%%%%%%%%%%%%%%%%%%%%%%%%%%%%%%%%%%%%%%%%%%%%%%%%
%\clearpage

%%%%%%%%%%%%%%%%%%%%%%%%%%%%%%%%%%%%%%%%%%%%%%%%%%%%%%%
%% APPENDIX %%
%%%%%%%%%%%%%%
\newpage
\section{Appendix}
\subsection{ADMM vs. Cyclic Coordinate Descend} \label{AlgoComp}
\begin{figure}
\centering
\caption{Computation Times for CyCoDe and ADMM.}\label{Comp_times}
\scalebox{0.9}{
\begin{tabular}{cc}
\includegraphics[scale=.58]{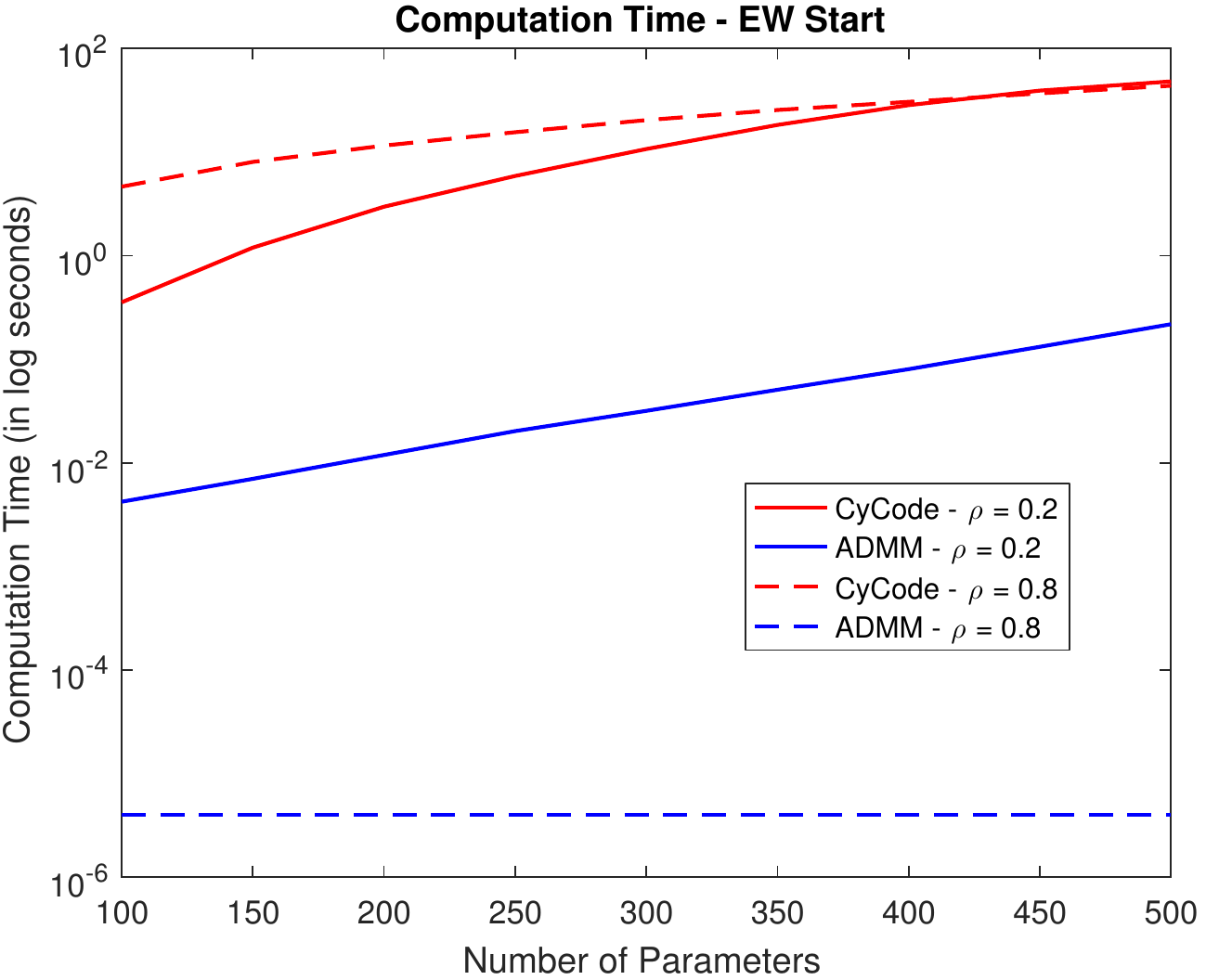} & \includegraphics[scale=.58]{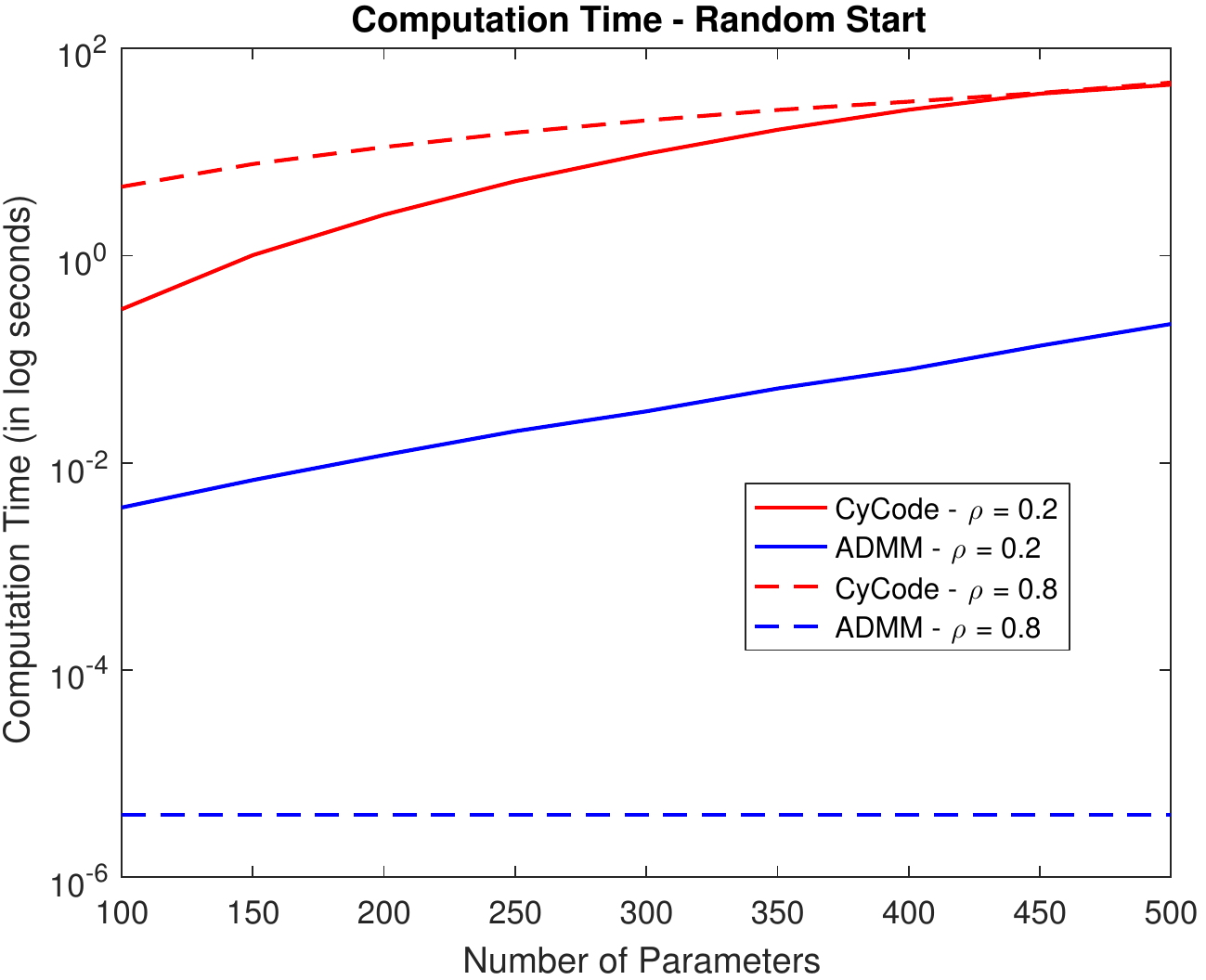}
\end{tabular}
}
\captionsetup{font=scriptsize,labelfont=scriptsize, width=\textwidth}
\caption*{The figure shows the average computation times needed for the CyCoDe and ADMM algorithm, depending on the correlation regime, the number of parameters and the soft start criterion. All values are based on 100 simulations, considering a constant correlation set-up.}
\end{figure}

In this section, we use the newly introduced ADMM algorithm for solving the minimum-variance optimization with an $\ell_{1}$ penalty (which is a specific instance of SLOPE penalty) and compare its performance to the one of the Cyclic Coordinate Descend algorithm (CyCoDe). CyCoDe algorithm is considered state-of-art and has found various applications in solving norm constrained optimization problems (see i.e. \cite{Fastrich2014}, \cite{Yen2015}).\\
The CyCoDe algorithm works by optimizing the weights along one coordinate direction, while holding all other weights constant. Although there is no general rule on how the CyCoDe updates the weight vector, we follow the procedure of \cite{Yen2015} and updating the weights cyclical, that is we first fix $w_i, \ i=2,...,k$ and find a new solution for $w_1$ that is closer to its optimal solution $w^{*}$. In a next step, we fix $w_i, \ i=1,3,...,k$ and find a value for $w_{2}$ that is again closer to the optimal one $w^{*}$. Given a starting criteria $\boldsymbol{w}^{0}$ for the weight vector, the Lagrange parameter for the budget constraint $\gamma$ and $\theta$, a trade-off parameter for $\mu$ and $\sigma^{2}$, Algorithm \ref{CyCoDe} shows the pseudo code for the CyCoDe.\\
\begin{algorithm}
\caption{Cyclic Coordinate Descend}\label{CyCoDe}
\begin{algorithmic}[1]
\STATE{Initialize $\boldsymbol{w}^{0}$ and $j=0$}
\WHILE{convergence criteria is not met}
\FOR{$i=1$ to $k$}\STATE
{$w_{i} = ST(\gamma-z_{i}, \lambda) \times (2 \times \sigma_{i}^{2})^{-1}$}
	\STATE{where $ST$ is the soft-thresholding function and $z_{i} = 2\sum_{j\neq i}^{k} w_{j}\sigma_{ij} - \theta \mu_{i}$}
\ENDFOR
\STATE{$j=j+1$}
\ENDWHILE
\end{algorithmic}
\end{algorithm}
To evaluate the performance of the two algorithms, we first draw a random sample of size $n$ for $k$ assets from a multivariate normal $\boldsymbol{X} \sim MVN(0, \boldsymbol{\Sigma})$, where $\boldsymbol{\Sigma}$:
\begin{gather}
\Sigma_{ij} =
 \begin{cases}
1, & i=j, \\
\rho, & i \neq j,
  \end{cases}
\end{gather}
and for which we choose $\rho=0.2$ and $0.8$, respectively. Then, we solve the minimum variance problem given in (\ref{eq:minreg}) and subject to the $\ell_{1}$- Norm on the weight vector, using as an input for $\Sigma$ the shrunken covariance matrix, introduced by \cite{Ledoit2004}.\\
We initialize both algorithms with a soft starting point $\boldsymbol{w}^{0}$, that is (1) $\boldsymbol{w}^{0}_{i} = \frac{1}{k}\ \forall \ i=1,...,k$, and (2) $\boldsymbol{w}^{0}_{i} = \frac{a_{i}}{\sum_{i=1}^{k} a_{i}}, \text{with}\ a_{i} \sim U(0, 1) \ \forall \ i=1,...,K$, and repeat the above procedure $100$ times, using for both algorithms a tolerance stopping point of $10^{-7}$. All computations are performed in Matlab 2016a on a Lenovo T430, with Windows 7, an Intel i7-3520M with 2.90 GHZ and 8 GB of RAM.\\
Table \ref{Simulation_EW} and \ref{Simulation_Random} display the minimum and the median of the objective function values, together with the median amount of shorting, the median time in seconds used for each algorithm to solve the $100$ simulations and the median absolute weight difference\footnote{The difference in the weights for is computed as: $\sum |\bfw^{ADMM} - \bfw^{CyCoDe}|$, where $\bfw^{ADMM}$ and $\bfw^{CyCoDe}$ are the optimal weights obtained with the ADMM and the CyCoDe algorithm, respectively.}, considering as soft starting point an equally weighted and a random portfolio weight vector, respectively.\footnote{Due to space limitations, we have restricted ourselves to report the above mentioned measures. Further results, including the standard deviation of the objective function value and the median number of active positions are available upon request to the authors.}
\noindent
%
%%%%%%%%%%%%%%%%%%%%%%%%%%%%%%%%%%%%%%%%%%%%%%%%%%%%%%%
\begin{table}[htbp]
  \centering
  \caption{Simulation Results - Equally Weighted.} \label{Simulation_EW}
  \scalebox{0.55}{
    \begin{tabular}{rccc|ccccc|ccccc|ccccc}
    \toprule
    \toprule
    \multicolumn{3}{c}{} &   & \multicolumn{5}{c|}{$\lambda = 4.03\times 10^{-6}$} & \multicolumn{5}{c|}{$\lambda= 5.65 \times 10^{-4}$} & \multicolumn{5}{c}{$\lambda = 7.91 \times 10^{-2}$} \\
    \midrule
    $\rho$ & $n$ & $p$ & Algo & Min & Med & Short & Time & W.Diff. & Min & Med & Short & Time & W.Diff. & Min & Med & Short & Time & W.Diff. \\
    \midrule
    \multicolumn{1}{c}{\multirow{6}[5]{*}{0.2}} & \multirow{2}[2]{*}{500} & \multirow{2}[2]{*}{100} & CyCoDe & 0.14 & 0.16 & 0.51 & 0.66 & \multirow{2}[2]{*}{$5 \times 10^{-7}$} & 0.14 & 0.16 & 0.49 & 0.62 & \multirow{2}[2]{*}{$5 \times 10^{-7}$} & 0.23 & 0.25 & 0.00 & 0.18 & \multirow{2}[2]{*}{$7 \times 10^{-8}$} \\
    \multicolumn{1}{c}{} &   &   & ADMM & 0.14 & 0.16 & 0.51 & 0.01 &   & 0.14 & 0.16 & 0.49 & 0.01 &   & 0.23 & 0.25 & 0.00 & 0.01 &  \\
    \multicolumn{1}{c}{} & \multirow{2}[2]{*}{500} & \multirow{2}[2]{*}{250} & CyCoDe & 0.09 & 0.11 & 2.13 & 13.63 & \multirow{2}[2]{*}{$8\times 10^{-6}$} & 0.09 & 0.11 & 2.02 & 12.87 & \multirow{2}[2]{*}{$6 \times 10^{-6}$} & 0.21 & 0.24 & 0.00 & 0.94 & \multirow{2}[2]{*}{$8 \times 10^{-8}$} \\
    \multicolumn{1}{c}{} &   &   & ADMM & 0.09 & 0.11 & 2.13 & 0.09 &   & 0.09 & 0.11 & 2.02 & 0.09 &   & 0.21 & 0.24 & 0.00 & 0.03 &  \\
    \multicolumn{1}{c}{} & \multirow{2}[1]{*}{1000} & \multirow{2}[1]{*}{500} & CyCoDe & 0.09 & 0.10 & 3.46 & 117.69 & \multirow{2}[1]{*}{$3 \times 10^{-5}$} & 0.09 & 0.11 & 3.23 & 116.29 & \multirow{2}[1]{*}{$2\times 10^{-5}$} & 0.22 & 0.24 & 0.00 & 5.58 & \multirow{2}[1]{*}{$1\times 10^{-7}$} \\
    \multicolumn{1}{c}{} &   &   & ADMM & 0.09 & 0.10 & 3.46 & 0.66 &   & 0.09 & 0.11 & 3.23 & 0.64 &   & 0.22 & 0.24 & 0.00 & 0.17 &  \\
    \midrule
    \multicolumn{1}{c}{\multirow{6}[5]{*}{0.8}} & \multirow{2}[1]{*}{500} & \multirow{2}[1]{*}{100} & CyCoDe & 0.55 & 0.64 & 3.39 & 11.67 & \multirow{2}[1]{*}{$2\times 10^{-3}$} & 0.55 & 0.65 & 3.30 & 11.23 & \multirow{2}[1]{*}{$2\times 10^{-3}$} & 0.73 & 0.83 & 0.00 & 1.37 & \multirow{2}[1]{*}{$8\times 10^{-7}$} \\
    \multicolumn{1}{c}{} &   &   & ADMM & 0.55 & 0.64 & 3.39 & 0.06 &   & 0.55 & 0.65 & 3.30 & 0.05 &   & 0.73 & 0.83 & 0.00 & 0.03 &  \\
    \multicolumn{1}{c}{} & \multirow{2}[2]{*}{500} & \multirow{2}[2]{*}{250} & CyCoDe & 0.34 & 0.42 & 10.98 & 35.33 & \multirow{2}[2]{*}{$8\times 10^{-1}$} & 0.35 & 0.43 & 10.46 & 34.75 & \multirow{2}[2]{*}{$8\times 10^{-1}$} & 0.67 & 0.82 & 0.00 & 6.03 & \multirow{2}[2]{*}{$1\times 10^{-6}$} \\
    \multicolumn{1}{c}{} &   &   & ADMM & 0.34 & 0.42 & 10.94 & 0.58 &   & 0.35 & 0.43 & 10.47 & 0.56 &   & 0.67 & 0.82 & 0.00 & 0.11 &  \\
    \multicolumn{1}{c}{} & \multirow{2}[2]{*}{1000} & \multirow{2}[2]{*}{500} & CyCoDe & 0.36 & 0.42 & 16.49 & 109.37 & \multirow{2}[2]{*}{$2.1$} & 0.38 & 0.44 & 15.44 & 107.64 & \multirow{2}[2]{*}{1.8} & 0.75 & 0.83 & 0.00 & 37.20 & \multirow{2}[2]{*}{$2\times 10^{-6}$} \\
    \multicolumn{1}{c}{} &   &   & ADMM & 0.36 & 0.42 & 16.34 & 3.96 &   & 0.38 & 0.43 & 15.33 & 3.76 &   & 0.75 & 0.83 & 0.00 & 0.61 &  \\
    \bottomrule
    \bottomrule
    \end{tabular}%
    }
\captionsetup{font=scriptsize,labelfont=scriptsize, width=\textwidth}
\caption*{The table reports for the Cyclic Coordinate Descend (CyCoDe) and the Alternating Direction Method of Multipliers(ADMM) the simulation results to the penalized minimum variance problem given in (\ref{eq:minreg}) considering six data sets drawn from a multivariate normal distribution, with $\rho =0.2$ and $\rho=0.8$, respectively, and using the equally weighted portfolio as a soft starting point. Stated are across all $100$ simulations: the minimum (Min) and median (Med) value of the objective function, the median value of the total amount of shorting (Short) the median time in seconds needed to compute the solution (Time) and the average weight difference (W.Diff.).}
\end{table}%
%%%%%%%%%%%%%%%%%%%%%%%%%%%%%%%%%%%%%%%%%%%%%%%%%%%%%%%
%
The tables show that both algorithms reach the same global minimum and median objective function value and the same amount of shorting for the low correlation environment, regardless of the chosen lambda value and whether we consider the equally weighted or a random weight vector as the soft starting point. The same also holds for the low dimensional data set, when the correlation is set to $\rho =0.8$. When $p=500$ for $\rho=0.8$, the ADMM reports a lower amount of shorting for the first two lambda values. This holds regardless how we choose the soft starting point. This difference might also explains the difference in the weight vectors, which is reported to be the highest for these two datasets. Still, the difference in the resulting weight vectors is modest and amounts to an average of $10^{-6}$ for both low correlation environments and $10^{-4}$, for the first two high correlation values and regardless on how we choose the soft starting point.\\
Most notably, the ADMM outperforms the CyCoDe, with regard to the median time in seconds used to compute the solution for all six data sets. This difference is not neglectable: the ADMM uses on average $0.265$ seconds in the low correlation environment across all lambdas and all starts, while the CyCoDe is slower by a factor of more than $100$, using on average $28.88$ seconds. The same holds for the high correlation environment, with the ADMM finding the solution in on average $2.65$ seconds and the CyCoDe using $38.98$ second.
Finally, and for both algorithms, selecting as a starting point the random weight vector results in longer computing times as opposed to using the equally weighted solution.
%
%%%%%%%%%%%%%%%%%%%%%%%%%%%%%%%%%%%%%%%%%%%%%%%%%%%%%%%
\begin{table}[htbp]
  \centering
  \caption{Simulation Results - Random Weights.}\label{Simulation_Random}
  \scalebox{0.55}{
    \begin{tabular}{rccc|ccccc|ccccc|ccccc}
    \toprule
    \toprule
    \multicolumn{3}{c}{} &   & \multicolumn{5}{c|}{$\lambda = 4.03\times 10^{-6}$} & \multicolumn{5}{c|}{$\lambda= 5.65 \times 10^{-4}$} & \multicolumn{5}{c}{$\lambda = 7.91 \times 10^{-2}$} \\
    \midrule
    $\rho$ & $n$ & $p$ & Algo & Min & Med & Short & Time & W.Diff & Min & Med & Short & Time & W.Diff & Min & Med & Short & Time & W.Diff \\
    \midrule
    \multicolumn{1}{c}{\multirow{6}[6]{*}{0.2}} & \multirow{2}[2]{*}{500} & \multirow{2}[2]{*}{100} & CyCoDe & 0.13 & 0.16 & 0.49 & 0.46 & \multirow{2}[1]{*}{$5\times 10^{-7}$} & 0.13 & 0.16 & 0.47 & 0.44 & \multirow{2}[1]{*}{$4\times 10^{-6}$} & 0.22 & 0.25 & 0.00 & 0.13 & \multirow{2}[2]{*}{$7\times 10^{-8}$} \\
    \multicolumn{1}{c}{} &   &   & ADMM & 0.13 & 0.16 & 0.49 & 0.01 &  & 0.13 & 0.16 & 0.47 & 0.01 &  & 0.23 & 0.25 & 0.00 & 0.01 &  \\
    \multicolumn{1}{c}{} & \multirow{2}[2]{*}{500} & \multirow{2}[2]{*}{250} & CyCoDe & 0.08 & 0.10 & 2.12 & 10.26 & \multirow{2}[1]{*}{$8\times 10^{-6}$} & 0.08 & 0.10 & 2.02 & 10.02 & \multirow{2}[2]{*}{$6\times 10^{-6}$} & 0.19 & 0.23 & 0.00 & 0.74 & \multirow{2}[2]{*}{$8\times 10^{-8}$} \\
    \multicolumn{1}{c}{} &   &   & ADMM & 0.08 & 0.10 & 2.11 & 0.07 &  & 0.08 & 0.10 & 2.02 & 0.07 &  & 0.19 & 0.23 & 0.00 & 0.02 &  \\
    \multicolumn{1}{c}{} & \multirow{2}[2]{*}{1000} & \multirow{2}[2]{*}{500} & CyCoDe & 0.08 & 0.10 & 3.50 & 111.66 & \multirow{2}[2]{*}{$3 \times 10^{-5}$} & 0.09 & 0.10 & 3.28 & 112.50 & \multirow{2}[2]{*}{$2 \times 10^{-5}$} & 0.22 & 0.24 & 0.00 & 5.31 & \multirow{2}[2]{*}{$1\times 10^{-7}$} \\
    \multicolumn{1}{c}{} &   &   & ADMM & 0.08 & 0.10 & 3.50 & 0.52 &  & 0.09 & 0.10 & 3.28 & 0.51 &  & 0.22 & 0.24 & 0.00 & 0.15 &  \\
        \midrule
    \multicolumn{1}{c}{\multirow{6}[6]{*}{0.8}} & \multirow{2}[2]{*}{500} & \multirow{2}[2]{*}{100} & CyCoDe & 0.55 & 0.64 & 3.30 & 8.02 & \multirow{2}[2]{*}{$2\times 10^{-3}$} & 0.56 & 0.64 & 3.21 & 7.86 & \multirow{2}[2]{*}{$2\times 10^{-3}$} & 0.72 & 0.82 & 0.00 & 0.89 & \multirow{2}[2]{*}{$8 \times 10^{-7}$} \\
    \multicolumn{1}{c}{} &   &   & ADMM & 0.55 & 0.63 & 3.30 & 0.03 & & 0.55 & 0.64 & 3.21 & 0.03 &  & 0.72 & 0.82 & 0.00 & 0.02 &  \\
    \multicolumn{1}{c}{} & \multirow{2}[2]{*}{500} & \multirow{2}[2]{*}{250} & CyCoDe & 0.33 & 0.41 & 10.77 & 31.54& \multirow{2}[2]{*}{$8\times 10^{-1}$} & 0.35 & 0.42 & 10.34 & 32.05 & \multirow{2}[2]{*}{$8 \times 10^{-1}$} & 0.68 & 0.81 & 0.00 & 5.35 & \multirow{2}[2]{*}{$1\times 10^{-6}$} \\
    \multicolumn{1}{c}{} &   &   & ADMM & 0.33 & 0.41 & 10.75 & 0.55 &  & 0.35 & 0.42 & 10.33 & 0.53 &  & 0.68 & 0.81 & 0.00 & 0.10 & \\
    \multicolumn{1}{c}{} & \multirow{2}[2]{*}{1000} & \multirow{2}[2]{*}{500} & CyCoDe & 0.36 & 0.40 & 16.42 & 111.10 & \multirow{2}[1]{*}{2.193} & 0.37 & 0.42 & 15.38 & 111.70 & \multirow{2}[2]{*}{1.99} & 0.76 & 0.82 & 0.00 & 38.7 & \multirow{2}[2]{*}{1.81} \\
    \multicolumn{1}{c}{} &   &   & ADMM & 0.36 & 0.40 & 16.37 & 3.89 &  & 0.37 & 0.42 & 15.36 & 3.69 &  & 0.76 & 0.82 & 0.00 & 0.60 & \\
    \bottomrule
    \bottomrule
    \end{tabular}%
    }
\captionsetup{font=scriptsize,labelfont=scriptsize, width=\textwidth}
\caption*{The table reports for the Cyclic Coordinate Descend (CyCoDe) and the Alternating Direction Method of Multipliers(ADMM) the simulation results to the penalized minimum variance problem given in (\ref{eq:minreg}) considering six data sets drawn from a multivariate normal distribution, with $\rho =0.2$ and $\rho=0.8$, respectively, and using the equally weighted portfolio as a soft starting point. Stated are across all $100$ simulations: the minimum (Min) and median (Med) value of the objective function, the median value of the total amount of shorting (Short) the median time in seconds needed to compute the solution (Time) and the average weight difference (W.Diff.).}
\end{table}%
%%%%%%%%%%%%%%%%%%%%%%%%%%%%%%%%%%%%%%%%%%%%%%%%%%%%%%%
%
Figure \ref{Comp_times} plots the computing times needed for the CyCoDe and the ADMM for both the EW and Random weight vector initialization, considering the two correlation regimes and varying the number of parameters that have to be estimated. Clearly the ADMM consistently shows a superior performance, by only using a fraction of the time of the CyCoDe. Furthermore, we can observe that both algorithms are also invariant to the selection of the soft starting point. Only the CyCoDe shows a slight difference for parameter values above $k=450$, signalling that an EW portfolio results in finding the optimal solution faster. The ADMM algorithm is able to reach the same performance as the currently state-of-the-art CyCoDe algorithm, by using only a fraction of the computing time.
%
%%%%%%%%%%%%%%%%%%%%%%%%%%%%%%%
%%%%%%%%%%%%%%%%%%%%%%%%%%%%%%%
%

\subsection{Portfolio Selection Models}
\subsubsection*{Equally Weighted Portfolio}
The equally weighted portfolio is considered as one of the toughest benchmarks to beat (see, i.e. \cite{DeMiguel2009}), and naively distributes the wealth equally among all constituents, such that with $k$ assets:
\begin{gather}
w_i = \frac{1}{k} \ \forall	\ i=\{1,...,k\},
\end{gather}
where $w_i$ is the weight of asset $i$. The EW ignores both the variances, the covariances and the return of the assets, and is the optimal portfolio on the mean-variance efficient frontier, when we assume that all three are the same.
\subsubsection*{Norm-Constrained Minimum Variance Portfolio}
Reconsider the formulation of the mean-variance problem in (\ref{minvar}). By disregarding the mean in the optimization, we obtain the Global Minimum Variance Portfolio (GMV), given by:
\begin{gather} \label{GMV}
\min_{\bfw \in \mathbb{R}^{k}} \sigma_{p}^{2} = \bfw'\Sigma\bfw \hspace{0.5cm}
s.t. \sum_{i=1}^{k} w_{i} = 1,\ \forall \ i=\{1,...,k\},
\end{gather}
However, this formulation is prone to estimation errors, and unstable portfolio weights. To circumvent these problems, we extend the framework in (\ref{GMV}) by adding a penalty function $\rho_{\lambda}(\bfw)$ on the weight vector. For LASSO, we add a $\ell_{1}$ - Norm to the formulation in (\ref{GMV}), such that:
\begin{gather}
\rho_{\lambda}(\bfw) = \lambda \times \sum_{i=1}^{k} |w_{i}|
\end{gather}
where $\lambda$ is a regularization parameter that controls the intensity of the penalty. Besides LASSO, we also consider the RIDGE penalty, which adds an $\ell_{2}$-Norm on the weight vector to the formulation in (\ref{GMV}), and that takes the form of:
\begin{gather}
\rho_{\lambda}(\bfw) = \lambda \times \sum_{i=1}^{k} w_{i}^{2}
\end{gather}
As opposed to the LASSO, the RIDGE is not singular at the origin and thus does not promote sparse solutions. Still, imposing the $\ell_{2}$ - Norm on the portfolio problem is equal to adding an identity matrix, weighted by the regularization parameter $\lambda$ to the inverse of the variance-covariance matrix, i.e. $(\bfSigma^{-1} + \lambda \boldsymbol{I})$, where $\boldsymbol{I}$ is the $k \times k$ identity matrix. This leads to more numerical stability and makes the RIDGE penalty especially appealing in environments that suffer from multicollinearity \citep{Zou2005}.
\subsubsection*{Equal Risk Contribution Portfolio}\label{ERC}
Finally, we consider the Equal Risk Contribution (ERC) portfolio, which aims to equalize the marginal risk contributions of the assets to the overall portfolio risk. That is, given that portfolio variance can be decomposed as:
\begin{gather}
\sigma_{p}^{2} = \sum_{i=1}^{k}\sum_{j=1}^{k} w_i w_{j}\sigma_{ij}=\sum_{i=1}^{k}w_{i}\sum_{j=1}^{k}w_{j}\sigma_{ij}
\end{gather}
the marginal contribution to the portfolio risk for asset $i$ is given as:
\begin{gather}
c_{i}^{var} = w_{i}\sum_{j=1}^{k}w_{j}\sigma_{ij} = w_{i}\ (\bfSigma \bfw)_{i} \ \  \text{with} \ \  \sum_{i=1}^{k}c_{i}^{var}=\sigma_{p}^{2}
\end{gather}
where $(\bfSigma \bfw)_{i}$ denotes the $i^{th}$ row of the product of $\bfSigma$ and $\bfw$ \citep{Roncalli2013}. As the marginal risk is dependent on the portfolio weight magnitude, the ERC portfolio has no analytically solution and must be obtained numerically,by solving:
\begin{gather}
\min_{\bfw \in \mathbb{R}^{N}} \sum_{i=1}^{k}(\frac{w_{i}\ (\bfSigma \bfw)_{i}}{\sigma_{p}^{2}}-\frac{1}{k})^{2} \ \ 
s.t. \ \ \sum_{i=1}^{k} w_{i}=1, \ \ 0 \leq w_{i} \leq 1 \ \ \forall \ i \in \{1, 2, ..., k \}
\end{gather}
The ERC favors assets with lower volatility, lower correlation with other assets, or both, and is less sensitive to small changes in the covariance matrix as compared to the GMV portfolio \citep{Kremer2017}. Furthermore, \citep{Maillard2010} show that the volatility of the ERC is between that of the EW and the GMV, and that it coincides with the latter, when both, correlations and SRs, are assumed to be equal \citep{Maillard2010}.

%\end{APPENDICES}

%%%%%%%%%%%%%%%%%
% Acknowledgments here
\section*{Acknowledgement}
Ma\l{}gorzata Bogdan acknowledges the grant of the Polish National Center of Science Nr 2016/23/B/ST1/00454, and together with Sandra Paterlini further acknowledge ICT COST Action IC1408 from CRoNoS. Sangkyun Lee acknowledges the support of the National Research Foundation of Korea (NRF) grant funded by the Korea government (MSIP; Ministry of Science, ICT \& Future Planning) (No. 2017R1C1B5018367).
%%%%%%%%%%%%%%%%%

%%%%%%%%%%%%%%%%%
% References
\bibliography{bibslope} % if more than one, comma separated
\bibliographystyle{informs2014} % outcomment this and next line in Case 1
%%%%%%%%%%%%%%%%%

%%%%%%%%%%%%%%%%%
\end{document}